\documentclass[11pt,a4paper]{article}

\pdfoutput=1

\usepackage{jheppub}

\usepackage[utf8]{inputenc}
\usepackage{booktabs}
\usepackage{amsmath} 
\usepackage{amssymb}
\usepackage{mathtools}
\usepackage{hhline}
\usepackage{url}
\usepackage{pdflscape}
\usepackage{caption}
\usepackage[numbers,sort&compress]{natbib}
\usepackage{epsfig}
\usepackage{multirow} 
\usepackage{color}
\usepackage{verbatim}
\usepackage{nicefrac}
\usepackage{upgreek}
\usepackage{bbm}
\usepackage{setspace}
\usepackage{pstricks}
\usepackage{psfrag}
\usepackage{color}
\usepackage{pdfrender}
\usepackage{subfigure}
\usepackage{enumerate}
\usepackage[mathscr]{euscript}

\definecolor{steelblue}{rgb}{0.27, 0.51, 0.71}
\definecolor{goldenrod}{rgb}{0.85, 0.65, 0.13}

\newcommand{\slashed}[1]{\displaystyle{\not}{#1}}

\newcommand{\GeV}{{\rm GeV}}

\newcommand{\eV}{{\rm eV}}

\usepackage{hyperref}
\newcommand{\ee}{e}

\renewcommand{\Re}{\text{Re}}
\newcommand{\im}{{\rm Im}}

\renewcommand{\Im}{{\rm Im}}
\renewcommand{\k}{\textbf{k}}

\newcommand{\ii}{{\rm i}}
\newcommand{\dd}{d}
\newcommand{\U}{\rm U}

\newcommand{\bse}{\begin{subequations}}
\newcommand{\ese}{\end{subequations}}

\definecolor{green}{rgb}{0,0.5,0}  

\title{Probing Leptogenesis at Future Colliders}

\author[a,b]{Stefan Antusch}
\author[a]{Eros Cazzato}
\author[c,d,e]{Marco Drewes}
\author[a,f]{Oliver Fischer}
\author[d]{Bj\"orn~Garbrecht}
\author[b,d,e]{Dario Gueter}
\author[d,e]{Juraj Klari\'{c}}

\affiliation[a]{Department of Physics, University of Basel,\\ Klingelbergstr.\ 82, CH-4056 Basel, Switzerland}
\affiliation[b]{Max-Planck-Institut f\"ur Physik (Werner-Heisenberg-Institut),\\
F\"ohringer Ring 6, 80805 M\"unchen, Germany}
\affiliation[c]{Centre for Cosmology, Particle Physics and Phenomenology, Universit\'{e} catholique de Louvain, Louvain-la-Neuve B-1348, Belgium}
\affiliation[d]{Physik Department T70, Technische Universit\"at M\"unchen,\\
James Franck Stra\ss e 1, 85748 Garching, Germany}
\affiliation[e]{Excellence Cluster Universe,\\
Boltzmannstra{\ss}e 2, 85748 Garching, Germany}
\affiliation[f]{Institute for Nuclear Physics, Karlsruhe Institute of Technology, \\
Hermann-von-Helmholtz-Platz 1, D-76344 Eggenstein-Leopoldshafen, Germany}

\emailAdd{stefan.antusch@unibas.ch}
\emailAdd{e.cazzato@unibas.ch}
\emailAdd{marco.drewes@uclouvain.be}
\emailAdd{oliver.fischer@kit.edu}
\emailAdd{garbrecht@tum.de}
\emailAdd{dario.gueter@tum.de}
\emailAdd{juraj.klaric@tum.de}

\abstract{
We investigate the question whether leptogenesis, as a mechanism for explaining the baryon asymmetry of the universe, can be tested at future colliders. Focusing on the minimal scenario of two right-handed neutrinos, we identify the allowed parameter space for successful leptogenesis in the heavy neutrino mass range between $5$ and $50\, \GeV$. Our calculation includes the lepton flavour violating contribution from heavy neutrino oscillations as well as the lepton number violating contribution from Higgs decays to the baryon asymmetry of the universe.
We confront this parameter space region with the discovery potential for heavy neutrinos at future lepton colliders, which can be very sensitive in this mass range via displaced vertex searches. Beyond the discovery of heavy neutrinos, we study the precision at which the flavour-dependent active-sterile mixing angles can be measured. The measurement of these mixing angles at future colliders can test whether a minimal type I seesaw mechanism is the origin of the light neutrino masses, and it can be a first step towards probing leptogenesis as the mechanism of baryogenesis.
We discuss how a stronger test could be achieved with an additional measurement of the heavy neutrino mass difference.
}

\arxivnumber{}

\begin{document}
\begin{flushright}
TUM-1160/18\\
CP3-17-48
\end{flushright}

\maketitle

\section{Introduction}
\label{Section1_Introduction}
\paragraph{Motivation.} The Standard Model (SM) of particle physics allows to describe almost all phenomena in nature at a fundamental level \cite{Patrignani:2016xqp}. Neutrino flavour oscillations, which clearly indicate the existence of neutrino masses, are the only phenomenon observed in the laboratory that points unambiguously towards the existence of new states beyond the SM.
If the neutrino masses are at least partly generated by the Higgs mechanism in the same way as the masses of all other fermions, then this necessarily implies the existence of \emph{right handed neutrinos} $\nu_{\rm R}$. 
Since the $\nu_{\rm R}$ are singlets under the gauge symmetries of the SM, they are allowed to have a Majorana mass term $-\frac{1}{2}\overline{\nu_{{\rm R}}^c}M\nu_{{\rm R}}$ in addition to the usual Dirac mass term that is generated by the Higgs mechanism. If there are $n_s$ right-handed neutrinos $\nu_{{\rm R}i}$, then $M$ is an $n_s\times n_s$ flavour matrix with eigenvalues $M_i$.

These Majorana masses $M_i$ are free parameters. Their magnitude cannot be fixed by neutrino oscillation data alone, since light neutrino oscillations are only sensitive to a particular combination of $M_i$ and the right-handed neutrinos' Yukawa couplings $Y_{ia}$ to SM leptons of flavour $a$, cf. eq.~(\ref{eq:seesaw_relation}). 
The range of masses allowed by neutrino oscillation data is in principle very large; 
even in the minimal model with $n_s=2$ it reaches from the eV scale \cite{deGouvea:2005er} up to values 1-2 orders of magnitude below the Planck mass \cite{Asaka:2015eda}.
The implications of the existence of right-handed neutrinos for particle physics and cosmology strongly depend on the magnitude of $M$, see e.g. ref.~\cite{Drewes:2013gca} for a review.
Traditionally it is often assumed that the $M_i$ are much larger than the electroweak scale.
In this case the smallness of the observed neutrino masses can be explained by the usual \emph{seesaw mechanism} \cite{Minkowski:1977sc,GellMann:1980vs,Mohapatra:1979ia,Yanagida:1980xy,Schechter:1980gr,Schechter:1981cv}, i.e., the smallness of $v/M_i$, where $v=174\, \GeV$ is the vacuum expectation value of the Higgs field.  
However, this choice is entirely based on theoretical arguments that e.g. relate $M_i$ to the scale of grand unification.

Alternatively one could e.g. argue that the $M_i$ and electroweak scale have a common origin \cite{Iso:2009ss,Iso:2012jn,Khoze:2013oga,Khoze:2016zfi}.  
A value of $M_i$ at (comparably) low scales is technically natural because $B-L$ becomes a symmetry of the model in the limit where all $M_i$ vanish, and large radiative corrections to the Higgs mass that plague high scale seesaw models can be avoided. 
Many \emph{low scale seesaw} models indeed involve an approximate ``lepton number''-like symmetry. 
The smallness of the light neutrino masses in these \emph{symmetry protected seesaw} scenarios does not primarily come from the seesaw suppression by the parameters $v/M_i$ (which are not small), but is due to the smallness of the symmetry violation, cf.\ eqs.~\eqref{B-L_parameters}.\footnote{One may argue that the name ``seesaw'' is not appropriate for scenarios with $M_i/v\lesssim 1$. However, it is common to refer to the Lagrangian (\ref{eq:Lagrangian}) with the parameter choice $M<$ TeV as \emph{low scale seesaw}, and we adopt this nomenclature throughout this paper.} A benchmark scenario can be found in \cite{Antusch:2015mia}.

The probably most studied model that involves a (sub-) electroweak scale seesaw is the \emph{Neutrino Minimal Standard Model} ($\nu$MSM) \cite{Asaka:2005an,Asaka:2005pn}.  This was indeed the model in which it was first discovered that leptogenesis is feasible for $M_i<v$ in the minimal scenario with $n_s=2$ \cite{Asaka:2005pn}, which is the setup that we are concerned with in the present paper. The $\nu$MSM realises the an approximate $B-L$ symmetry in part of its parameter space \cite{Shaposhnikov:2006nn}.
Other popular models of this type, are the ``inverse seesaw models'' \cite{Wyler:1982dd,Mohapatra:1986bd,Mohapatra:1986aw,GonzalezGarcia:1988rw}, ``linear seesaw models'' \cite{Akhmedov:1995vm,Akhmedov:1995ip,Barr:2003nn,Malinsky:2005bi} (see also \cite{Bernabeu:1987gr,Pilaftsis:1991ug,Abada:2007ux,Sierra:2012yy,Fong:2013gaa}), and ``minimal flavour violation'' \cite{Cirigliano:2005ck,Gavela:2009cd}, see also recent numerical implementations of different models that include radiative corrections~\cite{Ruiz:2015zca,Degrande:2016aje}. Related frameworks, which can be tested at future colliders, are left-right symmetric models \cite{Lindner:2016lxq}.
In the present paper we take an agnostic approach to the magnitude of the $M_i$ and focus on the mass range that is accessible to near future experiments.

The Yukawa couplings $Y$ of the right-handed neutrinos $\nu_R$ in general violate CP, while $M$ violates lepton number $L$. 
Since lepton number $L$ and baryon number $B$ can be converted into each other by electroweak sphaleron processes in the early universe \cite{Kuzmin:1985mm},
this opens up the possibility that they are the origin of the \emph{baryon asymmetry of the universe} (BAU) \cite{Fukugita:1986hr} and thereby solve one of the great mysteries in cosmology that cannot be explained within the SM.\footnote{An overview of the evidence for a matter-antimatter asymmetry in the observable universe is given in ref.~\cite{Canetti:2012zc}.} The experimentally observed BAU can be quantified by the baryon-to-entropy ratio~\cite{Ade:2015xua}
\begin{align}
\label{eq:Y_B}
Y_{B\,{\rm obs}} = (8.6 \pm 0.1) \times 10^{-11}\,.
\end{align}
While leptogenesis generically requires rather large $M_i$ \cite{Davidson:2002qv}, an approximate $B-L$ symmetry can alleviate this requirement in different ways \cite{Pilaftsis:2003gt,Asaka:2005pn,Asaka:2008bj,Racker:2012vw}.   
For $M_i$ below the TeV scale, and within the \emph{minimal seesaw model}, there are three different mechanisms that generate lepton asymmetries.
For $M_i$ above the electroweak scale, the baryon asymmetry can be generated in heavy neutrino decays via \emph{resonant leptogenesis} \cite{Pilaftsis:2003gt}. The lower bound on the mass comes from the requirement that the heavy neutrinos freeze out and decay before sphalerons freeze out at $T_{\rm sph}\sim 130\, \GeV$ \cite{DOnofrio:2014rug}. 
For masses $M_i$ below the electroweak scale, the baryon asymmetry can be produced via CP-violating oscillations of the heavy neutrinos during their production (instead of their decay) \cite{Akhmedov:1998qx,Asaka:2005pn}. This mechanism is also known as \emph{baryo- or leptogenesis from neutrino oscillations}.
Finally, there is a contribution to the asymmetries from the lepton number violating (LNV) decays of Higgs quasiparticles with large thermal masses into $\nu_{{\rm R}i}$ and SM leptons \cite{Hambye:2016sby,Hambye:2017elz}.

\paragraph{Goals of this work.} In the present paper we investigate the perspectives to probe low scale leptogenesis in the minimal seesaw model with $n_s=2$ at the proposed future lepton colliders, the electron-positron mode of the Future Circular Collider (FCC-ee) at CERN \cite{Gomez-Ceballos:2013zzn}, the Circular Electron Positron Collider (CEPC) in China \cite{CEPC-SPPCStudyGroup:2015csa} and the International Linear Collider (ILC) in Japan  \cite{Baer:2013cma,Brau:2015ppa}. 

We focus on the mass range $5\, {\rm GeV} < M_i < 80\, \GeV$, i.e., on heavy neutrinos that are heavier than b-mesons and lighter than $W$ bosons.\footnote{
With masses below $5\, \GeV$, the heavy neutrinos could be produced in the decay of mesons, and fixed target experiments like NA62 \cite{Talk:Spadaro,CortinaGil:2017mqf,Drewes:2018gkc} or SHiP \cite{Graverini:2015dka,
Graverini:2214085} and b-factories \cite{Antusch:2017hhu} have better chances to discover them.
The potential to search for heavy neutrinos with larger masses has recently e.g. been studied for both, hadron
~\cite{Kersten:2007vk,delAguila:2008cj,Atre:2009rg,BhupalDev:2012zg,Gago:2015vma,Das:2012ze,Anamiati:2016uxp,Dev:2015kca,Ruiz:2015gsa,Das:2015toa,Das:2017nvm,Das:2016hof,Das:2017pvt} and lepton colliders~\cite{Chen:2011hc,Asaka:2015oia,Das:2016hof,Das:2012ze,Banerjee:2015gca,Antusch:2015mia,Antusch:2015gjw,Abada:2015zea}.
}
In this regime, a lepton collider offers an ideal tool to search for heavy neutrinos \footnote{
Previous studies suggest that the LHC cannot probe the parameter region where leptogenesis is possible in the minimal model~ \cite{Helo:2013esa,Izaguirre:2015pga,Gago:2015vma,Dib:2015oka,Dib:2016wge}, or may just touch it \cite{Antusch:2017hhu}, but this statement has recently been put into question for two reasons. First, searching for a wider range of processes and improving the triggers may considerably increase the LHC sensitivity \cite{Cottin:2018nms,Abada:2018sfh}.
Further improvement could be achieved with additional detectors \cite{Kling:2018wct,Helo:2018qej,Curtin:2018mvb}.
Second, recent studies \cite{lucente_michele_2018_1289773} confirm earlier claims \cite{Canetti:2014dka} that leptogenesis is feasible for much larger heavy neutrino couplings (and hence larger production rates at colliders) if three of them participate in the process.
}
because large numbers of them can be produced (see e.g.\ \cite{Das:2012ze,
Blondel:2014bra,
Asaka:2015oia,
Antusch:2015mia,
Banerjee:2015gca,
Antusch:2015gjw,
Abada:2015zea,
Antusch:2015rma,
Antusch:2016vyf,
Antusch:2016ejd}) from on-shell $Z$ bosons at the so-called $Z$ pole run of a lepton collider, or from $W$ boson exchange at higher center-of-mass energies.
The potential to probe 
this mass range with lepton colliders has previously been studied in the literature, cf. e.g. refs.~\cite{Blondel:2014bra,Antusch:2015mia,Abada:2014cca,Asaka:2015oia,Graverini:2015dka,Antusch:2016vyf,Antusch:2016ejd,Caputo:2016ojx} and references therein, while the viable leptogenesis parameter region in the minimal model with $n_s=2$ has e.g. been studied in refs.~\cite{Canetti:2010aw,Canetti:2012vf,Canetti:2012kh,Canetti:2014dka,Hernandez:2015wna,Abada:2015rta,Hernandez:2016kel,Drewes:2016gmt,Drewes:2016jae,Abada:2017ieq,Hambye:2016sby,Hambye:2017elz}.
We improve past studies of both, the collider sensitivity and different aspects of the leptogenesis computation, in several ways:
\begin{itemize}
	\item In comparison to the studies in refs.~\cite{Canetti:2010aw,Canetti:2012vf,Drewes:2012ma,Canetti:2012kh,Canetti:2014dka,Hernandez:2015wna,Abada:2015rta,Hernandez:2016kel,Drewes:2016lqo,Drewes:2016gmt,Asaka:2016zib,Drewes:2016jae,Asaka:2017rdj,Abada:2017ieq}, we systematically include both, the lepton flavour violating thermal scatterings and the LNV decays and inverse decays in our scan of the leptogenesis parameter region during the symmetric phase of the SM. The importance of the LNV terms in the symmetric phase has recently been emphasised in refs.~\cite{Hambye:2016sby,Hambye:2017elz} and in the broken phase in refs.~\cite{Ghiglieri:2017gjz,Eijima:2017anv}.
\item The interaction strength of the heavy neutrinos with SM leptons of flavour $a$ can be characterised by an active-sterile mixing angle $U_a^2$. In previous studies, the potential of future experiments to discover leptogenesis has been estimated by comparing the projection of the viable leptogenesis parameter region on the $M_i - U_a^2$ planes to the projection of the experimental sensitivity on these planes.
This comparison of projections is not fully consistent because both, low scale leptogenesis and the experimental sensitivities, depend not only on the total interaction strength $U^2=\sum_a U_a^2$, but also on the relative size of $U_e^2$, $U_\mu^2$ and $U_\tau^2$.\footnote{In fact, the experimental sensitivities can only be calculated for fixed ratios of $U_e^2$, $U_\mu^2$ and $U_\tau^2$.}
For instance, the fact that a given combination of $M_i$ and $U_\mu^2$ is consistent with both (successful leptogenesis and an observable event rate at a collider) does not necessarily imply that those heavy neutrinos that are able to generate the BAU can actually be discovered at the collider because the parameter points for which one or the other can be realised may correspond to very different $U_e^2$. Using measurements of the $U_a^2$ at a lepton collider in order to
   determine  phases in $U_\nu$ has been investigated in~\cite{Caputo:2016ojx}. We  note that for leptogenesis, not only the relation between the $U_a^2$ and the amount of CP violation but also the washout of the different active lepton flavours is crucial.
In the present analysis, we calculate the BAU and the expected number of events for each point in the model parameters space to assess the question whether both requirements can be fulfilled simultaneously in a consistent manner.
\item We estimate how precisely the experiments can measure the magnitude of the individual $U_a^2$, extending previous studies \cite{Caputo:2017pit} that were focussed on SHiP and the FCC-ee. 
If any heavy neutral leptons are discovered in future experiments, the relative size of the $U_a^2$ provides a powerful test whether these particles can generate the light neutrino masses and the BAU in the minimal seesaw model with two RH neutrinos (cf.\ \cite{Hernandez:2016kel,Drewes:2016jae} for previous studies). 
The sensitivity of an experiment to individual $U_a^2$ is therefore important to assess the experiment's potential to not only discover heavy neutral leptons, but understand their role in particle physics and cosmology. 
\item In addition, we also investigate which values of the heavy neutrino mass splitting are consistent with leptogenesis and discuss strategies to measure them at future colliders. 
\end{itemize}

This article is organised as follows: In section \ref{Section2_TheModel} we recapitulate the basics of the seesaw mechanism and the symmetry protected seesaw scenario. 
In section \ref{Section3_Leptogenesis} we provide details of our scan of the leptogenesis parameter space. 
In section \ref{Section4_measurement} we specify our approach to assess the reach of future lepton colliders in terms of the $M_i$ and $U_a^2$.
In section \ref{Section5_Results} we present and discuss the results of our numerical analysis.
We conclude in section \ref{Section6_Conclusions}. In appendix \ref{Appendix_1} we give details on the
expected number of events at the given lepton collider, while the statistics behind the precision of the measurements of the mixings is explained in appendix \ref{Appendix_2}. A derivation of the evolution equations for the heavy neutrinos including lepton number violating processes is given in appendix \ref{Appendix_3}. We briefly overview the numerical methods used to deal with stiff differential equations in \ref{Appendix_4}.


\section{The model}
\label{Section2_TheModel}

\subsection{The type-I seesaw model}  
The Lagrangian of the type-I seesaw model, 
\begin{align}
\label{eq:Lagrangian}
\mathscr{L}=\mathscr{L}_{\rm SM}
+{\rm i} \overline{\nu_{{\rm R} i}}\partial\!\!\!/\nu_{{\rm R} i}
-\frac{1}{2}(\overline{\nu_{{\rm R} i}^c}M_{ij}\nu_{{\rm R} j} +  \overline{\nu_{{\rm R} i}}M_{ji}^*\nu_{{\rm R} j}^c)-Y_{ia}^*\overline{\ell_a}\varepsilon\phi \nu_{{\rm R} i}
-Y_{ia}\overline{\nu_{{\rm R} i}}\phi^\dagger \varepsilon^\dagger \ell_a\,,
\end{align}
is obtained by extending the SM Lagrangian $\mathscr{L}_{\rm SM}$ by $n_s$ right handed (RH) spinors $\nu_{{\rm R} i}$, which  represent the RH neutrinos. The interaction with the SM only happens via the Yukawa interactions $Y_{ia}$ with the SM lepton doublets $\ell_a$ for $a=e,\mu,\tau$ and the Higgs field $\phi$. The superscript $c$ appearing on the RH neutrinos denotes charge conjugation and $\varepsilon$ is the antisymmetric SU(2)-invariant tensor with $\varepsilon_{12}=1$. Further $M_{ij}$ are the entries of the Majorana mass matrix $M$ of the RH neutrinos with eigenvalues $M_i$.

At tree level the connection between the Lagrangian~(\ref{eq:Lagrangian}) and the low energy oscillation data can be provided by the Casas-Ibarra parametrisation \cite{Casas:2001sr}
\begin{align}\label{CasasIbarraDef}
Y^\dagger=\frac{1}{v}U_\nu\sqrt{m_\nu^{\rm diag}}\mathcal{R}\sqrt{M^{\rm diag}}\,.
\end{align}
Here $(m_\nu^{\rm diag})_{ab} = \delta_{ab} m_a$ denotes the light neutrino mass matrix in the mass basis, i.e., $m_a$ are the light neutrino masses.  The mixing matrix of the light neutrinos is given by
\begin{eqnarray}\label{VnuDef}
V_\nu= \left(\mathbbm{1}-\frac{1}{2}\theta\theta^{\dagger}+\mathcal{O}\left(\theta^4\right)\right)U_\nu\,,
\end{eqnarray} 
where
\begin{align}
\theta = v Y^\dagger M^{-1},
\end{align}
with $v$ the temperature dependent Higgs expectation value evaluated at zero temperature: $v = 174\,\GeV$.
The unitary matrix $U_\nu$  diagonalises the light neutrino mass matrix
\begin{align}
\label{eq:seesaw_relation}
m_\nu = v^2 Y^\dagger M^{-1} Y^*\,. 
\end{align}
as $m_\nu=U_\nu m_\nu^{\rm diag} U_\nu^T$. $U_\nu$ can be expressed as
\begin{align}
\label{PMNS}
U_\nu=V^{(23)}U_\delta V^{(13)}U_{-\delta}V^{(12)}{\rm diag}(e^{\ii \alpha_1/2},e^{\ii \alpha_2 /2},1)\,,
\end{align}
with $U_{\pm \delta}={\rm diag}(1,\ee^{\mp \ii \delta /2},\ee^{\pm \ii \delta /2})$. The non-vanishing entries of $V^{(ab)}$ for $a=e,\mu,\tau$ are given by
\begin{eqnarray}
V^{(ab)}_{aa}=V^{(ab)}_{bb}=\cos \uptheta_{ab} \ , \
V^{(ab)}_{ab}=-V^{(ab)}_{ba}=\sin \uptheta_{ab} \ , \
V^{(ab)}_{cc}=1 \quad \text{for $c\neq a,b$}\,,
\end{eqnarray}
where $\uptheta_{ab}$ are the neutrino mixing angles. In order to fix these light neutrino oscillation parameters in $U_\nu$, we neglect the non-unitarity in eq.~(\ref{VnuDef}) and use the parameters given in table~\ref{tab:active_bounds}. 
\begin{table}[t]
  \centering
  \begin{tabular}{l||c|c|c|c|c|r}
 &$m_1^2$ & $m_2^2$ & $m_3^2$  & $\sin^2\uptheta_{12}$ & $\sin^2\uptheta_{13}$ & $\sin^2\uptheta_{23}$ \\
  \hline\hline
   NH & $0$ & $m^2_{\rm sol}$ & $\Delta m^2_{31}$ & $0.308$ & $0.0219$ & $0.451$\\
      \hline
  IH & $-\Delta m^2_{32}-m^2_{\rm sol}$ & $-\Delta m^2_{32}$ & $0$ & $0.308$ & $0.0219$ & $0.576$\\
  \end{tabular}
    \caption{Recently updated best fit values for the active neutrino masses $m_a$ and mixings $\uptheta_{ab}$ in case of NH in the top row and IH in the bottom row taken from ref.~\cite{Gonzalez-Garcia:2014bfa}. The smaller measured mass difference (solar mass difference) is for both hierarchies given by $m^2_{\rm sol}=m_2^2-m_1^2=7.49\times 10^{-5}\,\eV^2$, while the larger mass differences are given by $\Delta m^2_{31}=m_3^2-m_1^2=2.477\times 10^{-3}\,\eV^2$ for NH and $\Delta m^2_{32}=m_3^2-m_2^2=-2.465\times 10^{-3}\,\eV^2<0$ for IH. The $n_s=2$ flavour case requires lightest neutrino to be massless ( $m_1=0$ for NH and $m_3=0$ for IH). The atmospheric mass difference $m^2_{\rm atm}$ is often referred to as the absolute value $|m_3^2-m_1^2|$. Thus, $n_s=2$ directly sets $m^2_{\rm atm}=m_3^2$ for NH and $m^2_{\rm atm}=m_1^2= m_2^2+\mathcal{O}(m^2_{\rm sol}/m^2_{\rm atm})$ for IH. Even though these values for $m^2_{\rm atm}$ slightly differ for the two hierarchies the errors are of order $m^2_{\rm sol}/m^2_{\rm atm}$ and can be neglected.}
     \label{tab:active_bounds}
\end{table}
Further, $\delta$ is the so-called Dirac phase, while $\alpha_{1,2}$ are referred to as the Majorana phases. 
In the case of two generations of sterile neutrinos only one combination of the Majarona phases is physical, i.e.~for normal hierarchy (NH) it is $\alpha_2$, while for  inverted hierarchy (IH) it is $\alpha_2-\alpha_1$. Without loss of generality we can set $\alpha_2 \equiv \alpha$ and $\alpha_1 = 0$. The mixing of the doublet state $\nu_{\rm L}$ with the singlet state $\nu_{\rm R}$ implies a deviation of $V_\nu$ from unitary.  The complex matrix $\mathcal{R}$ in eq.~(\ref{CasasIbarraDef}) fulfils the orthogonality condition $\mathcal{R}^T \mathcal{R}=1$.

The number $n_s$ of RH neutrinos is unknown and not restricted by anomaly requirements on SM interactions because they carry no SM charges.
The fact that two non-zero mass splittings $m_{\rm sol}^2$ and $m_{\rm atm}^2$ have been observed implies that $n_s\geq2$ if the seesaw mechanism is the sole origin of light neutrino masses
because the number of non-vanishing eigenvalues $m_a$ in equation (\ref{eq:seesaw_relation}) has to be smaller than or equal to the number of generations of heavy neutrinos. 
In the following we focus on the minimal scenario with $n_s=2$, in which the lightest neutrino is massless ($m_{\rm lightest}=0$).
This effectively also describes leptogenesis and neutrino mass generation in the $\nu$MSM.\footnote{There are $n_s=3$ heavy neutrinos in the $\nu$MSM, but one of them is required to compose the Dark Matter (DM). Existing observational constraints \cite{Adhikari:2016bei,Boyarsky:2018tvu} imply that the contribution of the DM candidate to leptogenesis and neutrino mass generation can be neglected, so that these phenomena can effectively be described within the $n_s=2$ scenario.}
For $n_s=2$ one can parametrise the matrix $\mathcal{R}$ by via a complex mixing angle $\omega = \Re\omega + \ii \Im \omega$
\begin{align}
\mathcal{R}^{\rm NH}=
\begin{pmatrix}
0 && 0\\
\cos \omega && \sin \omega \\
-\xi \sin \omega && \xi \cos \omega
\end{pmatrix}\,,\quad \quad 
\mathcal{R}^{\rm IH}=
\begin{pmatrix}
\cos \omega && \sin \omega \\
-\xi \sin \omega && \xi \cos \omega \\
0 && 0
\end{pmatrix}
\,,
\end{align}
where $\xi = \pm 1$.
After electroweak symmetry breaking there are two distinct sets of mass eigenstates, which we represent by Majorana spinors $\upnu_i$ and $N_i$. 
The three lightest can be identified with the familiar light neutrinos,
\begin{align}\label{LightMassEigenstates}
\upnu_i=\left[ V_\nu^{\dagger}\nu_{\rm L}-U_\nu^{\dagger}\theta\nu_{\rm R}^c + V_\nu^{T}\nu_{\rm L}^c-U_\nu^{T}\theta\nu_{\rm R} \right]_i\,.
\end{align}
In addition, there are $n_s$ heavy neutrinos
\begin{align}
N_i=\left[V_N^\dagger\nu_{\rm R}+\Theta^{T}\nu_{\rm L}^{c} +  V_N^T\nu_{\rm R}^c+\Theta^{\dagger}\nu_{\rm L}\right]_i\,.
\end{align}
These expressions are valid up to second order in $|\theta_{ai}|\ll 1$.
The heavy neutrino mass matrix
\begin{eqnarray}
M_N=M + \frac{1}{2}(\theta^{\dagger} \theta M + M^T \theta^T \theta^{*})\label{MNDef}
\end{eqnarray}
gets diagonalised by the unitary matrix $U_N$, and we can define $V_N =(1-\frac12 \theta^T\theta^*)U_N$ in analogy to $V_\nu$.
Even though the heavy neutrinos are gauge singlets they are able to interact with the ordinary matters due to their quantum mechanical mixing. Therefore, any process that involves ordinary neutrinos can also occur with heavy neutrinos if it is kinematically allowed, but the amplitude is suppressed by the mixing angle
\begin{equation}
\Theta_{a i}=(\theta U_N^*)_{a i}\approx \theta_{a i}\ \,.
\end{equation}
Thus, it is convenient to express the branching ratios in terms of
\begin{eqnarray}
\label{eq:U_theta}
U_{a i}^2 = |\Theta_{a i}|^2\approx |\theta_{a i}|^2\,.
\end{eqnarray}
It is well known that low scale leptogenesis with only two heavy neutrinos requires a mass degeneracy $|\Delta M| \ll \bar{M}$, where
\begin{equation}
\bar{M}=\frac{M_2 + M_1}{2} 
\, , \quad 
\Delta M = \frac{M_2 - M_1}{2}\,. 
\end{equation}
The physical masses observed at colliders are given by the eigenvalues of $M_N$ in eq.~(\ref{MNDef}) and includes a contribution $\mathcal{O}[\theta^2]$ from the Higgs mechanism.
Their splitting $\Delta M_{\rm phys}$ can be expressed in terms of the model parameters as
\begin{equation}\label{eq:DeltaMphys}
\Delta M_{\rm phys} = \sqrt{\Delta M^2 + \Delta M_{\theta\theta}^2 - 2\Delta M\Delta M_{\theta\theta}\cos(2{\rm Re}\omega)}
\end{equation}
with $\Delta M_{\theta\theta}=m_2-m_3$ for normal ordering and $\Delta M_{\theta\theta}=m_1-m_2$ for inverted ordering.
If the mass splitting $\Delta M_{\rm phys}$ is too tiny to be resolved experimentally, experiments are only sensitive to the quantities
\begin{eqnarray}
U_a^2=\sum_i U_{a i}^2\,.
\end{eqnarray} 
The overall coupling strength of the heavy neutrinos
\begin{equation}
U^2=\sum_a U_a^2
\end{equation}
can be expressed in terms of the Casas-Ibarra parameters as
\begin{eqnarray}\label{U2NH}
U^2&=&\frac{M_2-M_1}{2M_1 M_2} (m_2-m_3)\cos(2 {\rm Re}\omega)+\frac{M_1+M_2}{2M_1 M_2}(m_2+m_3)\cosh(2 {\rm Im}\omega)
\end{eqnarray}
in case of normal hierarchy and 
\begin{eqnarray}\label{U2IH}
U^2&=&\frac{M_2-M_1}{2M_1 M_2} (m_1-m_2)\cos(2 {\rm Re}\omega)+\frac{M_1+M_2}{2M_1 M_2}(m_1+m_2)\cosh(2 {\rm Im}\omega)
\end{eqnarray}
for inverted hierarchy.

\subsection{Symmetry protected scenario}\label{Sec:Sym} 

Many models that motivate a low scale seesaw exhibit additional ``lepton number''-like symmetries that make small $M_i$ with comparatively large neutrino Yukawa couplings technically natural.
Such \emph{symmetry protected scenarios}, see e.g.\ \cite{Antusch:2015mia} for a benchmark scenario, are phenomenologically very interesting because they allow to make mixings $U_{ai}^2$ much larger than suggested by the ``naive estimate'' 
\begin{equation}\label{NaiveSeesaw}
U^2 \sim \sqrt{m_{\rm atm}^2 + m_{\rm lightest}^2}/M\,.
\end{equation} 
In the symmetry protected scenario it is convenient to use the parameters 
\begin{eqnarray}\label{B-L_parameters}
\upepsilon \equiv  e^{-2 {\rm Im}\omega } \ &{\rm and}& \ \upmu=\frac{M_1-M_2}{M_2+M_1} 
\end{eqnarray}
instead of $\Delta M$ and ${\rm Im}\omega$.
For $\bar{M}$ near the electroweak scale, experimental constraints allow individual Yukawa couplings $|Y_{ia}|$ that are larger than the electron Yukawa coupling \cite{Drewes:2015iva,Drewes:2016jae}, and the smallness of the $m_i$ is primarily a result of the smallness of $\upepsilon$ and $\upmu$.
Specific models that invoke $\upmu, \upepsilon \ll 1$ typically predict a relation between these parameters, i.e., specify a path in the $\upepsilon$-$\upmu$ plane along which the limit $\upmu, \upepsilon \rightarrow 0$ should be taken.\footnote{While there is nothing that forbids setting $\upmu=0$, $\upvarepsilon_a$ must remain finite to ensure that  the neutrino masses $m_i > 0$.}
In the present work we take an agnostic approach and treat $\upepsilon$ and $\upmu$ as independent parameters.

An approximate $B-L$ symmetry emerges in the limit where these parameters are small.
This can be made explicit by applying the rotation matrix
\begin{eqnarray}
\U=
\frac{1}{\sqrt{2}}
\left(
\begin{tabular}{c c}
$1$ & $\ii$ \\
$1$ & $-\ii$
\end{tabular}
\right)
\end{eqnarray}
to the fields $\nu_{R i}$ in eq.~(\ref{eq:Lagrangian}), which brings $M$ and $Y$ into the form
\begin{align}
M =
\bar{M}\left(
\begin{tabular}{c c}
$\upmu$ & $1$ \\
$1$ & $\upmu$
\end{tabular}
\right)  
\,, \quad \quad   
Y= \left(
\begin{tabular}{c c c}
$Y_e$ & $Y_\mu$ & $Y_\tau$ \\
$\upvarepsilon_e$ & $\upvarepsilon_\mu$ & $\upvarepsilon_\tau$
\end{tabular}
\right)\label{definitionderepsilona} 
\end{align}
with
\begin{align}
Y_a = \frac{1}{\sqrt{2}}(Y_{1a} + \ii Y_{2a}) \,, \quad \quad  \upvarepsilon_a = \frac{1}{\sqrt{2}}(Y_{1a} - \ii Y_{2a})\,,
\end{align}
where $\upvarepsilon_a \ll Y_a <1$\footnote{Note that the entries $\upvarepsilon_a$ are actually of order $\mathcal{O}[\sqrt{\upepsilon}]$ in the parameter $\upepsilon$ defined in eq.~(\ref{B-L_parameters}). We use these conventions to be consistent with the notation commonly used in the literature.}.
When setting $\upvarepsilon_a\rightarrow 0$, one can assign lepton number $+1$ and $-1$ to the states 
\begin{align}\label{OnlyForTheWeak}
\nu_{R {\rm s}}= \frac{1}{\sqrt{2}}(\nu_{R 1} + \ii \nu_{R 2}) \,, \quad \quad  
\nu_{R {\rm w}}= \frac{1}{\sqrt{2}}(\nu_{R 1} - \ii \nu_{R 2})\,,
\end{align}
where $\nu_{R i}$ refer to the flavour eigenstates of $M$.
In the limit $\upmu,\upepsilon\rightarrow 0$ the state $\nu_{R {\rm w}}$ decouples. 
This scenario is often called pseudo-Dirac scenario because the Lagrangian can be expressed in terms of a single Dirac spinor 
$\psi_N=(\nu_{R {\rm s}} + \nu_{R {\rm w}}^c)$,
\begin{eqnarray}\label{PseudoDiracL}
\mathscr{L}&=&\mathscr{L}_{\rm SM} + \overline{\psi_N}(\ii\slashed{\partial} - \bar{M})\psi_N 
- Y_a\overline{\psi_N}\phi^\dagger\varepsilon^\dagger P_L\ell_a - Y_a^* \overline{\ell}_a \varepsilon\phi P_R \psi_N\nonumber\\
&&-\upvarepsilon_a\overline{\psi^c_N}\phi^\dagger\varepsilon^\dagger P_L\ell_a 
- \upvarepsilon_a^* \overline{\ell}_a \varepsilon\phi P_R \psi_N^c
-\frac{1}{2}\upmu\bar{M} \left(\overline{\psi_N^c}\psi_N + \overline{\psi_N}\psi_N^c\right)\,.
\end{eqnarray}
The second line in eq.~(\ref{PseudoDiracL}), which summarises the LNV terms, vanishes in the limit $\upmu,\upepsilon\rightarrow0$.
In the mass basis we find we find $Y_{1a}=\ii Y_{2a}=Y_a/\sqrt{2}$ in this limit, and therefore 
\begin{equation}
U_{a1}^2 = U_{a2}^2 = \frac{1}{2} U_a^2\,.
\end{equation}
The ratios $U_a^2/U^2$ are mostly determined by the parameters in $U_\nu$ \cite{Shaposhnikov:2008pf,Gavela:2009cd,Ibarra:2010xw,Ibarra:2011xn,Lopez-Pavon:2015cga,Hernandez:2016kel,Drewes:2016jae}.
To leading order in $\sqrt{m_{\rm sol}/m_{\rm atm}}$, and $\theta_{13}$ ~\cite{Gavela:2009cd,Ruchayskiy:2011aa,Asaka:2011pb,Hernandez:2016kel,Drewes:2016jae} they can be approximated as
\begin{align}
	\begin{rcases}
	U_e^2/U^2 &\approx \left|s_{12} \sqrt{\frac{m_2}{m_3}} \ee^{\ii \alpha_2/2} - \ii\,s_{13} \ee^{-\ii \delta} \xi \right|^2\\
	U_\mu^2/U^2 &\approx \left|c_{12} c_{23} \sqrt{\frac{m_2}{m_3}} \ee^{\ii \alpha_2/2} - \ii\,s_{23} \xi \right|^2\\
	U_\tau^2/U^2 &\approx \left|c_{12} s_{23} \sqrt{\frac{m_2}{m_3}} \ee^{\ii \alpha_2/2} + \ii\,c_{23} \xi \right|^2
\end{rcases}\text{for NO,}\\
\begin{rcases}
	U_e^2/U^2 &\approx \frac12 \left|c_{12} - \ii s_{12} \ee^{\ii (\alpha_2-\alpha_1)/2}\right|^2\\
	U_\mu^2/U^2 &\approx \frac12 \left| s_{12} c_{23} + c_{12} s_{13} s_{23} \ee^{\ii \delta}
	+ \ii (c_{12} c_{23} - \ee^{\ii \delta} s_{12} s_{13} s_{23} )\ee^{\ii (\alpha_2-\alpha_1)/2} \xi \right|^2\\
	U_\tau^2/U^2 &\approx \frac12 \left|s_{12} s_{23} - c_{12} s_{13} c_{23} \ee^{\ii \delta}
	+ \ii (c_{12} s_{23} + \ee^{\ii \delta} s_{12} s_{13} c_{23} )\ee^{\ii (\alpha_2-\alpha_1)/2} \xi\right|^2
\end{rcases}\text{for IO.}
\end{align}
This is the limit in which we perform the collider analysis in section~\ref{Section4_measurement}.

Strictly speaking we should distinguish between the lepton number $L$ carried by the SM fields and a generalised lepton number $\bar{L}$ that includes the lepton charges associated with some ``lepton number''-like symmetry.  
$\bar{L}$ is conserved in the limit $\upmu,\upepsilon\rightarrow 0$, while $L$ is still violated by active-sterile oscillations in this limit. $L$ is only conserved if one in addition sets $\bar{M}=0$, in which case the seesaw approximations that we use cannot be applied.
To simplify the notation and stick to the commonly used terminology, we in the following refer to the $B-\bar{L}$ conserving limit  $\upmu,\upepsilon\rightarrow 0$ as ``approximate $B-L$ conservation''.


\section{Leptogenesis}
\label{Section3_Leptogenesis}

In leptogenesis, the matter-antimatter asymmetry of the universe is generated in the lepton sector and transferred to the baryonic sector via weak sphalerons \cite{Kuzmin:1985mm}.
These processes are only efficient for temperatures above $T_{\rm sph} = 130\,\GeV$, below which the baryon number $B$ is conserved.  In the standard Leptogenesis scenario \cite{Fukugita:1986hr}, often referred to as \emph{thermal leptogenesis}, the baryon asymmetry is generated due to the decay of the heavy neutrinos with $M_i$ above the electroweak scale such that the $N_i$ are not light enough to be produced efficiently at lepton colliders in the decay of real weak gauge bosons. Instead, the baryon asymmetry can be generated via CP-violating flavour oscillations of the $N_i$ (\emph{baryogenesis from neutrino oscillations}) \cite{Akhmedov:1998qx} or the decay of Higgs bosons \cite{Asaka:2005pn}.
The main difference between standard leptogenesis and these mechanisms lies in the way how the deviation from thermal equilibrium, which is a necessary condition for baryogenesis \cite{Sakharov:1967dj}, is realised. Superheavy $N_i$ come into equilibrium at temperatures $T\gg T_{\rm sph}$. In this case the deviation from equilibrium that is responsible for the asymmetry we observe today is caused by their freezeout and decay, which must occur at $T>T_{\rm sph}$ in order to be transferred from a lepton into a baryon asymmetry ("freeze out scenario").
For $M_i$ below the electroweak scale, on the other hand, the relation  (\ref{eq:seesaw_relation}) implies that the $Y_{ia}$ can be small enough that the $N_i$ do not reach thermal equilibrium at $T<T_{\rm sph}$ ("freeze in scenario").

\subsection{Full system of differential equations}

\paragraph{Characterisation of the asymmetries.} 
It is convenient to describe the asymmetries by the quantities
\begin{align}
\Delta_a = \frac{B}{3}-L_a\,,
\end{align}
that are conserved by all SM processes including the weak sphaleron transitions. Here $B$ is the baryon number and $L_a=(g_w q_{\ell a} + q_{{\rm R}a})$ are the individual lepton flavour charges stored in the right handed leptons $q_{{\rm R}a}$ and the doublet leptons $q_{\ell a}$. 
The total SM lepton number is given by
\begin{align}\label{LSMDef}
L= \sum_a L_a = \sum_a \left(g_w q_{\ell a} + q_{{\rm R}a}\right)\,.
\end{align} 
The $L_a$ and $L$ are violated by the Yukawa couplings $Y_{ia}$.
The smallness of the $Y_{ia}$ implies that the $\Delta_a$ evolve very slowly compared to the rate of gauge interactions that keep the SM fields in kinetic equilibrium, so that they can effectively be described by chemical potentials, cf. eq.~(\ref{ChargeDefs}).
Due to their Majorana nature the $N_i$ do not carry any charge in the strict sense, but in the limit $T \gg M_i$ the helicity states of the heavy neutrinos effectively act as particle and antiparticles. This allows to define the sterile neutrino charges in terms of the difference between the occupation numbers of $N_i$ with positive and negative helicity, i.e., 
 the diagonal elements of the matrices $n_{h ij}$ defined in eq.~(\ref{deltanij}) in the heavy neutrino mass basis,
\begin{align}\label{eq:qNdef}
q_{Ni} \equiv \delta n_{+ii}-\delta n_{-ii} = 2 \delta n_{ii}^{\rm odd}\,.
\end{align}
It is useful to introduce the generalised lepton number
\begin{align}\label{LtildeDef}
\tilde{L} = L + \sum_i q_{Ni}\,.
\end{align}
This quantity is approximately conserved in the temperature regime $T\gg M_i$ because helicity conserving processes are suppressed when the $N_i$ are relativistic. It is in general not equivalent to the lepton number $\bar{L}$ that is conserved in the limit $\upepsilon, \upmu\rightarrow 0$. 
Though the violation of $\bar{L}$ is parametrically small, it can be significant in the early universe because heavy neutrino oscillations generally convert $\nu_{R {\rm s}}$ and $\nu_{R {\rm w}}$ into each other, and $\bar{L}$ is therefore not a useful quantum number to describe leptogenesis. Only in the highly \emph{overdamped regime} these conversions are suppressed, and one can identify $\nu_{R{\rm w}}$ with the slowly evolving state.

\paragraph{Quantum kinetic equations.}
The evolution equations for the deviations of the heavy neutrino number densities from equilibrium $\delta n_{h}$ form a coupled system of equations together with the evolution equations for the asymmetries. 
A consistent set of equations to describe low scale leptogenesis has first been formulated in ref.~\cite{Asaka:2005pn} as a  generalisation of the \emph{density matrix equations} that are commonly used in neutrino physics \cite{Sigl:1992fn}.
In this work we employ a modified version of the quantum kinetic equations  used in ref.~\cite{Drewes:2016jae}, which have been derived in ref.~\cite{Drewes:2016gmt}.
The main improvement in the present work is the inclusion of LNV processes, which were neglected in most previous studies. 
Here we only present the quantum kinetic equations, the modifications to the derivation in ref.~\cite{Drewes:2016gmt} that are necessary to derive them are presented in appendix~\ref{Appendix_3}. 
We use the time variable $z=T_{\rm ref}/T$, where $T_{\rm ref}$ is an arbitrarily chosen reference scale. It is convenient to set this to the sphaleron freeze out temperature $T_{\rm sph}$ such that the freeze out happens at $z=1$.
In terms of $z$, the evolution equations for the deviations of the $N_i$ occupation numbers from equilibrium read
\begin{align}
\label{sec3:eq:diff_eq_RHN}
\frac{\dd}{\dd z}\delta n_{h} &= -\frac{\ii}{2}[H_N^{\rm th}+z^2 H_N^{\rm vac},\delta n_{h}]-\frac{1}{2}\{\Gamma_N,\delta n_{h}\}+\sum_{a,b=e,\mu,\tau}\tilde{\Gamma}_N^a (A_{ab} + C_b/2)\Delta_b\,,
\\
\label{sec3:eq:diff_eq_SM}
\frac{\dd}{\dd z} \Delta_a&=
\frac{\tilde{\gamma}_++\tilde{\gamma}_-}{g_w} \frac{a_{\rm R}}{T_{\rm ref}}\sum_{i} |Y_{ia}|^2 \,\left(\sum_b (A_{ab} + C_b/2)\Delta_b\right)
-\frac{S_a(\delta n_{hij})}{T_{\rm ref}}\,,
\end{align}
where $a_{\rm R}=m_{\rm Pl}\sqrt{45/(4\pi^3 g_*)}=T^2/H$
can be referred to as the comoving temperature in a radiation dominated universe,
$H$ is Hubble parameter,
$g_*$ is the number of relativistic degrees of freedom in the primordial plasma and $m_{\rm Pl}$ the Planck mass. 
The collision term for the $N_i$ is conventionally decomposed into two contributions: The \emph{damping term} $\Gamma_N$ and the \emph{backreaction term} $\tilde{\Gamma}_N$.
The latter pursues chemical equilibration of the heavy neutrinos with the Higgs field and the SM lepton doublets. 
The \emph{source term} for the lepton charges $S_a \equiv S_{aa}$ can be obtained from the more general expression
\begin{align}
S_{ab} 
&= 2\frac{a_{\rm R}}{g_w}\sum_{i, j} Y_{ia}^*Y_{jb} \sum_{s=\pm}  \gamma_s \left[\ii {\rm Im}(\delta n_{ij}^{\rm even})+s  {\rm Re}(\delta n_{ij}^{\rm odd})\right]\,,
\end{align}
and is responsible for the creation of the lepton doublet charges due to the off-diagonal correlations $\delta n_{ij}$.
The momentum independent rates $\gamma_\pm$ with $\gamma_+= \gamma_+^{\rm av}$ and $\gamma_-= \gamma_-^{|\k_{\rm av}|}$ that appear in $S_a$ are discussed in appendix \ref{Appendix_3}.
The term in brackets in eq.~(\ref{sec3:eq:diff_eq_SM}) is the \emph{washout term}. 
The matrices $A$, $D$ as well as the vector $C$ account for the fact that spectator effects redistribute the charges \cite{Buchmuller:2001sr}. 
They are given by
\begin{align}
A=
\frac{1}{711}
\left(
\begin{array}{ccc}
-221 & 16 & 16\\
16 & -221  & 16\\
16 & 16 & -221
\end{array}
\right)
\,,
\quad
C=
-\frac{8}{79}
\left(
\begin{array}{ccc}
1 & 1 & 1
\end{array}
\right)
\,,
\quad
D=
\frac{28}{79}
\left(
\begin{array}{ccc}
1 & 1 & 1
\end{array}
\right)\,.
\end{align}
$H_N^{\rm vac}$ is the effective Hamiltonian in vacuum. It is responsible for the oscillations of the heavy neutrinos due to the misalignment between the mass and the flavour basis. $H_N^{\rm th}$ corresponds to the Hermitian part of its finite temperature correction and effectively acts as the thermal mass of the heavy neutrinos. The individual components are given by 
\begin{align}
H^{\rm vac}_N &=
\frac{\pi^2 }{18 \zeta(3)}\frac{a_{\rm R}}{T_{\rm ref}^3}
\left(\Re[M^\dagger M] + \ii h  \Im[M^\dagger M]\right)\,,\label{HeffT}
\\
H^{\rm th}_N&=\frac{a_{\rm R}}{{T_{\rm ref}}}
\left(\mathfrak{h}_{+}^{\rm th} \Upsilon_{+h} +\mathfrak{h}_{-}^{\rm th} \Upsilon_{-h}\right)+\mathfrak{h}^{\rm EV}\frac{a_{\rm R}}{{T_{\rm ref}}}\Re[Y^* Y^t]\,,
\\
\Gamma_N &=\frac{a_{\rm R}}{{T_{\rm ref}}}\left(\gamma_{+} \Upsilon_{+h} +\gamma_- \Upsilon_{-h}\right)\,,
\\
\tilde{\Gamma}^a_N&=h \frac{a_{\rm R}}{T_{\rm ref}}
\left(\tilde{\gamma}_+ \Upsilon_{+h}^a -\tilde{\gamma}_- \Upsilon_{-h}^a\right)\,,\label{TildeGammaa}
\end{align}
where the following notations are used:
\begin{align}
\Upsilon^a_{hij}&=\Re[Y_{ia}Y^\dagger_{aj}]+\ii h \im[Y_{ia}Y^\dagger_{aj}]\,,
\\
\Upsilon_{hij}&=\sum_a\left(\Re[Y_{ia}Y^\dagger_{aj}]+\ii h \im[Y_{ia}Y^\dagger_{aj}]\right)\,.
\end{align}
Further, the momentum independent rates $\tilde{\gamma}_\pm$ with $\tilde{\gamma}_+= \gamma_+^{\rm av}$ and $\tilde{\gamma}_-= \gamma_-^{\rm av}$ that appear in the term $\tilde{\Gamma}^a_N$ are discussed in appendix \ref{Appendix_3}. These expressions for $H^{\rm th}_N$, $\Gamma_N$ and $\tilde{\Gamma}_N^a$ include LNV effects only at leading order in $\bar{M}/T$ and are valid in the case of two relativistic heavy neutrinos with kinematically negligible mass splitting. 
This improves the accuracy of our treatment compared to the previous analysis in ref.~\cite{Drewes:2016jae}, as explained in section~\ref{LNVprocesses}.
There are, however, several effects that we still neglect. 
This includes the kinematic effect of particle masses (gauge boson, top quark and $N_i$), the temperature dependent continuous freeze out of the weak sphalerons (we assume an instantaneous freeze out at $T = T_{\rm sph}$\footnote{The effect of the gradual sphaleron freezeout has recently been studied in ref.~\cite{Eijima:2017cxr}.
Based on those results, we do not expect a large change in the largest allowed $U_a^2$ from this effect.
}) and the error that occurs due to the momentum averaging. The latter is briefly discussed after eq.~(\ref{deltanij}).
We expect that these effects are subdominant in most of the parameter region we study here. 
However, they may become important if either the baryon asymmetry is generated shortly before the sphaleron freeze out or if the heavy neutrinos have masses comparable to the $W$ boson. 

\paragraph{Collision terms.
} 
The lepton number conserving and violating contributions to the collision terms exhibit a different dependence on the Yukawa couplings and on $T$. The different dependency on the $Y_{ia}$ can  be expressed in terms of the quantities $\Upsilon_{+h}$ and $\Upsilon^a_{+h}$ (lepton number conserving) or 
$\Upsilon_{-h}$ and $\Upsilon^a_{-h}$ (lepton number violating).
The exact $T$ dependence is in principle rather complicated because various different processes contribute to the equilibration rates. Here we employ the commonly used linear approximation for the $T$ dependence of the momentum averaged rates, which hides all the complications in the numerical coefficients $\gamma_+^{\rm av}$, $\gamma_-^{\rm av}$ and $\gamma_-^{|\k_{\rm av}|}$.

For the lepton number conserving rates we use the values $\gamma_+^{\rm av} = 0.012$ given in ref.~\cite{Garbrecht:2014bfa}, based on the results obtained in refs.~\cite{Besak:2012qm,Garbrecht:2013urw}. 
Lepton number conserving processes are highly dominated by hard momenta $|\k|\sim T$ of the heavy neutrinos, such that there is no significant difference between evaluating the rates at the average momentum $|\k_{\rm av}|\approx 3.15 T$ and momentum averaging the rates. 
Note that the production rate $\gamma_+$ and the rate for the backreaction term $\tilde{\gamma}_+$ are in general different since they come with different powers of the equilibrium distribution function before momentum averaging. Equating them induces an error of roughly $30\,\%$, an approximation that is still sufficient for our discussion.

The effect of momentum averaging is different for the lepton number violating processes because the corresponding rates come with an additional factor of $M^2/|\k|^2$. This results in an infrared enhancement of these rates. Consequently, we have to distinguish between the lepton number violating rate that is momentum averaged $\gamma_-^{\rm av}$ and the one that is evaluated at the averaged momentum $|\k_{\rm av}|$. Approximate numerical results are derived in appendix~\ref{Appendix_3}. We use 
\begin{align}
\gamma_-^{\rm av} = 1.9 \times 10^{-2} \times z^2 \frac{\bar{M}^2}{T_{\rm ref}^2}
\end{align}
for the backreaction term $\tilde{\Gamma}^a_N$, as it depends on quantities that are in equilibrium.
In contrast to that, we use 
\begin{align}
\gamma_-^{|\k_{\rm av}|} = 9.7 \times 10^{-4} \times z^2 \frac{\bar{M}^2}{T_{\rm ref}^2}
\end{align}
for the terms that depend on the deviations $\delta n$, such as the source term $S_a$ and the production term $\Gamma_N$.

\paragraph{Thermal corrections to the effective Hamiltonian.}
In the absence of helicity flips, the thermal correction to the heavy neutrino masses is simply given by the term involving $\mathfrak{h}^{\rm th}_+=0.23$, as mentioned in ref.~\cite{Drewes:2016gmt}, plus a contribution arising from the expectation value of the Higgs field
\begin{align}
\mathfrak{h}^{\rm EV}=\frac{2\pi^2}{18\zeta(3)}\frac{z^2 v^2(z)}{T_{\rm ref}^2}.
\end{align}
We evaluate the latter term with the approximation~(\ref{app3:eq:Higgs_EV}) that has already been used in ref.~\cite{Drewes:2016gmt}. 
Lepton number violating forward scatterings generate an additional correction
\begin{align}
\mathfrak{h}^{\rm th}_- = \left[3.50-0.47\log\left(z^2\frac{\bar{M}^2}{T_{\rm ref}^2}\right)+3.47 \log^2\left(z \frac{\bar{M}}{T_{\rm ref}}\right)\right] \times 10^{-2} \times z^2 \frac{\bar{M}^2}{T_{\rm ref}^2}.
\end{align}
The derivation of this expression is sketched in appendix~\ref{Appendix_3}.

\subsection{The role of lepton number violating processes}\label{LNVprocesses}
The lepton number $\tilde{L}$ is violated by the Majorana mass term $M$. For $M_i$ below the $W$ mass, the BAU is generated when the $N_i$ are relativistic. 
The rates of  $\tilde{L}$-violating processes in this regime are suppressed by $M_i^2/T^2$ and have therefore been neglected in most previous studies.  This is, however, not justified in general.
	There are two important effects arising from the $\tilde{L}$-violating processes, the first is a lepton number violating source term, the second is the enhanced equilibration of the right handed neutrino states in the symmetry protected regime.

	To understand the importance of the $\tilde{L}$ violating source let us first consider the case $\upepsilon \sim 1$, i.e., the ``naive seesaw limit'' (\ref{NaiveSeesaw}). There are two competing sources of lepton asymmetries, the CP-violating oscillations of the $N_i$ \cite{Akhmedov:1998qx,Asaka:2005pn} and the decay of Higgs bosons with large thermal masses \cite{Hambye:2016sby} (cf. also ref.~\cite{Giudice:2003jh} for an earlier discussion).
The heavy neutrino oscillations do not directly change $L$. This can be understood intuitively in terms of the suppression of LNV by $M_i^2/T^2$ and is shown in detail in appendix D.3 of ref.~\cite{Drewes:2016gmt}. A total $L\neq 0$ (and hence $B\neq 0$) is generated with the help of the washout, which erases the individual $L_a$ at different rates. This generates a total $L\neq0$ (and hence $B-L\neq 0$) even if $\tilde{L}$ is conserved because it stores part of the asymmetry in the sterile flavours, where it is hidden from the sphalerons.
Since the washout is also mediated by the Yukawa couplings $Y_{ia}$, the final baryon asymmetry in the regime $\upepsilon \sim 1$ is $\mathcal{O}[Y^6]$, cf. also ref.~\cite{Shuve:2014zua} for a pedagogical discussion. 
In contrast to that, the Higgs decays directly violate $L$ and $\tilde{L}$, which leads to a contribution $\mathcal{O}[Y^4 M_i^2/T^2]$ to the baryon asymmetry.\footnote{In addition to the different dependence on the $Y_{ia}$, the $N_i$ mass spectrum also affects the two contributions in a different way. A more detailed comparison is e.g. given in ref.~\cite{Hambye:2017elz}. }
In the regime $\upepsilon\sim 1$  the seesaw relation (\ref{eq:seesaw_relation}) predicts a $|Y_{ia}|^2 \sim  m_a M_i/v^2$. Hence, the Higgs decays can dominate the baryon asymmetry if it is primarily generated at temperature scales lower than $\mathcal{O}(v \sqrt{\bar{M}/m_a})$.
Furthermore, as the asymmetry generated this way is a total $\tilde{L}$ asymmetry it cannot be erased by the usual $\tilde{L}$-conserving rate and can therefore dominate the baryon asymmetry even for $\upepsilon \ll 1$.

\begin{figure}
        \centering
        \begin{tabular}{cc}
\textpdfrender{
    TextRenderingMode=FillStroke,
    LineWidth=.1pt,
    FillColor= goldenrod,
  }{$\bigstar$}	
			\textbf{BAU suppressed by LNV} &
			\textpdfrender{
    TextRenderingMode=FillStroke,
    LineWidth=.1pt,
    FillColor=steelblue,
  }{$\bigstar$}
			\textbf{BAU enhanced by LNV} \\\\
            \includegraphics[width=0.43\textwidth]{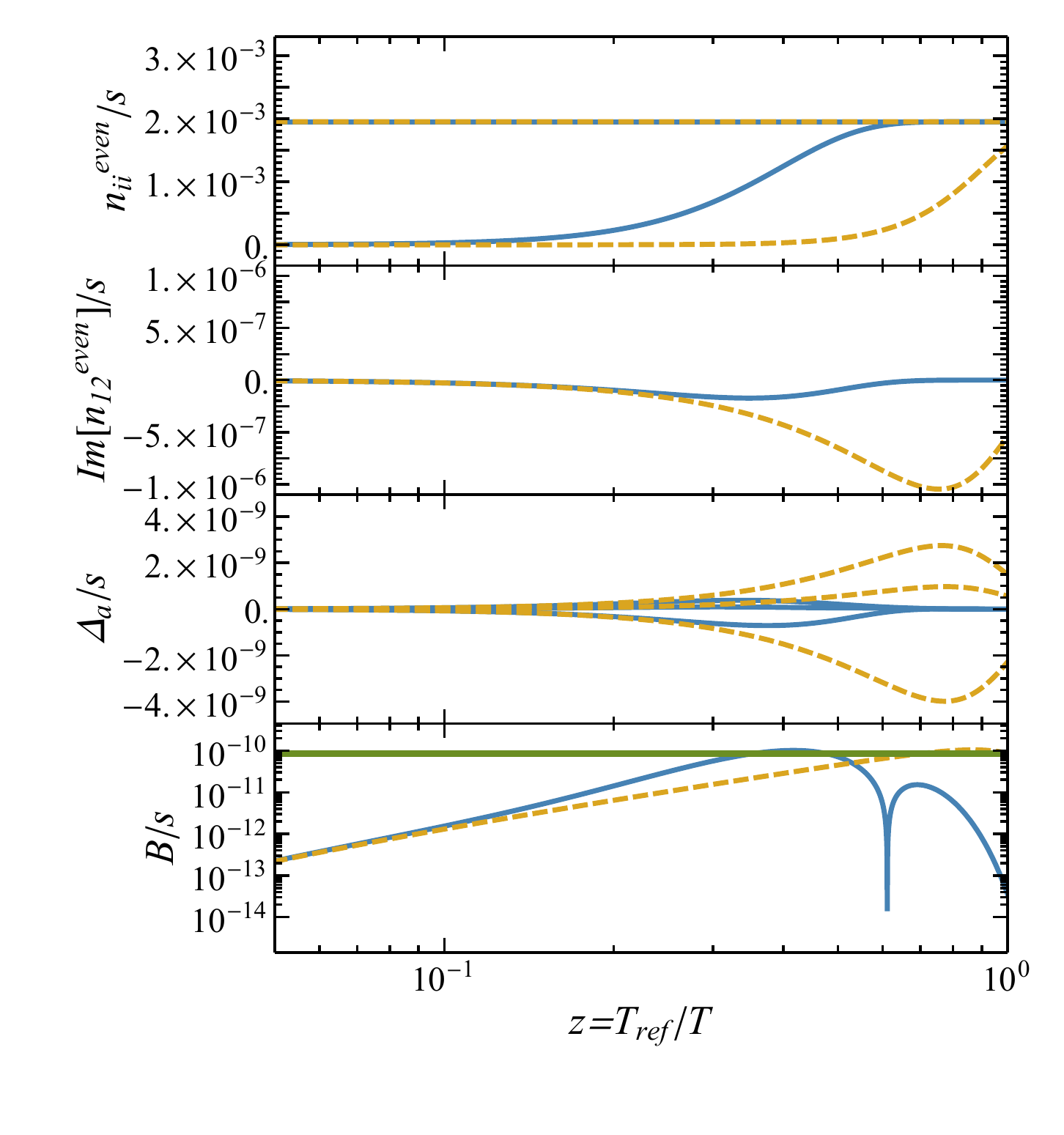} &
			\includegraphics[width=0.43\textwidth]{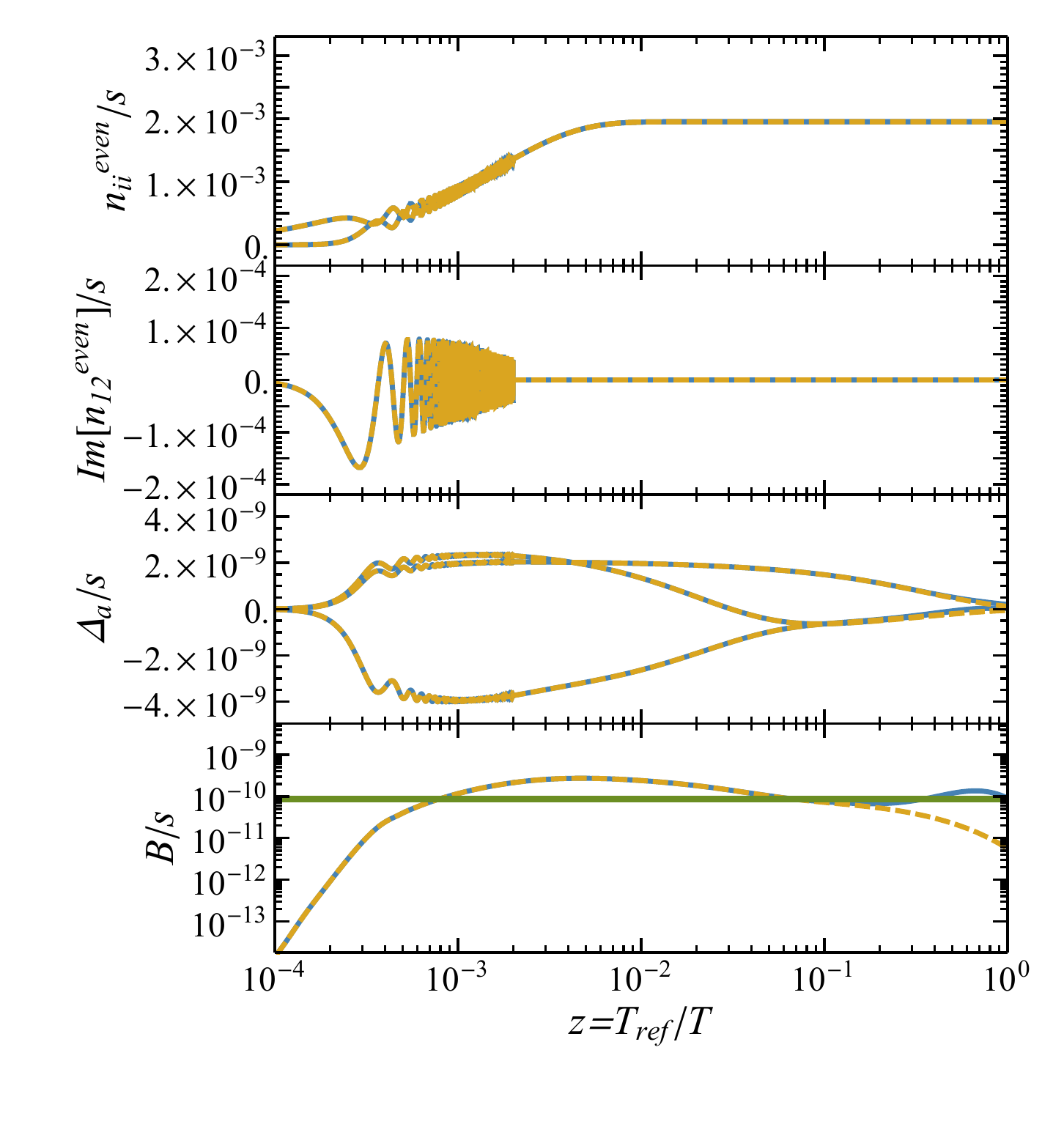}
        \end{tabular}
\caption{\label{fig:wwoLNVcomparison}
The evolution of the number densities with (blue, full), and without (yellow, dashed) the lepton number violating processes.
In the left panel we present a point in parameter space where the lepton number violating processes suppress the final BAU.
Without the LNV processes, this point is allowed because the equilibration of the RHN number can be postponed arbitrarily by adjusting the mass splitting.
This is, however, not possible in the presence of the lepton number violating processes, as they enhance the equilibration rate of the weakly coupled RHN flavour.
In the right panel we present an example where the BAU is enhanced through the presence of the lepton number violating processes. Note that while the evolution
of the active and sterile charges with and without lepton number violation is almost identical at early times, it changes dramatically at late times where the
lepton number violating term becomes large. If the LNV processes are neglected,
the washout of the RHN charges suppresses the total lepton number asymmetry.
In the presence of LNV processes this suppression is not as effective, leading to a larger final BAU.
}
\end{figure}

The second effect is the enhanced equilibration of the heavy neutrino eigenstates, which is most important in the symmetry protected limit, where the matrix $YY^\dagger$ has two vastly different eigenvalues, the magnitudes of which scale as $\sum_a \upvarepsilon_a^2 \sim \upepsilon$ and $\sum_a Y_a^2 \sim 1/\upepsilon$.
The damping of deviations in the heavy neutrino occupation numbers from equilibrium is governed by the eigenvalues of $\Gamma_N$.
Let us for a moment assume that there are no $\tilde{L}$-violating processes. Then we can define the heavy neutrino interaction eigenstates as the eigenvectors of the helicity dependent flavour matrices $\Upsilon_{+h}$. In the limit $\upmu,\upepsilon\rightarrow 0$, they can be identified with the states $\nu_{R{\rm s}}$ and $\nu_{R{\rm w}}$ that carry the generalised lepton charge $\bar{L}$, cf. eq.~(\ref{OnlyForTheWeak}).
One pair of interaction eigenstates (approximately $\nu_{R{\rm s}}$) has comparably strong couplings $Y_a^2\propto 1/\upepsilon$, while the couplings of the other pair (approximately $\nu_{R{\rm w}}$) are suppressed by $\upvarepsilon_a \propto \sqrt{\upepsilon}$.
This leads to an overdamped behaviour of the $N_i$ oscillations because the more strongly coupled states come into equilibrium before the heavy neutrinos have performed a single oscillation. The deviations from equilibrium which drive baryogenesis are then given by the slow evolution of the feebly coupled states. The feebly coupled state instead reaches equilibrium through the mixing with the strongly coupled state.
In the presence of  $\tilde{L}$-violating processes, both eigenvalues of $\Gamma_N$ receive corrections from the terms involving $\Upsilon_{-h}$.
For the larger eigenvalue, these can be neglected.
However, the correction $\sim Y_a^2 \bar{M}^2/T^2\propto 1/\upepsilon\times\bar{M}^2/T^2$ to the smaller eigenvalue, which governs the damping of the feebly coupled interaction eigenstate, is not necessarily negligible compared to the terms $\sim \upvarepsilon_a^2 \propto \upepsilon $ in $\Upsilon_{+h}$.
For sufficiently small $\upepsilon$ they dominate over the lepton number conserving terms involving $\Upsilon_{+h}$, which are suppressed by $\upepsilon$. 
Since the deviations from equilibrium in the overdamped regime are mostly determined by the occupation numbers of the feebly coupled state, this modification of the damping rates has a strong effect on the behaviour of the entire system of equations and affects the BAU.
One can roughly estimate that this can affect the asymmetry if this occurs before sphaleron freezeout ($\upepsilon < \bar{M}/T_{\rm sph}$) if the BAU is generated in the overdamped regime.
An example of such an evolution is presented in the left panel of figure~\ref{fig:wwoLNVcomparison}.

\begin{figure}
        \centering
        \begin{tabular}{cc}
			\textbf{Normal Ordering} & \textbf{Inverted Ordering} \\\\
            \includegraphics[width=0.43\textwidth]{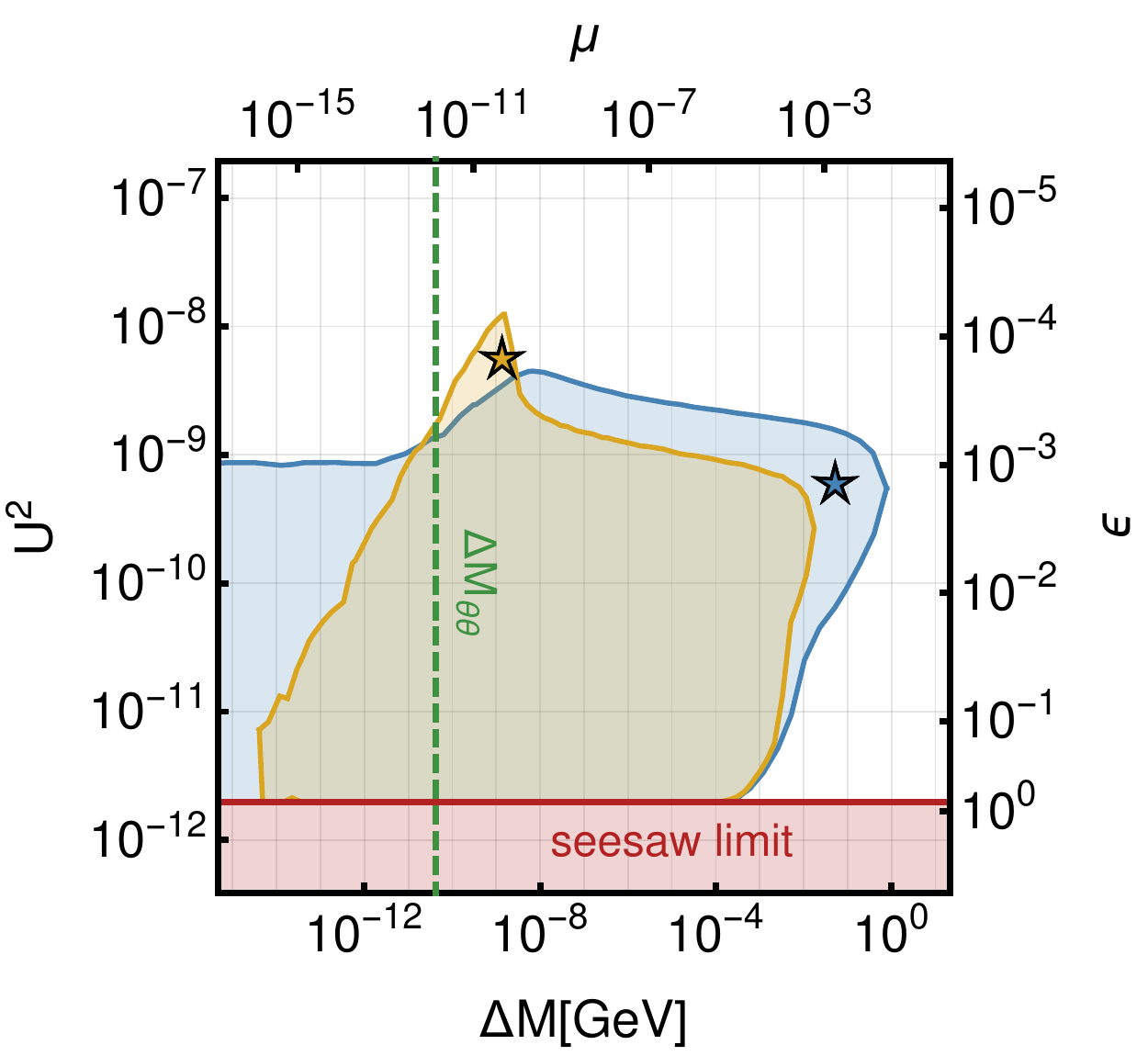} &
			\includegraphics[width=0.43\textwidth]{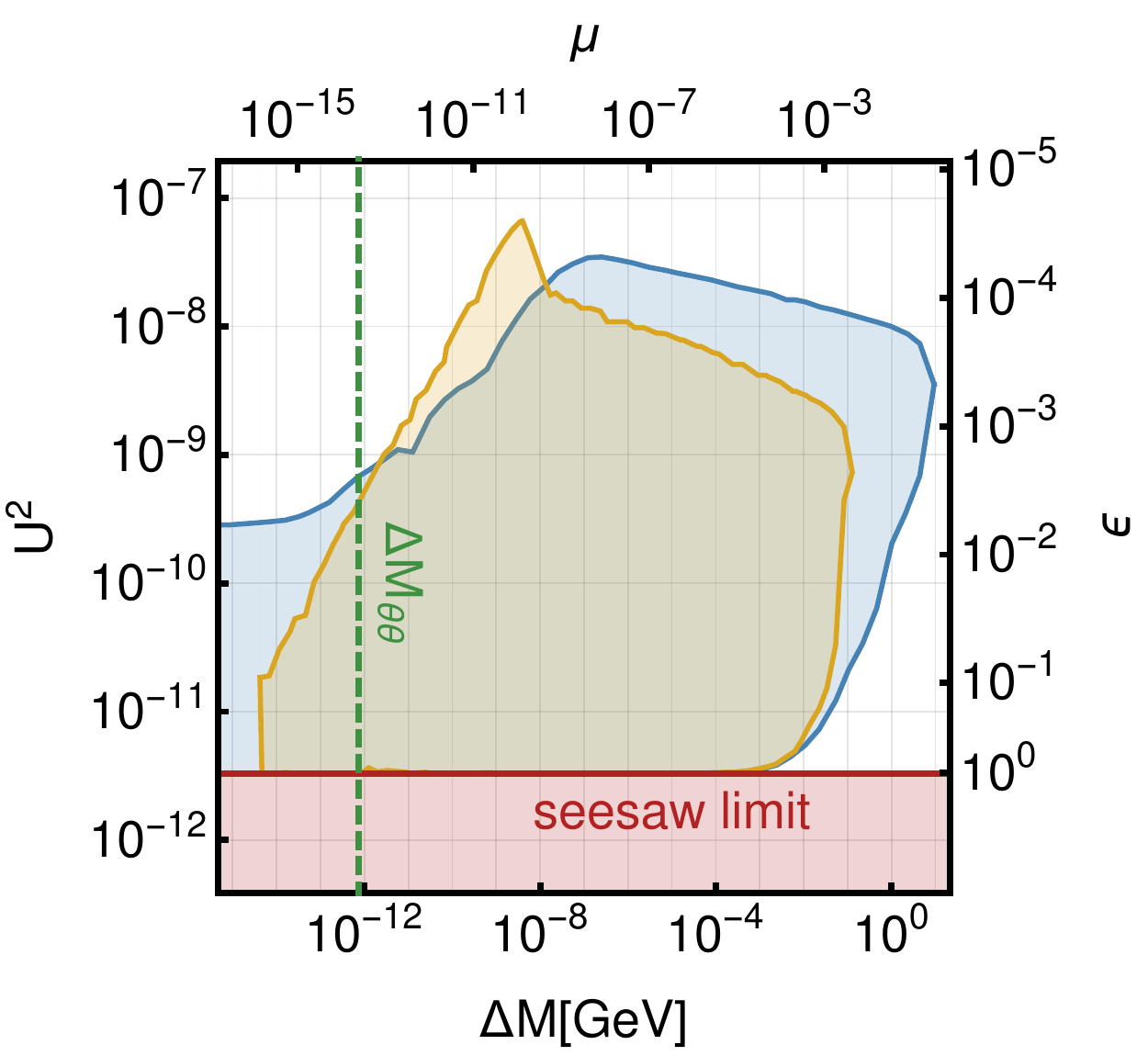}
        \end{tabular}
\caption{\label{fig:dmU2LNVLNC}
Comparison of the allowed parameter range in the $\Delta M$ - $U^2$ plane with (blue) and without (yellow) the lepton number violating processes for normal and inverted hierarchies with the benchmark mass of $\bar{M}=30 \GeV$. One can see that the range of the allowed mass splitting increases when lepton number violation is included by two orders of magnitude, reaching sizes that can be resolved in experiments. The blue and yellow stars correspond to benchmark points for which we present a comparison between the evolution with and without the lepton number violating processes in figures~\ref{fig:wwoLNVcomparison}.
The yellow star corresponds to a point in parameter space that can reproduce the correct the observed BAU only if one neglects the lepton number violating processes, while the blue star corresponds to a point in parameter space that can only reproduce the correct BAU if one includes the LNV processes.}
\end{figure}

Finally, the LNV processes can also enhance the asymmetry if the mass splitting between the right-handed neutrinos is large. Without the LNV processes, the washout of the charges in the RHN sector would supress the lepton nubmer asymmetry. However, if one includes LNV processes, this supression is not as effective, leading to a larger final BAU cf. the right panel of~\ref{fig:wwoLNVcomparison}.
Together these effects imply that for larger masses the parameter space consistent with leptogenesis may be quite different from the one with LNV effects neglected. From the results of our scan we have observed that while the the allowed region in the $M-U^2$ plane appear quite similar, the maximal mass splittings can change by up to two orders of magnitude as shown in figure~\ref{fig:dmU2LNVLNC}, thereby making the model more testable.

\paragraph{The generation of the asymmetry in the degenerate limit $\Delta M \rightarrow 0$.}
It is standard lore that leptogenesis is not possible if the vacuum masses of the heavy neutrinos are exactly degenerate $\Delta M \rightarrow 0$. The effective Hamiltonian in this case commutes with the matrix of damping rates, meaning that the effective mass basis and interaction basis for heavy neutrinos coincide and no oscillations take place.
The inclusion of the LNV processes allows us to bypass this constraint because the effective Hamiltonian and matrix of damping rates in general do not commute due to their different dependence on flavour and helicity.
To illustrate this effect explicitly, we constrain ourselves to the weak washout regime, where a perturbative expansion in the Yukawa couplings is applicable. The solution to Eq.~\ref{sec3:eq:diff_eq_RHN} in this regime can be approximated by:
\begin{align}
	\delta n_h(z) &= - n^{\rm eq} +
	n^{\rm eq} \frac{a_{\rm R}}{T_{\rm ref}}
	\int^z_0 \dd z^\prime \left( \gamma_+(z^\prime) \Upsilon_{+h} + \gamma_-(z^\prime) \Upsilon_{-h}\right)\\\notag
	&- n^{\rm eq} \left(\frac{a_{\rm R}}{T_{\rm ref}}\right)^2
	\frac{\ii}{2} \left[ \Upsilon_{+h}, \Upsilon_{-h} \right]
	\int^z_0 \dd z^\prime \int^{z^\prime}_0 \dd z^{\prime\prime}\left(
	\mathfrak{h}_+(z^\prime) \gamma_-(z^{\prime \prime})
	-\mathfrak{h}_-(z^\prime) \gamma_+(z^{\prime \prime}) \right)	
	\,.
\end{align}
Inserting the solutions into the source term:
\begin{align}
\label{eq:dm0src}
	S_{aa} &= \frac{a_{\rm R}}{g_w} \left[ \sum_h  h \gamma_+ \mathrm{Tr} \left( \Upsilon_{+h}^a \delta n_h \right) + h \gamma_- \mathrm{Tr}\left( \Upsilon_{-h}^a \delta n_h \right)\right]\\\notag
	&\sim \ii \mathrm{Tr}\left( \Upsilon^a_+ \left[ \Upsilon_+, \Upsilon_- \right] \right)
	\int^z_0 \dd z^\prime \int^{z^\prime}_0 \dd z^{\prime\prime}\left(
	\mathfrak{h}_+(z^\prime) \gamma_-(z^{\prime \prime})
	-\mathfrak{h}_-(z^\prime) \gamma_+(z^{\prime \prime}) \right)
	 \neq 0\,,
\end{align}
we obtain a non-vanishing result. Note that the total lepton asymmetry generated this way vanishes as $S= \sum_a S_{aa} \sim \ii \mathrm{Tr}(\Upsilon_+ \left[ \Upsilon_+,\Upsilon_-\right])=0$. Therefore in this scenario, one still
has to rely on washout to convert the lepton flavour asymmetry into a total lepton number asymmetry. It can be shown that Eq.~\eqref{eq:dm0src} gives a vanishing source in the usual ARS mechanism where only one helicity interacts with the medium and $\gamma_- = \mathfrak{h}_- =0$, which leads to a vanishing RHS of Eq.~\eqref{eq:dm0src}.
The same result is true in standard leptogenesis where helicity effects are typically neglected, $\gamma_-=\gamma_+$ and $\mathfrak{h}_-= \mathfrak{h}_+$,
once more leading to a vanishing integral on the RHS of~\eqref{eq:dm0src}.

\subsection{How to perform the parameter scan}\label{HowToDo}
We perform a parameter scan by numerically solving the evolution equations (\ref{sec3:eq:diff_eq_RHN}) and  (\ref{sec3:eq:diff_eq_SM}) in order to identify the range of heavy neutrino parameters for which the observed BAU can be explained. A brief explanation of the treatment of the numerically stiff equations is presented in appendix~\ref{Appendix_4}.
We limit ourselves to masses in the range between $5\,{\rm GeV} < \bar{M}  <50 \,\GeV$. 
At smaller $M_i$, fixed target experiments offer much better perspectives to search for heavy neutrinos than high energy lepton colliders.
At larger $M_i$, the approximations in the derivation of the expressions (\ref{HeffT})-(\ref{TildeGammaa}) are not justified.
The scan is performed as follows.
We first randomly choose a value for $\bar{M}$ between $5 \, \GeV$ and $50 \, \GeV$ with a logarithmic prior and a value for ${\rm Im}\,\omega$ with a flat prior between $0<{\rm Im} \omega<7$.
This  corresponds to a logarithmic prior in $U^2\sim {\rm e}^{2{\rm Im}\omega}$.
The upper limit ${\rm Im} \omega < 7$ does not limit the scan; in practice we find that the observed BAU cannot be produced for larger values of ${\rm Im} \omega$ due to a stronger washout.

After fixing $\bar{M}$ and ${\rm Im}\omega$, we perform a simple Markov-chain-Monte-Carlo (MCMC) search using the Metropolis-Hastings algorithm over the remaining parameters $\alpha$, $\delta$, $\Delta M$  and ${\rm Re} \omega$. 
The proposal distribution is a multivariate Gaussian distribution in $\alpha\,,\delta\,,{\rm Re}\,\omega$ and $\log{\Delta M/\bar{M}}$, while the acceptance distribution is given by
\begin{align}
	A(\alpha,\delta,{\rm Re} \omega, \Delta M | \alpha^\prime,\delta^\prime,{\rm Re} \omega^\prime, \Delta M^\prime)=
	\min\left\{1,\exp\left[-\frac{(|Y_B^\prime|-\bar{Y}_{B\,{\rm obs}})^2-(|Y_B|-\bar{Y}_{B\,{\rm obs}})^2}{2 \sigma_{\rm obs}^2}\right]\right\}\,,
\end{align}
where $|Y_B|$ is obtained by numerically solving the evolution equations (\ref{sec3:eq:diff_eq_RHN}, \ref{sec3:eq:diff_eq_SM}).
In the final analysis we accept all parameter choices that give a BAU within the $5 \sigma_{\rm obs}$ region of the observed BAU.
As the largest mixing angles require a hierarchical flavour pattern $U_a^2 \ll U^2$, we in addition perform targeted scans in which the initial values of parameters $\alpha$ and $\delta$ for the MCMC scan are chosen to minimise the ratio $U_a^2/U^2$. These points yield the largest numbers of events one can expect at an experiment. 

The upper bound on the mixing angle $U^2$ in figure~\ref{fig:SensitivityU2M} is determined by binning the data points consistent with the BAU according to the logarithm of the mass
$\log{\bar{M}}$ into $60$ bins, and in each bin choosing the point with the largest mixing angle $U^2$.
If the $N_i$ are produced in the decay of $Z$ bosons in the $s$ channel, then the total number of expected collider events (to be explained in the next section) in a given experiment and for fixed $\bar{M}$ in good approximation only depends on $U^2$, cf. figures~\ref{fig:TotalU2M_FCC} and \ref{fig:TotalU2M_ILC_CECP}. For the $Z$ pole runs, we can therefore uniquely define the area where one can expect more than 4 events in figure~\ref{fig:SensitivityU2M}.
If the $N_i$ are produced via $t$-channel exchange of $W$ bosons, then the mixing in at least one of the vertices must be $U_e^2$ because the experiment collides electrons and positrons. The total event rate therefore depends on $U_e^2$ in a different way than on $U_\mu^2$ and $U_\tau^2$, and the number of events cannot be determined by fixing $\bar{M}$ and $U^2$ alone, cf. figure~\ref{fig:TotalU2M_nonZ}.
In figure~\ref{fig:SensitivityU2M} we therefore indicate two regions: The one where the expected number of events exceeds four under the most pessimistic assumptions about the relative size of the $U_a^2$ for fixed $U^2$ (``guaranteed discovery''), and the one where it exceeds four under the most optimistic assumptions (``potential discovery'').
The lines corresponding to a guaranteed discovery are obtained by picking the smallest mixing angle $U^2$ in each bin. 
To obtain the lines with a potential discovery, we instead select the points where the number of expected events is $N<4$, and from the bins select those with the largest mixing angle $U^2$.
For the plots showing the maximal/minimal number of expected events in appendix \ref{Appendix_1}, we divide the points into $60$ bins according to the logarithm of the mass $\bar{M}$, as well as $60$ bins according to the logarithm of their mixing angle $U^2$. From each bin we then select the points with the maximal and minimal numbers of events expected at the future collider considered.


\section{Measurement of the leptogenesis parameters at colliders}
\label{Section4_measurement}

In this section we discuss the possibility of measuring the neutrino parameters at possible future high-precision lepton colliders.
 The precise knowledge of the mixing quantities $U_a^2$ and the flavour mixing ratios $U_a^2/U^2$ is crucial to test whether the properties of a hypothetical heavy neutral lepton that is discovered at a future collider are compatible with leptogenesis and the generation of light neutrino masses.

\begin{figure}
\begin{center}
\includegraphics[width=0.7\textwidth]{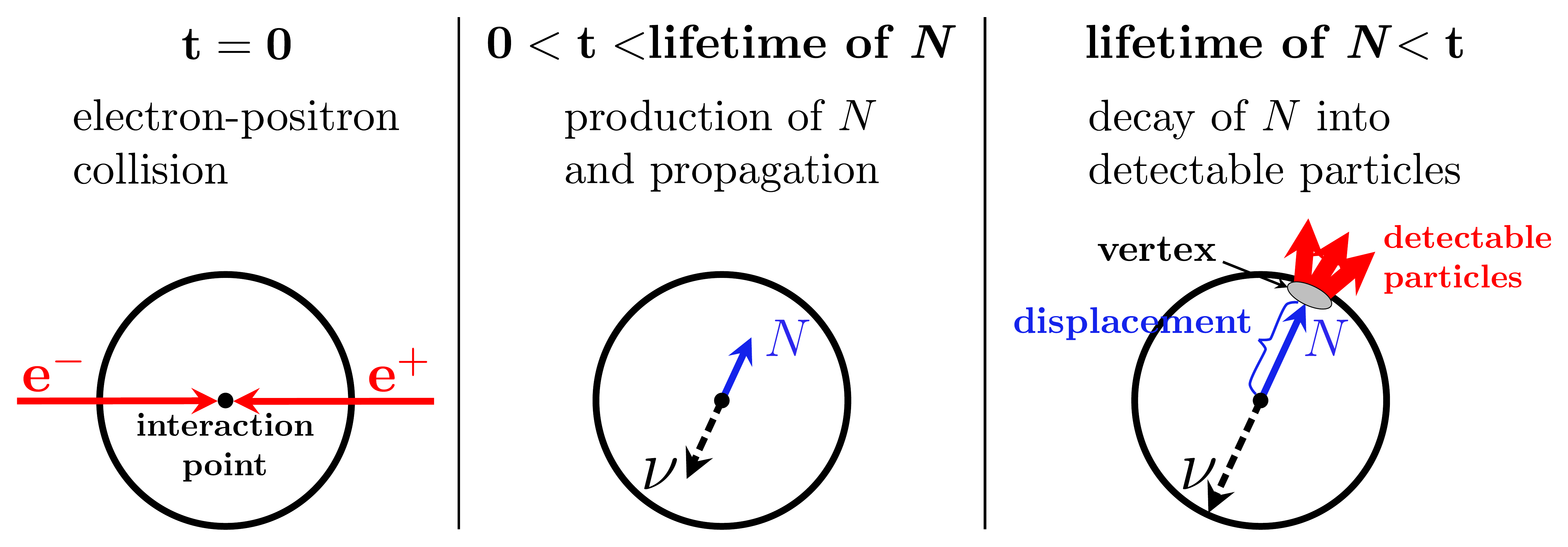}
\end{center}
\caption{Heavy neutrinos with long lifetimes yield the exotic signature of a displaced vertex, which is the visible displacement of the vertex from the interaction point.
This signature allows to search for heavy neutrinos down to tiniest active-sterile mixings.}
\label{fig:displacedvertex}
\end{figure}

The upper bound on the active-sterile mixing angles $U_a^2$ that is consistent with the observed baryon asymmetry of the universe in the minimal seesaw model results in comparatively long lifetimes in the range of picoseconds to nanoseconds for the heavy neutrinos with masses between a few GeV and the $W$ boson mass \cite{Gronau:1984ct}. 
Therefore, the heavy neutrinos produced in the particle collisions have a long enough lifetime to travel a finite distance before they decay inside the detector, giving rise to a displaced vertex.
The origin of the displaced vertex signature by heavy neutrinos\footnote{See e.g.\ also ref.\ \cite{Curtin:2017quu} for a discussion of this signature in other theoretical frameworks.} at particle colliders is shown schematically in figure~\ref{fig:displacedvertex}.
Such an exotic signature allows for extremely sensitive tests of the active-sterile mixing angles for heavy neutrino masses below $\sim 80\, \GeV$, especially for future lepton colliders with their high integrated luminosities, see e.g.\ ref.~\cite{Blondel:2014bra,Antusch:2016vyf}.

The part of the viable leptogenesis parameter region that can be accessed by colliders corresponds to the symmetry protected scenario described in section~\ref{Sec:Sym} \cite{Antusch:2015mia,Drewes:2016gmt,Drewes:2016jae}.\footnote{This statement is clearly true for the region that can be accessed by ILC and CEPC under the assumptions about the center-of-mass energies and luminosities used here. It may be violated for the smallest mixing angles that can be probed with FCC-ee. We postpone a detailed analysis of this specific region to future work.} 
Mixing angles much larger than the estimate (\ref{NaiveSeesaw}) require $\upepsilon\ll1$, and explaining the BAU with $n_s=2$ requires  $\upmu\ll1$.
For the collider phenomenology it is sufficient to consider the model (\ref{PseudoDiracL}) with $\upvarepsilon_a=\upmu=0$. Small deviations from this limit have a negligible impact on the production and decay rates and we use the results from ref.~\cite{Antusch:2016vyf}, wherein displaced vertex searches are discussed in the symmetric limit. 
As will be discussed in more details in section \ref{sec:5.3}, due to the small mass splittings the heavy neutrinos can oscillate into their antiparticles and vice versa. This has no effect on the sensitivity of the considered displaced vertex searches, since we do not distinguish events from heavy neutrinos and antineutrinos anyways. 

We focus on the following future lepton colliders with these specific physics programs for our study: 
\begin{itemize}
\item FCC-ee: The Future Circular Collider in the electron positron mode with its envisaged high integrated luminosity of $\mathcal{L}=10\,\text{ab}^{-1}$ for the $Z$ pole run\footnote{It also features a physics run at $240\, \GeV$ center-of-mass energy with an integrated luminosity of $\mathcal{L}=5\,\text{ab}^{-1}$ same as the CEPC however the $Z$ pole run is more competitive at the FCC-ee.}.
\item CEPC: The Circular Electron Positron Collider running at the $Z$ pole and $240\, \GeV$ center-of-mass energy with an integrated luminosity $\mathcal{L}=0.1\,\text{ab}^{-1}$ and $5\,\text{ab}^{-1}$, respectively.
\item ILC: The International Linear Collider running at the $Z$ pole and $500\, \GeV$ center-of-mass energy with an integrated luminosity of $\mathcal{L}=0.1\,\text{ab}^{-1}$ and $\mathcal{L}=5\,\text{ab}^{-1}$, respectively. 
\end{itemize}

Lepton colliders produce heavy neutrinos primarily by the process $e^+e^- \to \nu N$. 
At the $Z$ pole, the production process is dominated by the $s$-channel exchange of a $Z$ boson while at both center-of-mass energies of $240$ and $500\, \GeV$ the production process is dominated by the $t$-channel exchange of a $W$ boson. 
For its cross section $\sigma_{\nu N}$ we can thus take the following mixing angle dependency for the two cases: $\sigma_{\nu N}(U^2)$ at the $Z$ pole and $\sigma_{\nu N}(U_e^2)$ for the center-of-mass energies above the $Z$ pole. 

\begin{figure}
\begin{center}
\includegraphics[width=0.75\textwidth]{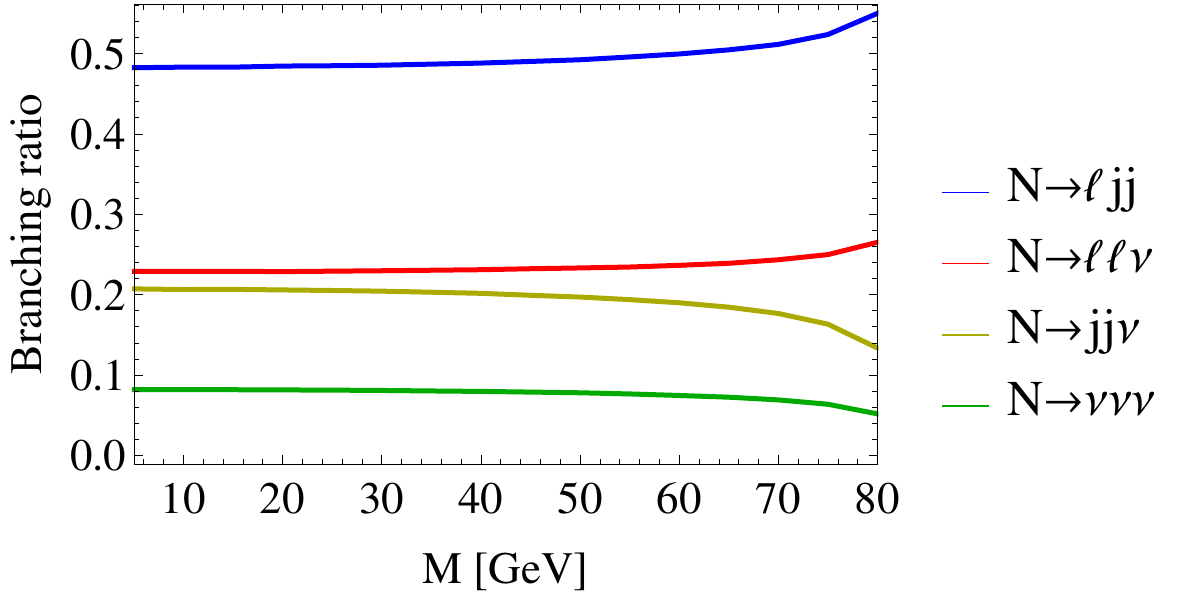}
\end{center}
\caption{Branching ratios of heavy neutrino decays. The colour code denotes the different possible final states, namely the semileptonic lepton-dijet (``$\ell jj$'', blue line), the dilepton (``$\ell\ell \nu$'', red line), the dijet (``$jj\nu$'', yellow line), and the invisible decays (``$\nu\nu\nu$'', green line). The semileptonic and leptonic branching ratios are summed over all lepton flavors.}
\label{fig:branchings}
\end{figure}

The produced heavy neutrinos decay into four different classes of final states: semileptonic ($N\to\ell jj$), leptonic ($N\to\ell\ell\nu$), hadronic ($N\to jj\nu$), and invisible ($N\to\nu\nu\nu$). 
We display the branching ratios\footnote{We note that for heavy neutrino masses around and below 5 GeV, the here employed parton picture is not sufficient. See e.g.\ refs.~\cite{Gorbunov:2007ak,Atre:2009rg} for a discussion of heavy neutrinos into vector and scalar mesons.} for the four classes with varying heavy neutrino mass in figure~\ref{fig:branchings}. 
We note that for the branching ratios in the figure we summed over all the lepton flavours. 
We are mainly interested in semileptonic decays of the heavy neutrino, which provide a charged lepton of flavour $a$ from which one can probe the squared mixing angle $U_a^2$ and thus test the flavour mixing pattern. 
The resulting branching ratios of the semileptonic decays with a charged lepton $\ell_a$ are given by Br$(N\to\ell_a jj)\simeq 0.5\times U_a^2/U^2$. 
In the narrow width approximation, the expected number of the displaced decay events of the heavy neutrino that decay semileptonic with a charged lepton of flavour $a$ can be expressed as
\begin{equation}
N_a = \sigma_{\nu N}(\sqrt{s},\bar M,U_e,U_\mu,U_\tau)\,\times\text{Br}(N \to \ell_a j j)\,\times \mathcal{L}\,\times P(x_1,x_2,\uptau)\,.
\end{equation}
Here $\mathcal{L}$ denotes the integrated luminosity of the experiment, and $P(x_1,x_2,\uptau)$ denotes the fraction of the displaced decays of the heavy neutrino with proper lifetime $\uptau$ to take place between the detector-defined boundaries $x_1$ and $x_2$. 
The lifetime is given by the inverse of the total decay width, which is proportional to $U^2\bar{M}^5$ if the masses of final state particles are neglected.
The probability of particle decays follows an exponential distribution such that $P$ can be written for $x_2\geq x_1$ as
\begin{equation}
P(x_1,x_2,\uptau)=\exp\left(-\frac{x_1}{\beta\gamma c\uptau}\right)-\exp\left(-\frac{x_2}{\beta\gamma c\uptau}\right)
\end{equation}
with the relativistic $\beta=v/c$ and Lorentz factor $\gamma$. 
We assume that the displaced vertex signature is free from SM background (see ref.\ \cite{Antusch:2016vyf}) for the boundaries as given by an SiD-like detector \cite{Behnke:2013lya}, given by the inner region ($x_1=10\,\mu\text{m}$) and the outer radius of the tracker ($x_2=1.22 \, \text{m}$). The numerical calculation of the cross section for the different discussed performance parameters of the considered colliders is done in \texttt{WHIZARD}    \cite{Kilian:2007gr, Moretti:2001zz} by including initial state radiation and only for the ILC by including also a (L,R) initial state polarisation of (80\%,20\%) and beamstrahlung effects. 

For the expected number of semileptonic events, $N_{\rm sl}=\sum_a N_a$, we demand at least four events over the zero background hypothesis to establish a signal above the 2$\sigma$ level.
In the case of the $Z$ pole run, the active-sterile mixing dependence of $N_{\rm sl}$ is given by $U^2$, which allows to uniquely infer $U^2$ for $N_{\rm sl}$ exceeding four events. In the case of the center-of-mass energies above the $Z$ pole run, however, $U^2$ cannot be determined uniquely since $N_{\rm sl}$ depends differently on $U^2_e$ than on $U^2_\mu$ and $U^2_\tau$ due to the production cross section being dependent on $U^2_e$. This ambiguity leads to the ``guaranteed discovery'' and ``potential discovery'' regions discussed at the end of section~\ref{HowToDo}.

Moreover, the heavy neutrino mass $\bar M$ could be measured from the invariant mass of the semileptonic final states $M_{\ell jj}$. Its precision can be assumed to be of the same order as the jet-mass reconstruction, which is $\sim 4\,\%$ for jet energies of $45\, \GeV$ with the {\tt Pandora Particle Flow Algorithm} \cite{Marshall:2013bda}. If a sizeable number of events is present, a more precise mass resolution may come from the $\nu \mu^-\mu^+$ final states. For a displaced vertex the neutrino momentum can be inferred from the requirement of pointing back to the primary vertex, which yields the invariant mass $M_{\nu \mu \mu}$ \cite{mogens_dam}.

In order to determine the achievable precision on measuring the flavour pattern realised by leptogenesis we consider only the statistical uncertainties on the flavour mixing ratios $U_a^2/U^2$.
The observable random variables are $\hat N_{\rm sl}$ which is Poisson distributed with mean $N_{\rm sl}$, and the $\hat N_a$ which follow a multinomial distribution with probability $p_a=U_a^2/U^2$. 
The expected number of semileptonic decays with a lepton of flavour $a$ is given as above by $N_a = N_{\rm sl} U_a^2/U^2$. 
The precision on measuring $U_a^2/U^2$, expressed as $\frac{\delta(U_a^2/U^2)}{U_a^2/U^2}$ with $\delta$ being the standard deviation for $U_a^2/U^2$, comes from the statistical uncertainty of the ratio $ N_a/ N_{\rm sl}$. 
Since $N_a$ is not independent on $N_{\rm sl}$ the following precision for the flavour mixing ratio $U_a^2/U^2$ results in
\begin{equation}
\frac{\delta(U_a^2/U^2)}{U_a^2/U^2} \approx \sqrt{\frac{1}{N_a}-\frac{1}{N_{\rm sl}}}\,,
\end{equation}
unlike for the usual propagation of error where the uncertainties add. This point is discussed in more detail in appendix \ref{Appendix_2}.
We confront the leptogenesis parameter space region with the achievable statistical precision of the flavour-dependent mixing $U_a^2/U^2$ for the different lepton colliders in section~\ref{Section5.2}.


\section{Results}
\label{Section5_Results}
 
\subsection{Sensitivity in the $\bar{M}-U^2$ plane}
\label{Section5.1}

\begin{figure}
        \centering
        \begin{tabular}{cc}
			\textbf{Normal Ordering} & \textbf{Inverted Ordering} \\\\
            \includegraphics[width=0.43\textwidth]{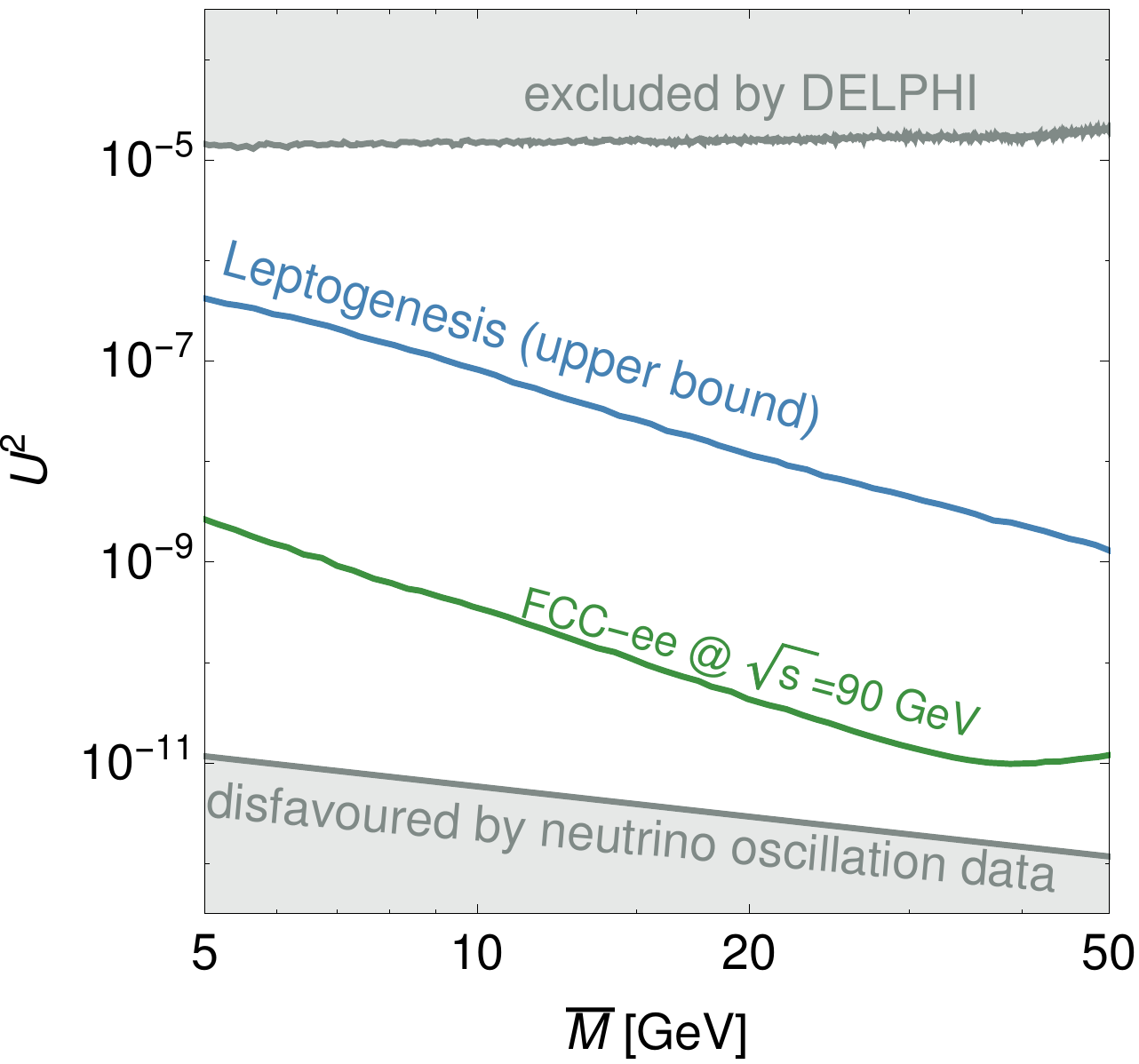} &
			\includegraphics[width=0.43\textwidth]{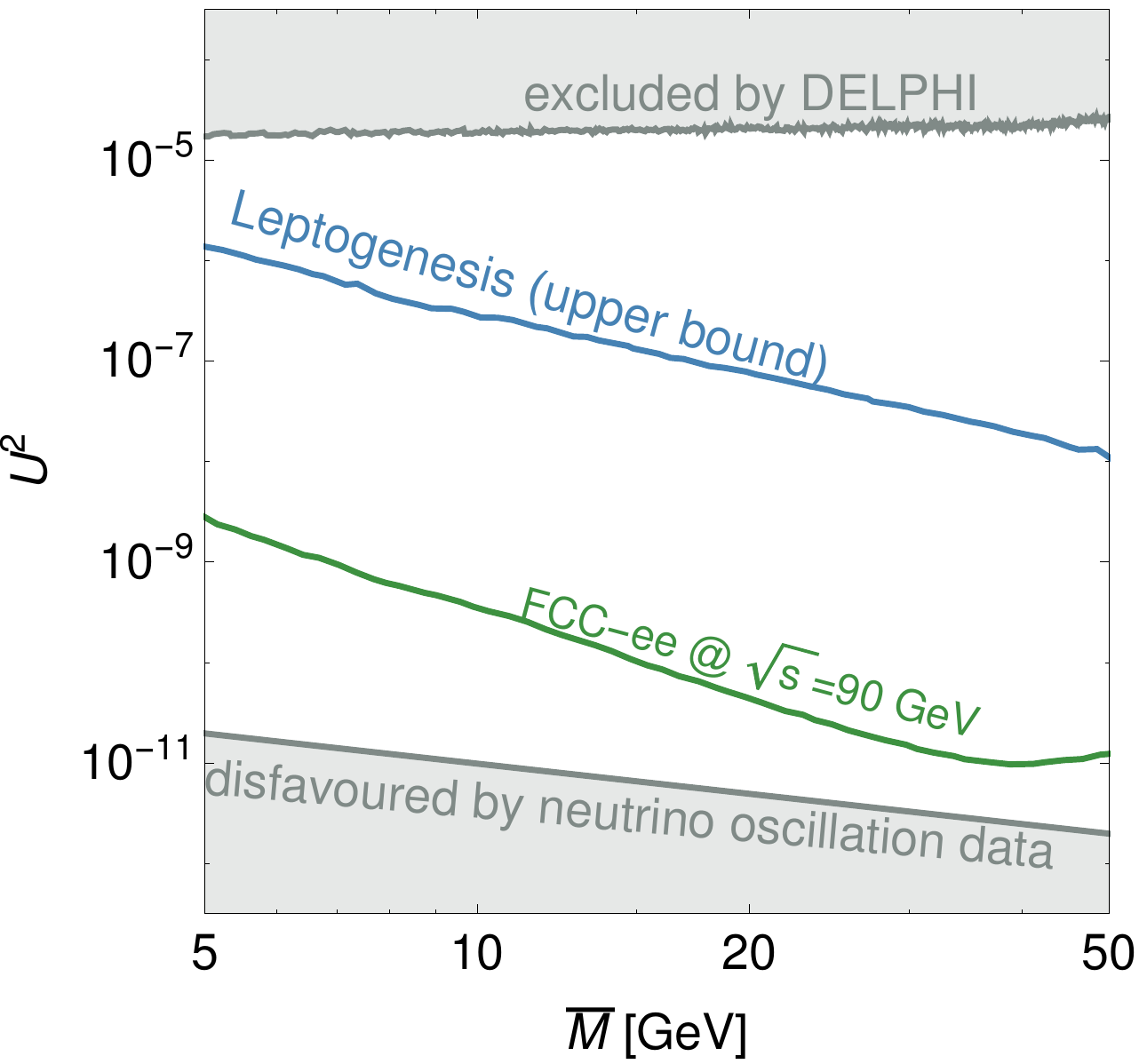}\\
			\includegraphics[width=0.43\textwidth]{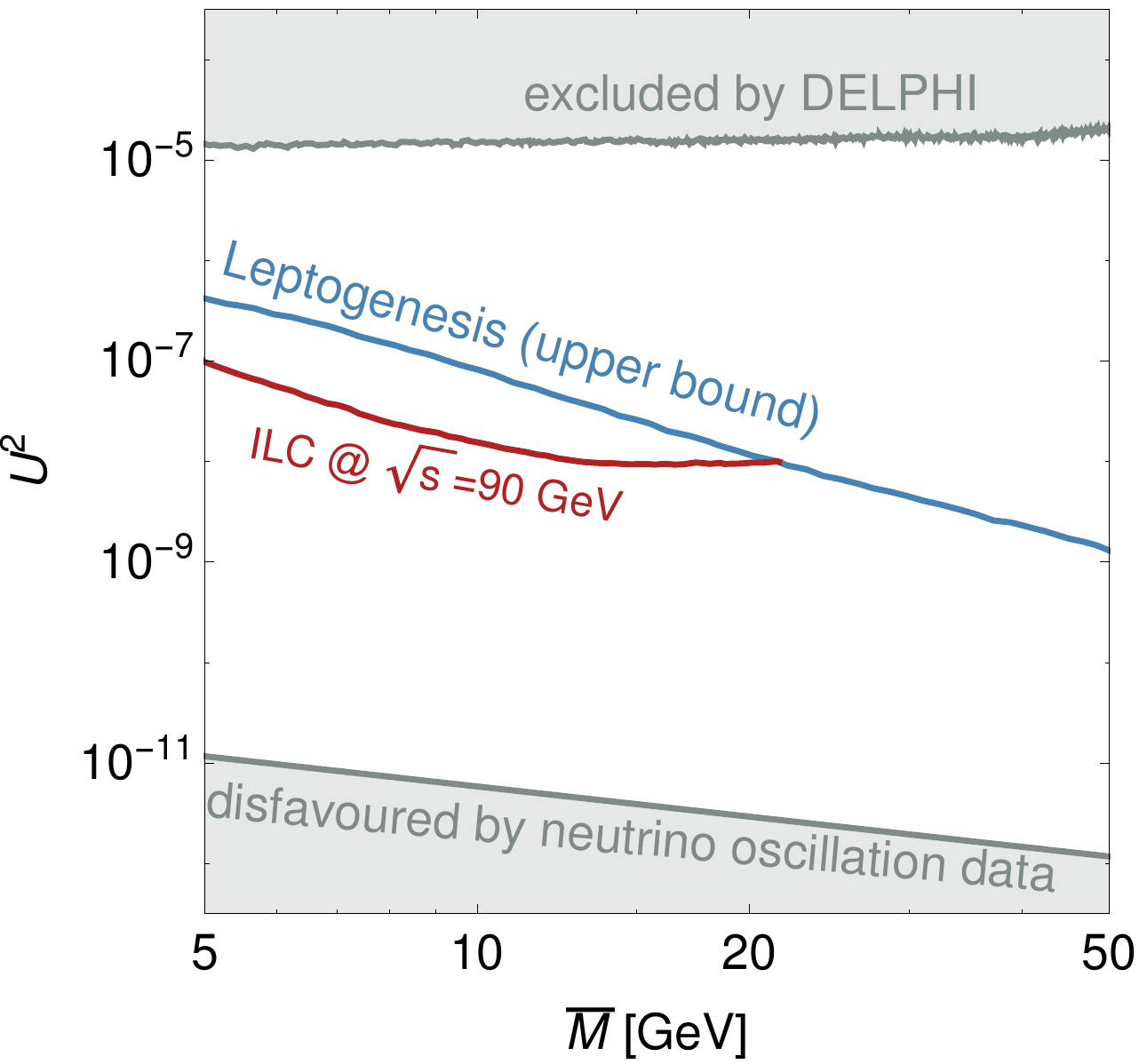} &
			\includegraphics[width=0.43\textwidth]{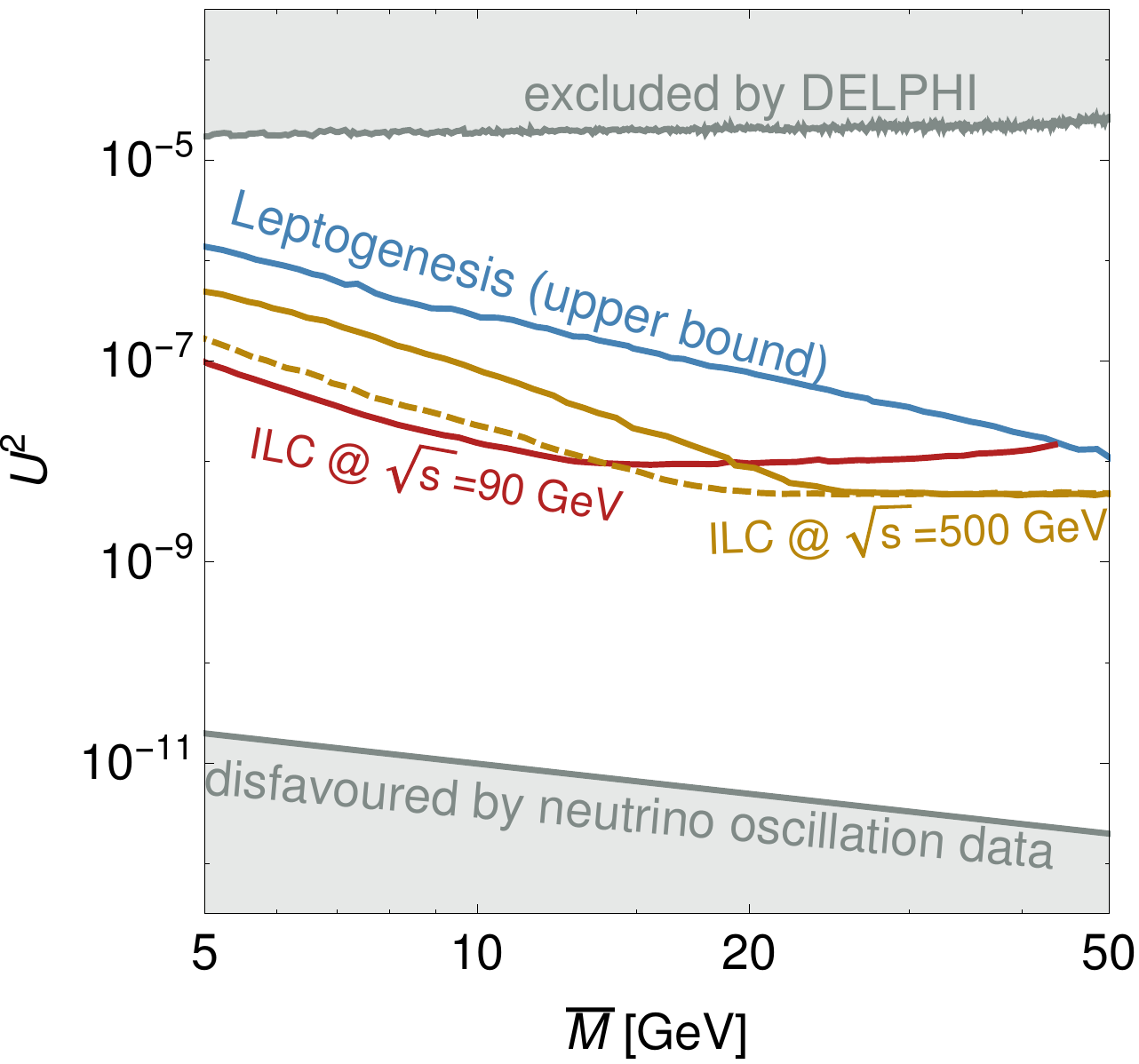}\\
            \includegraphics[width=0.43\textwidth]{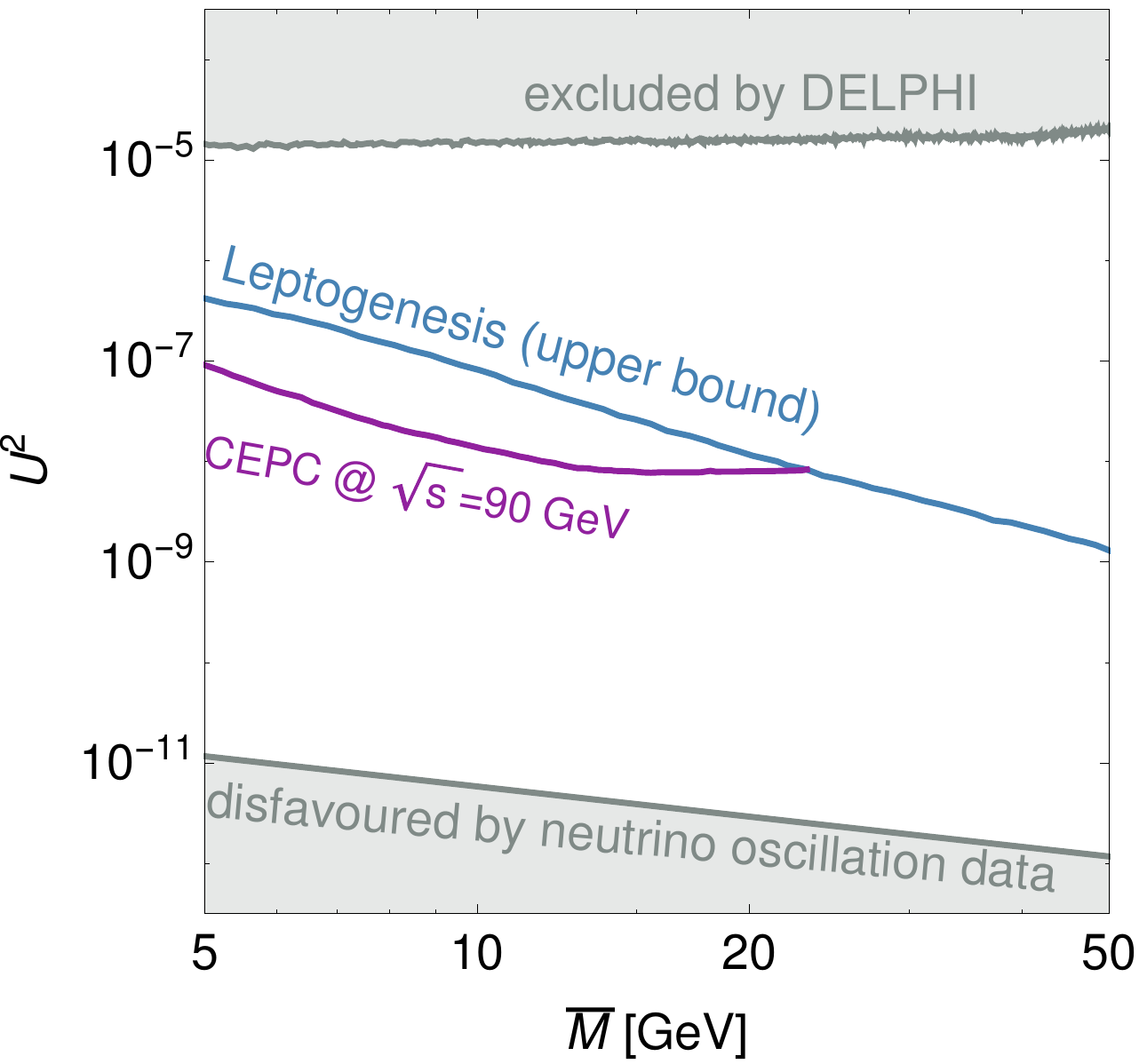} &
			\includegraphics[width=0.43\textwidth]{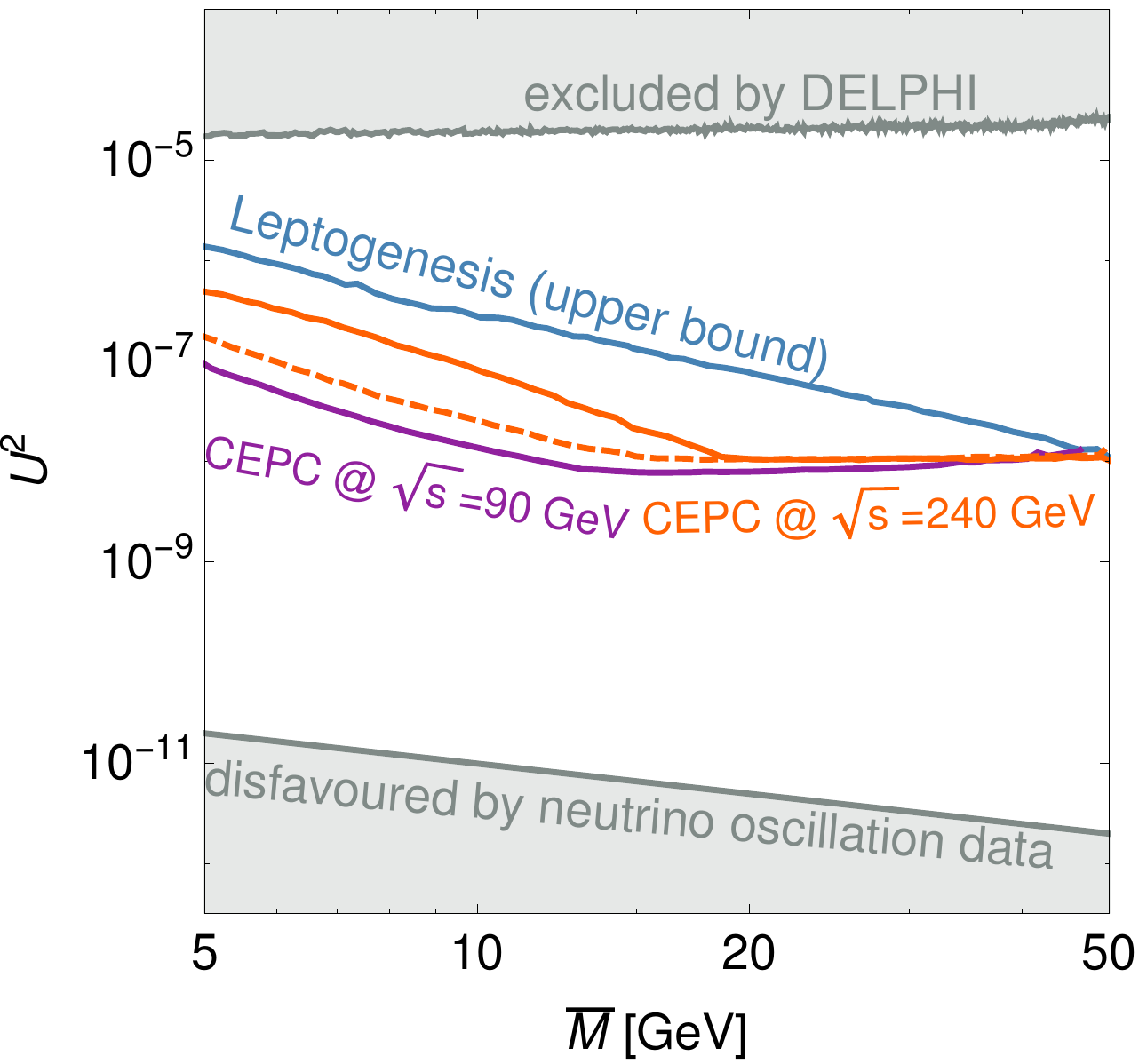}
        \end{tabular}
\caption{\label{fig:SensitivityU2M}
The blue ``BAU'' line shows the largest possible $U^2$ for which the BAU can be generated for given $\bar{M}$, as found in the parameter scan described in section~\ref{HowToDo}. 
The other coloured lines mark the parameter regions in which future lepton colliders can 
observe at least four expected displaced vertex events  
from $N_i$ with properties that are consistent with successful leptogenesis. 
The solid and dashed lines correspond to the ``guaranteed discovery area'' and ``potential discovery area'' discussed in section~\ref{HowToDo}.
The grey area is disfavoured by DELPHI (on the top) and the neutrino oscillation data (at the bottom). We show no lower bound on $U^2$ from leptogenesis because it is lower than the constraint from neutrino oscillation data in this mass range.
More details are given in the main text, cf. section \ref{Section5.1}.}
\end{figure}

\begin{figure}
        \centering
        \begin{tabular}{cc}
			\textbf{Normal Ordering} & \textbf{Inverted Ordering} \\\\
            \includegraphics[width=0.43\textwidth]{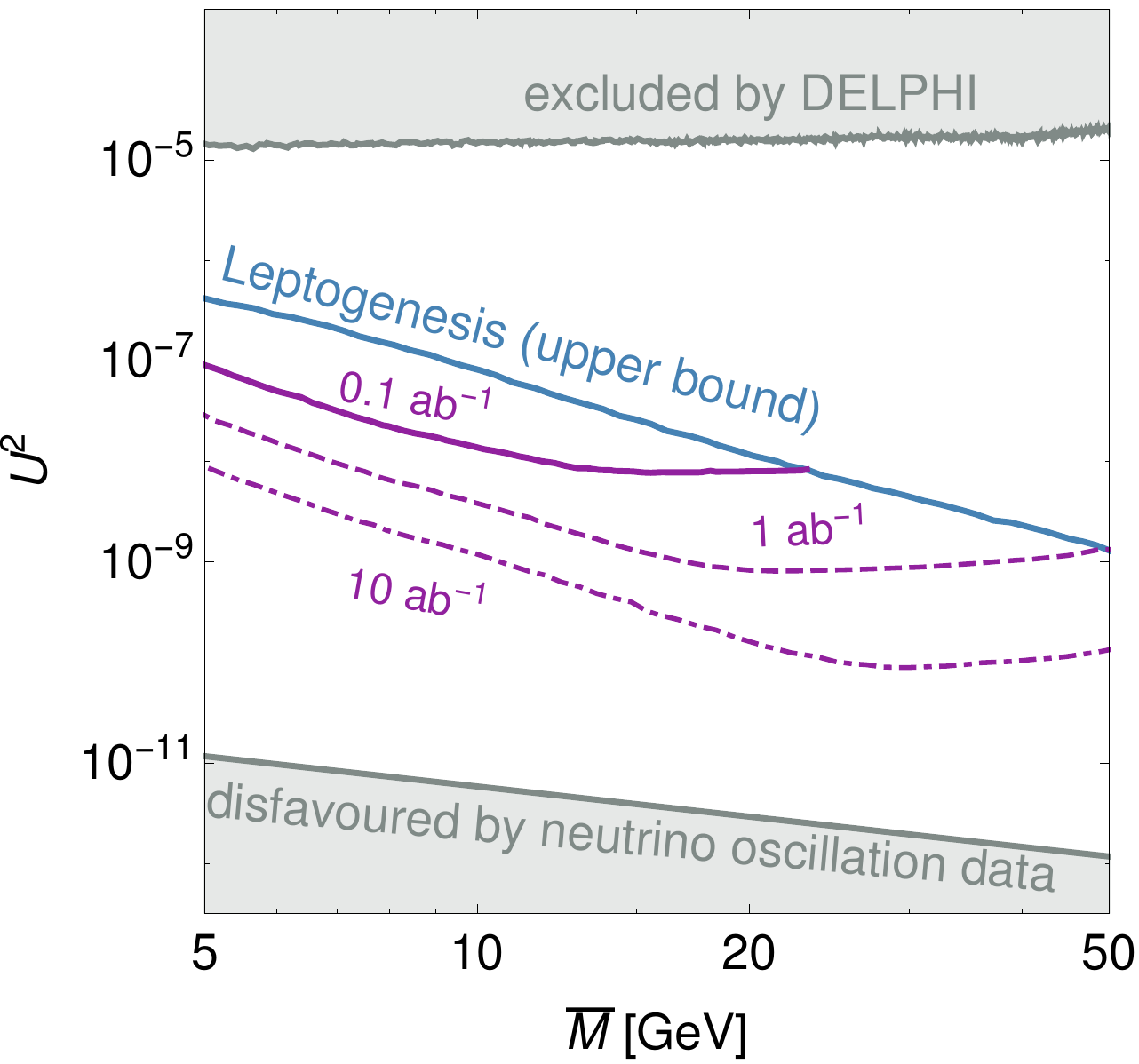} &
			\includegraphics[width=0.43\textwidth]{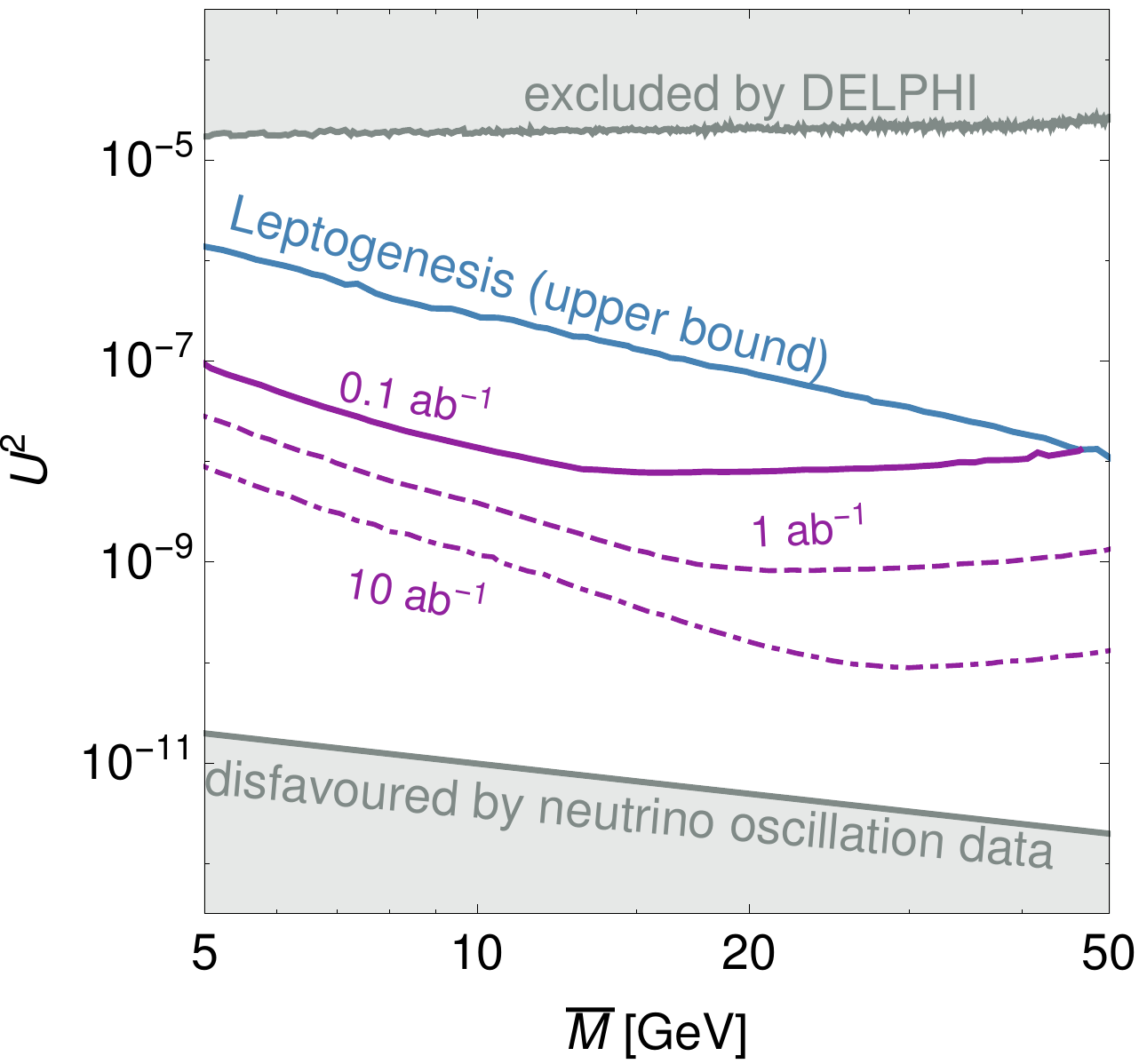}
        \end{tabular}
\caption{\label{fig:CEPCcomparison}
The blue ``BAU'' line shows the largest possible $U^2$ for which the BAU can be generated for given $\bar{M}$, as found in the parameter scan described in section~\ref{HowToDo}. 
The purple lines lines mark the parameter regions in which the CEPC experiment can observe at least four expected displaced vertex events  from $N_i$ with properties that are consistent with successful leptogenesis. The solid line corresponds to the currently planned run, the dashed line corresponds to the equal Z-pole running time as is currently planned by FCC-ee, while the dot-dashed line corresponds to what is possible with the crab waist technology. The grey area is disfavoured by DELPHI (on the top) and the neutrino oscillation data (at the bottom). We show no lower bound on $U^2$ from leptogenesis because it is lower than the constraint from neutrino oscillation data in this mass range.
More details are given in the main text, cf. section \ref{Section5.1}.
}
\end{figure}

In figure~\ref{fig:SensitivityU2M} we show the region in the $\bar{M}-U^2$ plane consistent with leptogenesis and the potential of future collider experiments from the displaced vertex search to probe sterile (right-handed) neutrinos with this mass and active-sterile mixing.

The grey region at the top of the plots corresponds to the experimental constraint on $U^2$ from DELPHI \cite{Abreu:1991pr,Abreu:1996pa}. We have not included the constraint from displaced vertex searches from LHCb (cf.~ref.~\cite{Antusch:2017hhu}), which are slightly more sensitive in the range between $5$ and $10\,\GeV$ but do not directly probe only $U^2$ (it probes mainly $U_\mu^2$). The grey area at the bottom is excluded since in this region, assuming two right-handed neutrinos, the observed two light neutrino mass squared differences cannot be generated. The region below the blue line indicates the parameter space for which the baryon asymmetry can successfully be generated by leptogenesis.

The left column of the figure shows the case of normal ordering (NO) for the light neutrino masses, and the right column the case of inverse ordering (IO). We show the results for $\bar{M} < 50\,\GeV$ since above $50\,\GeV$ the uncertainty for the leptogenesis calculation increases significantly. Regarding the collider sensitivities, we show the lines for four expected events. The FCC-ee with $\sqrt{s}=90\,\GeV$ (solid, green) is displayed in the top row, the ILC with $\sqrt{s}=90\,\GeV$ (solid, red) and with $\sqrt{s}=500\,\GeV$ (solid, brown) in the middle row, while the discovery line of the CEPC with $\sqrt{s}=90\,\GeV$ (solid, purple) and with $\sqrt{s}=240\,\GeV$ (solid, orange)\footnote{Since the FCC-ee also features a physics run at $\sqrt{s}=240\,\GeV$ with the same integrated luminosity this result is also valid for the FCC-ee.} is shown in the bottom row. 

With the performance parameters and run times considered, the FCC-ee clearly shows the best prospects of probing the right-handed neutrinos involved in the leptogenesis mechanism. It can cover a large part of the parameter region consistent with leptogenesis. 

The ILC and the CEPC both have significantly better sensitivity for the IO case, especially regarding the runs with higher center-of-mass energy. In fact, there are no expected events consistent with leptogenesis for ILC with $\sqrt{s}=500\,\GeV$ and CEPC with $\sqrt{s}=240\,\GeV$ in case of normal ordering. The reason is that for the NO case the electron mixing, which mediates the dominant production channel for the higher energy runs, is suppressed compared to the mixing to the other flavours through the requirement of reproducing low energy neutrino parameters.  

For the ILC with $\sqrt{s}=500\,\GeV$ and the CEPC with $\sqrt{s}=240\,\GeV$ there are two lines shown, a dashed line and a solid line. This takes into account that the sensitivity for these runs depends not only on $U^2$, but also on $U_e^2$. The solid line means that in the parameter space for consistent leptogenesis there are parameter points which can be probed by the experiment, whereas the dashed line means that for all the leptogenesis parameter points with this $U^2$ the right-handed neutrinos can be discovered. These two cases correspond to the ``potential discovery'' and ``guaranteed discovery'' regions discussed in section~\ref{HowToDo}.

We note that compared to the FCC-ee, CEPC plans a much shorter run time for the $90\,\GeV$ run, since the current plans focus on Higgs measurements (and therefore on $240\,\GeV$). A longer CEPC run time at $90\,\GeV$ could strongly improve the discovery potential for right-handed neutrinos, up to sensitivities close to the ones of the FCC-ee, a comparison plot is provided in figure~\ref{fig:CEPCcomparison}.

Finally we remark that further above the four-event lines shown in figure~\ref{fig:SensitivityU2M}, large numbers of displaced vertex events from the long-lived heavy neutrinos could be observed, especially at the FCC-ee, as shown in figures \ref{fig:TotalU2M_FCC} and \ref{fig:TotalU2M_ILC_CECP}. As we will discuss in the next subsection, this large number of events can even allow for precise measurement of the flavour composition of the right-handed neutrinos.

\subsection{Precision for $U_a^2/U^2$ in the $U_a^2/U^2-U^2$ plane for different flavours}
\label{Section5.2}

\begin{landscape}
\begin{figure}
\begin{center}
\begin{minipage}{1.25\textheight}
	\includegraphics[width=0.33\textwidth]{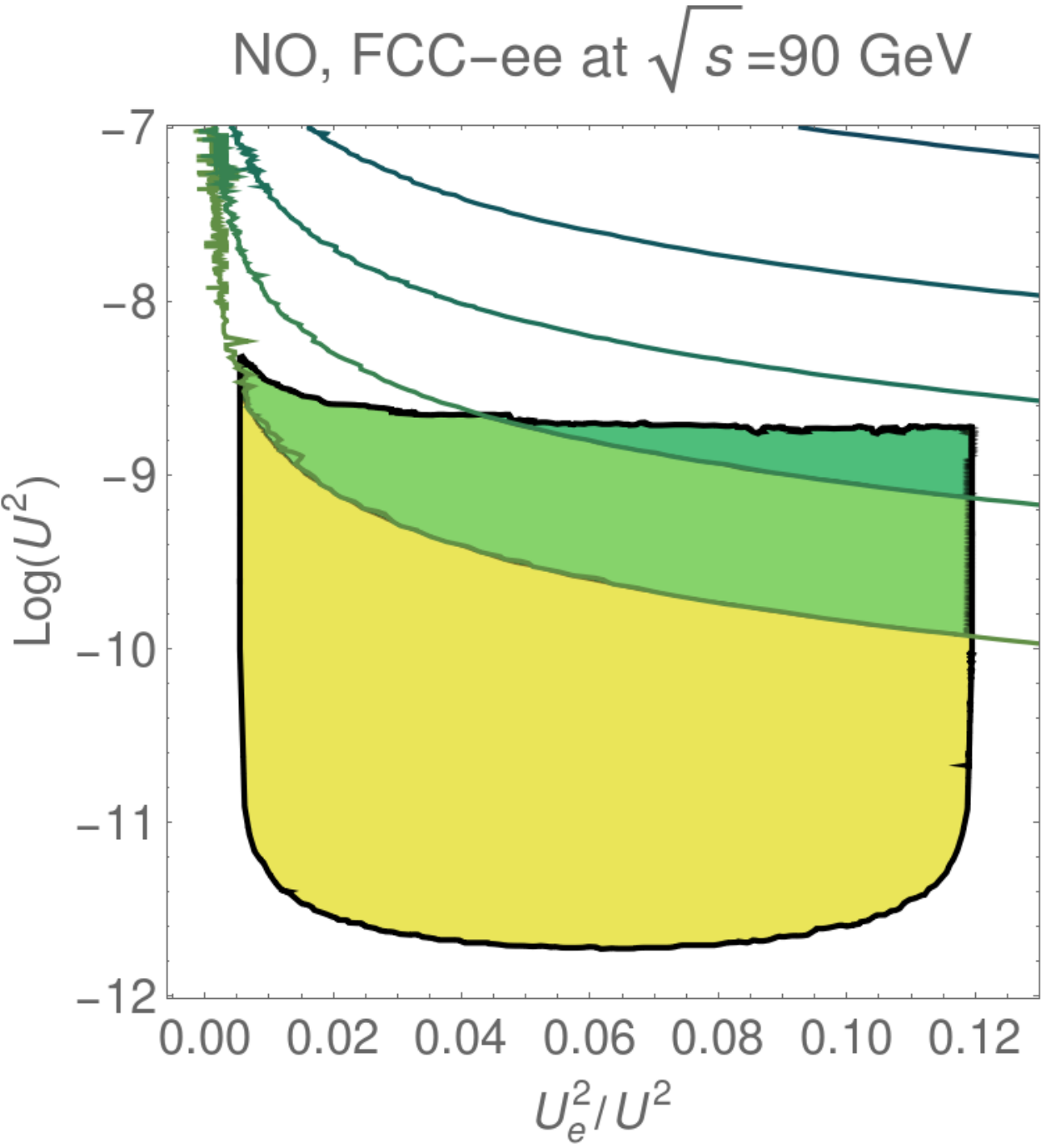}
	\quad
	\includegraphics[width=0.33\textwidth]{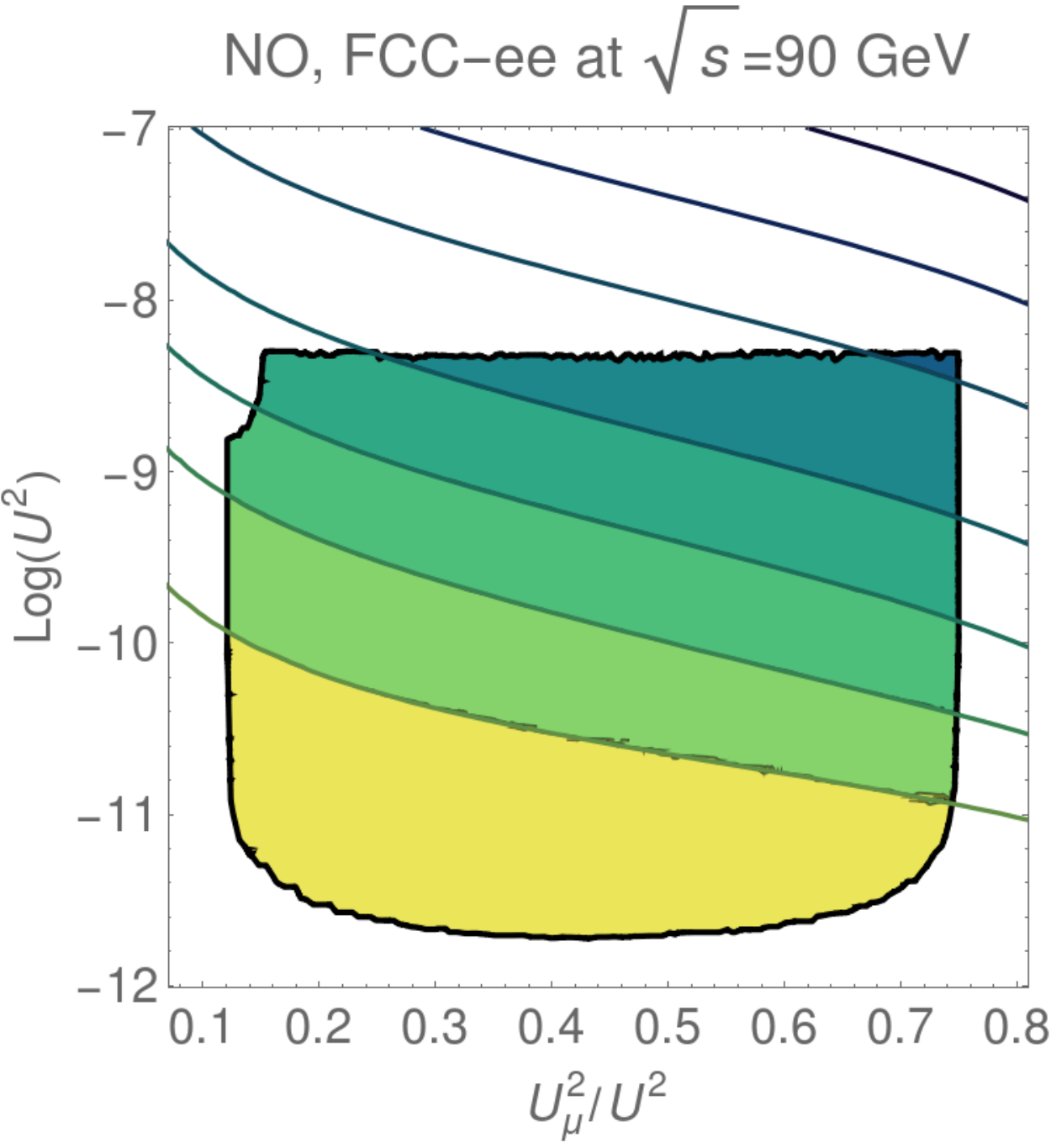}
	\quad
	\includegraphics[width=0.33\textwidth]{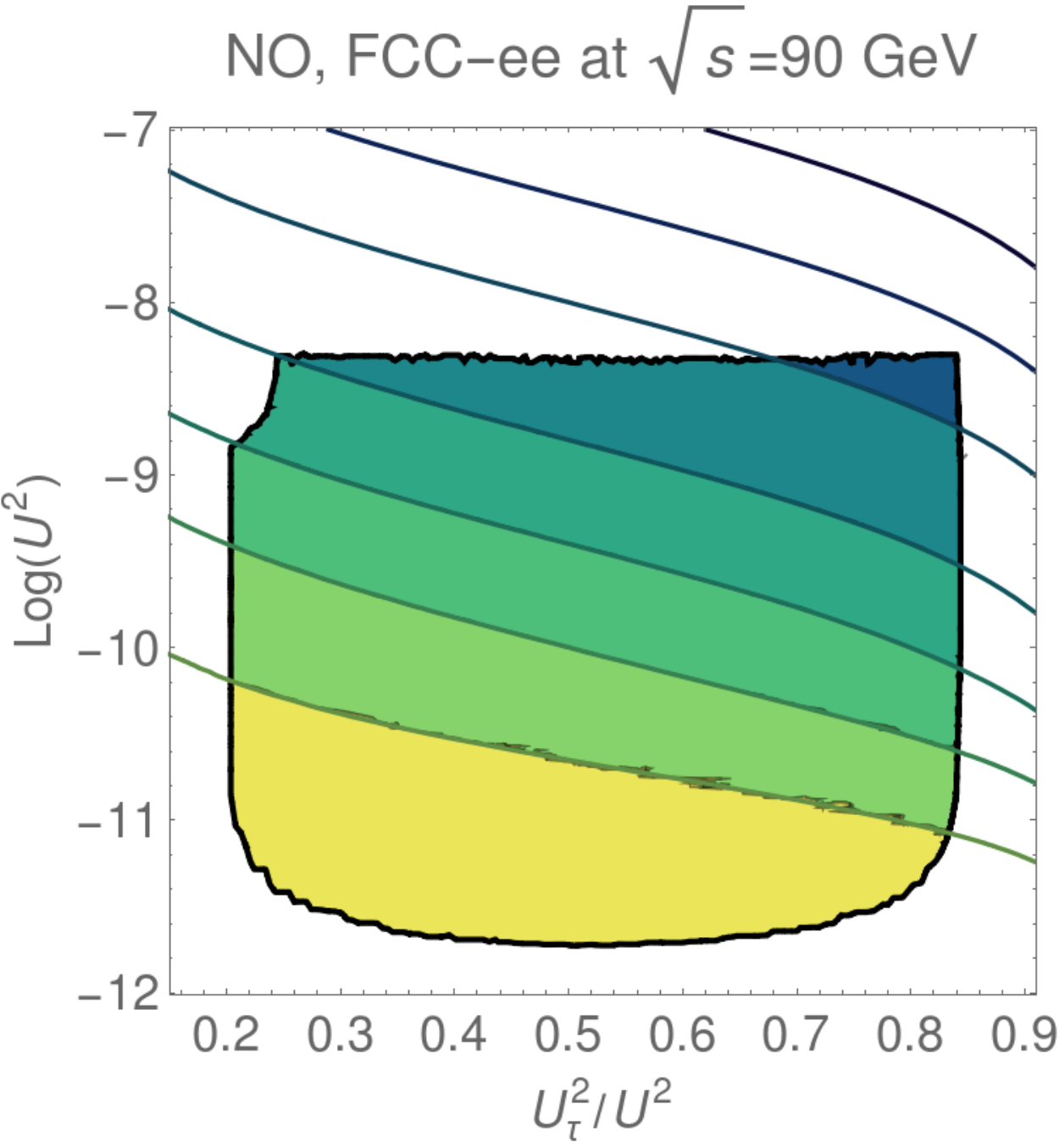}
\end{minipage}
\qquad
\begin{minipage}{0.15\textheight}
	\includegraphics[width=0.75\textwidth]{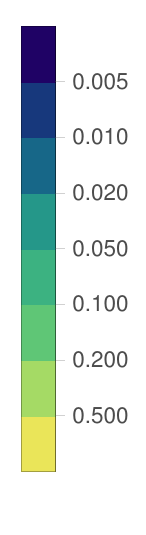}
\end{minipage}
	\\[1.5cm]
\begin{minipage}{1.25\textheight}
	\includegraphics[width=0.33\textwidth]{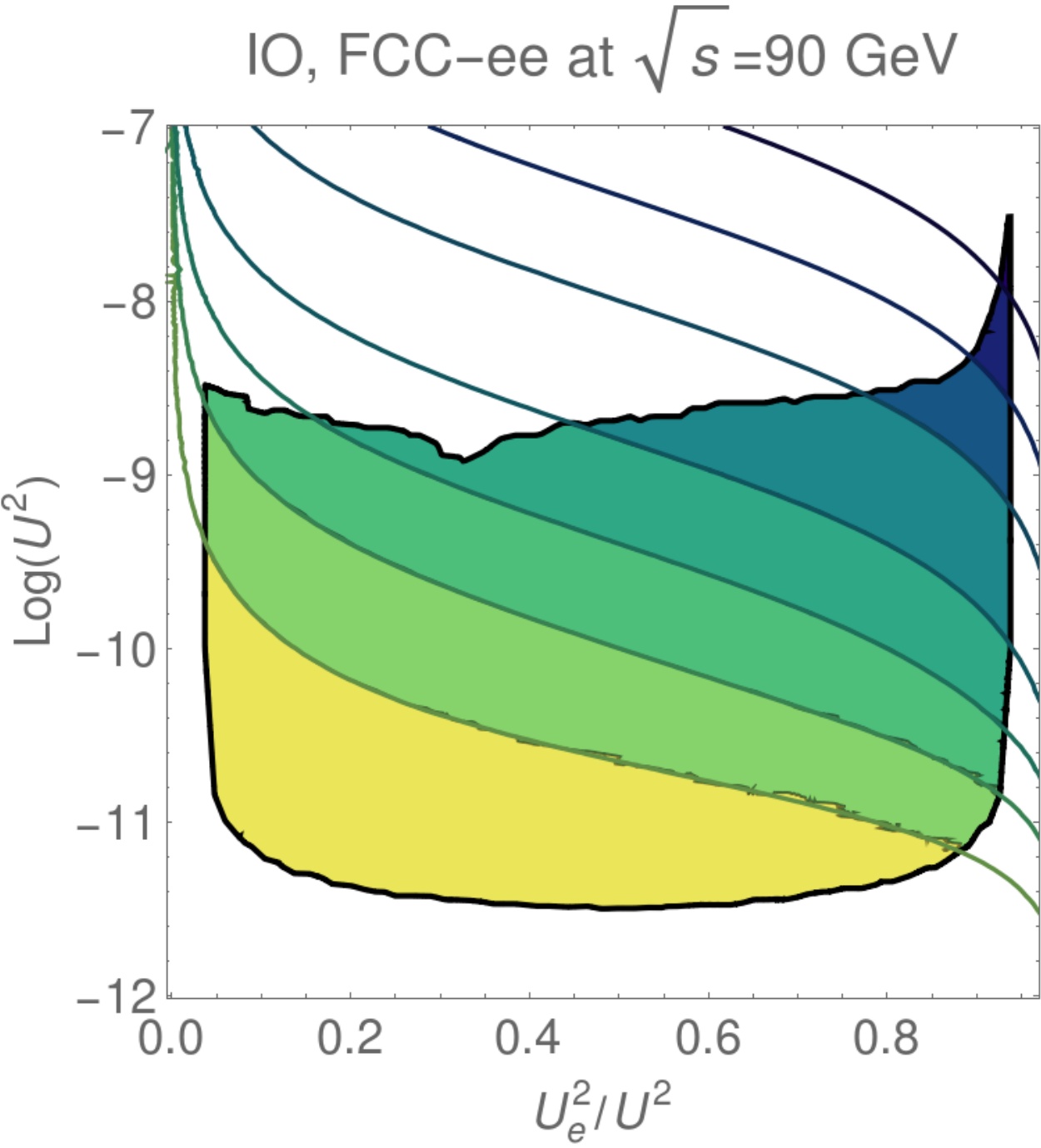}
	\quad
	\includegraphics[width=0.33\textwidth]{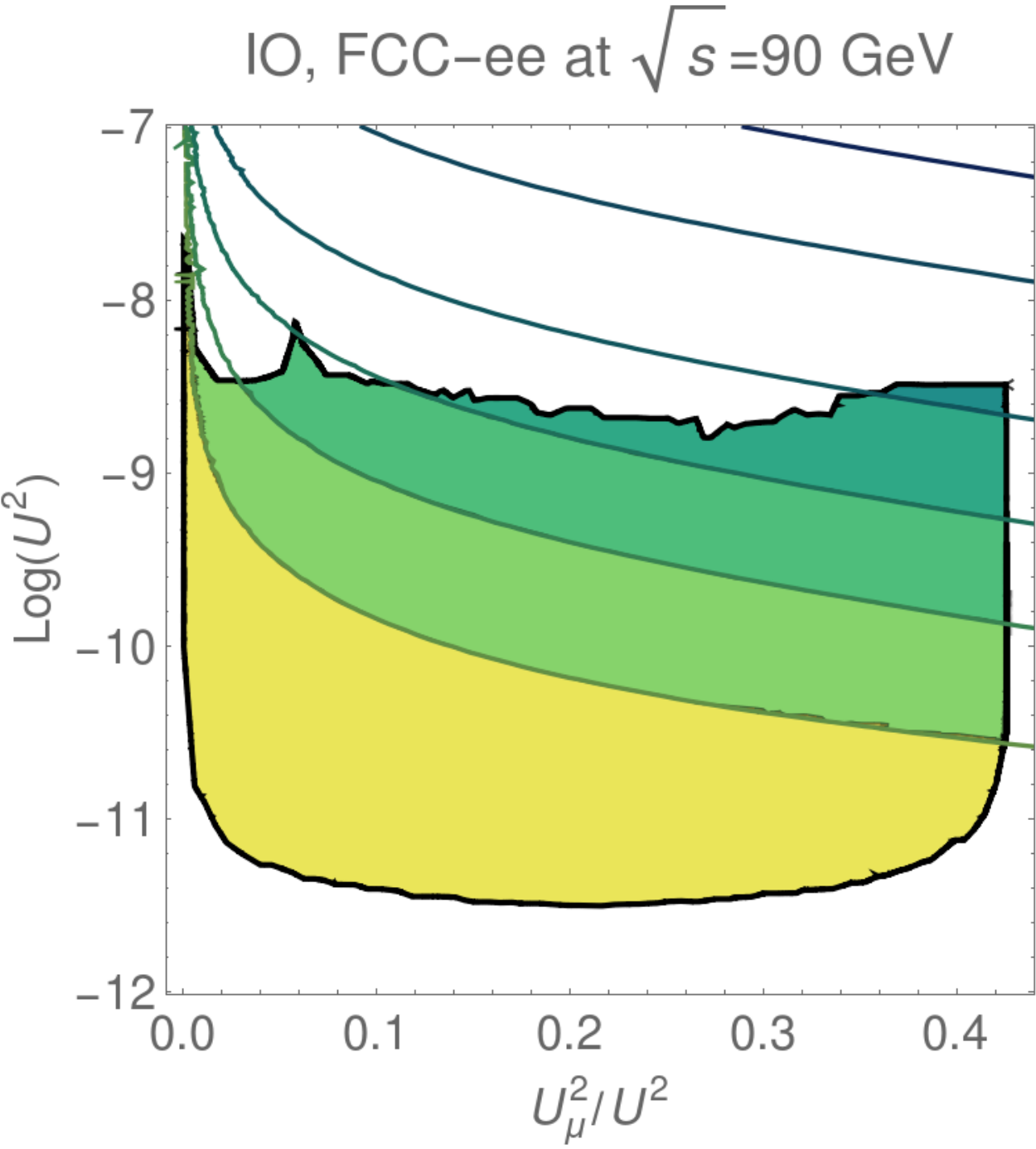}
	\quad
	\includegraphics[width=0.33\textwidth]{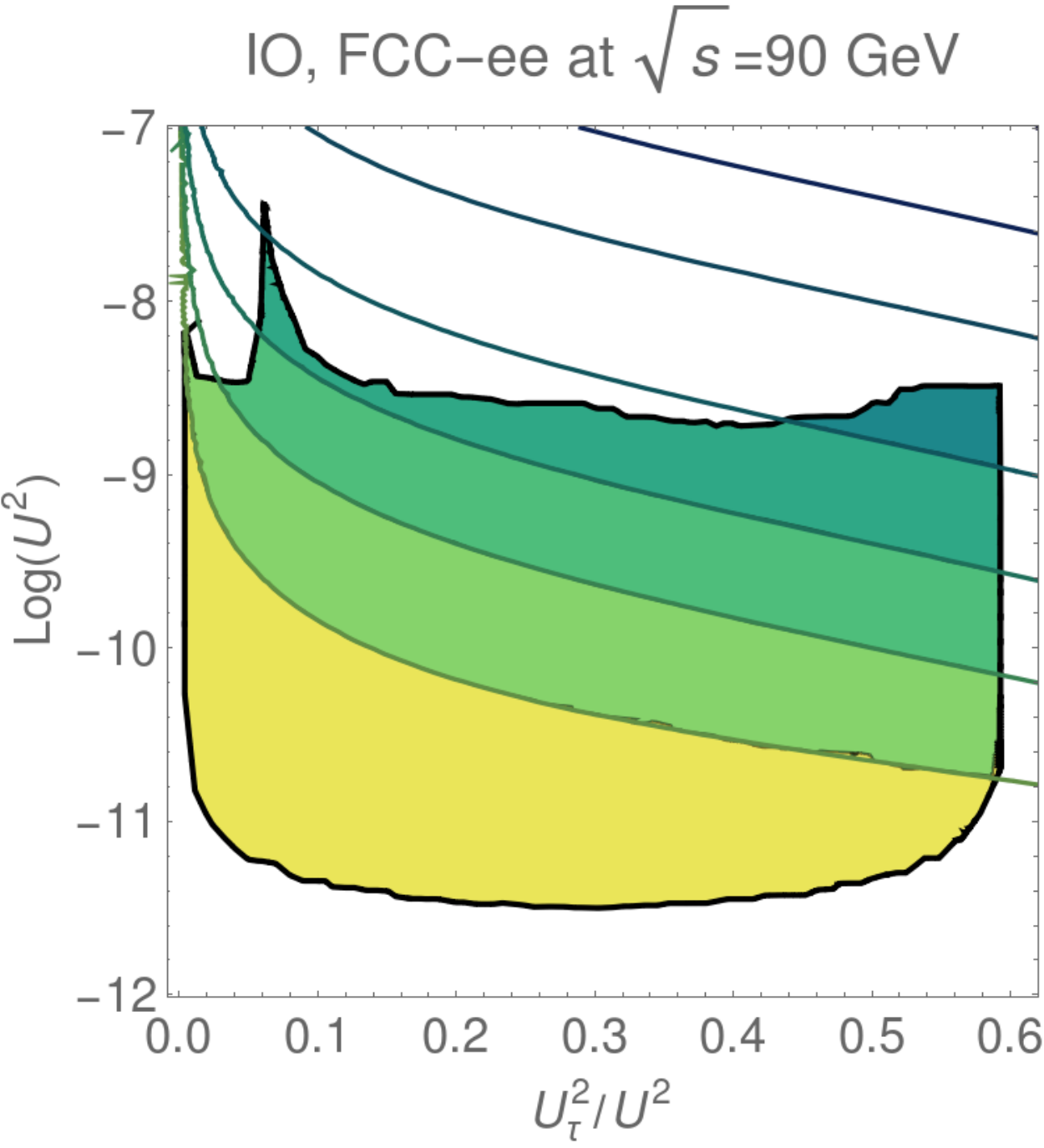}
\end{minipage}
\qquad
\begin{minipage}{0.15\textheight}
	\includegraphics[width=0.75\textwidth]{Figures/U2_Ua2_Final/BarLegend.pdf}
\end{minipage}	
\end{center}
\caption{\label{fig:Precision-FCC}
The colour indicates the precision that can be achieved for measuring $U_a^2/U^2$ with $a=e,\mu,\tau$ at the FCC-ee with $\sqrt{s}=90\,\text{GeV}$ for both normal ordering (NO) and inverse ordering (IO) of the light neutrino masses. Leptogenesis is viable in the regions with solid colour. The two heavy neutrinos are assumed to be almost degenerate in mass at $M=30\,\text{GeV}$. Details are given in the main text, cf. section \ref{Section5.2}.}
\end{figure}
\end{landscape}

\begin{figure}[t]
        \centering
        \begin{tabular}{ccc}
	\includegraphics[width=0.4\textwidth]{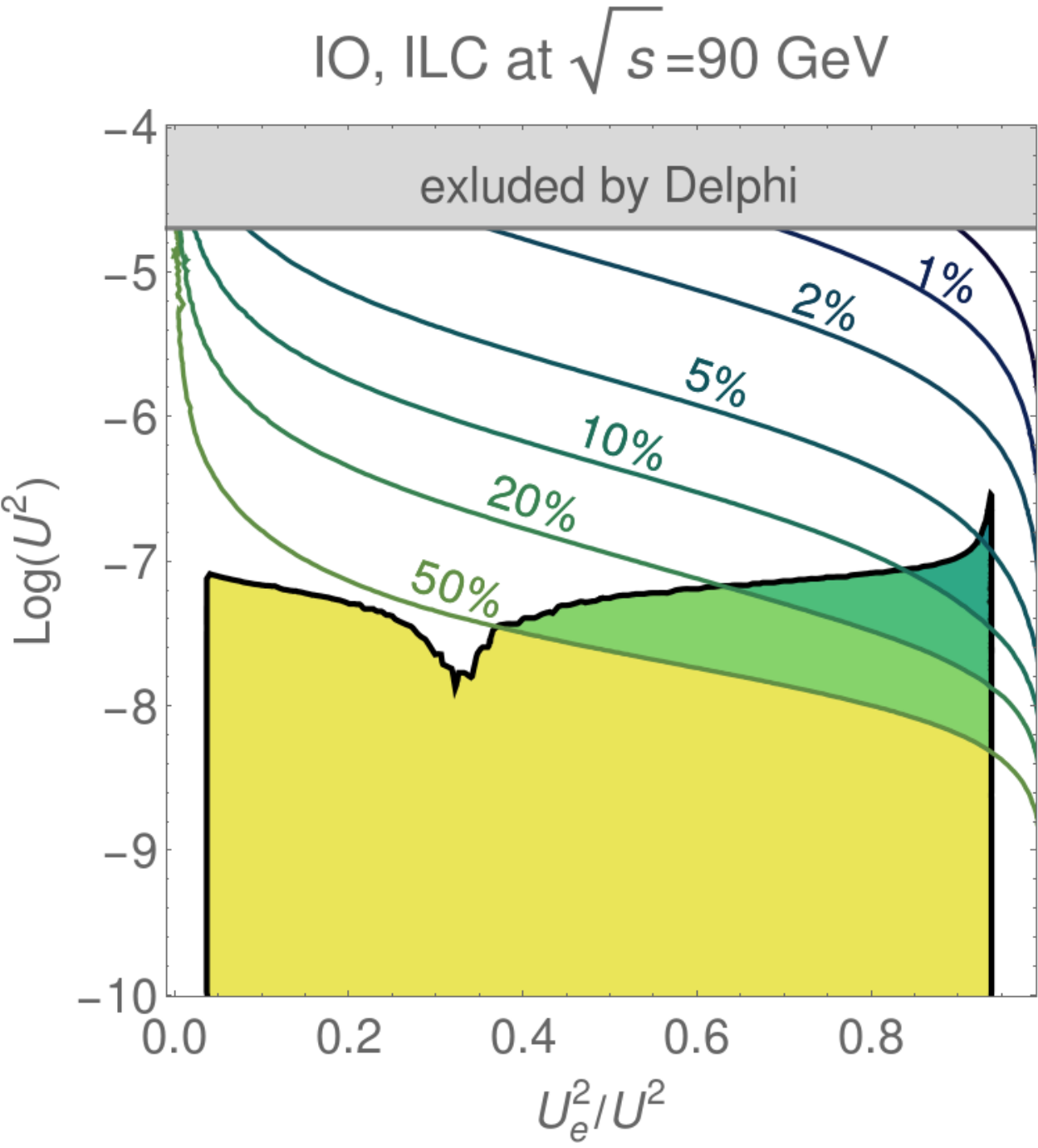} 
			&
			\includegraphics[width=0.4\textwidth]{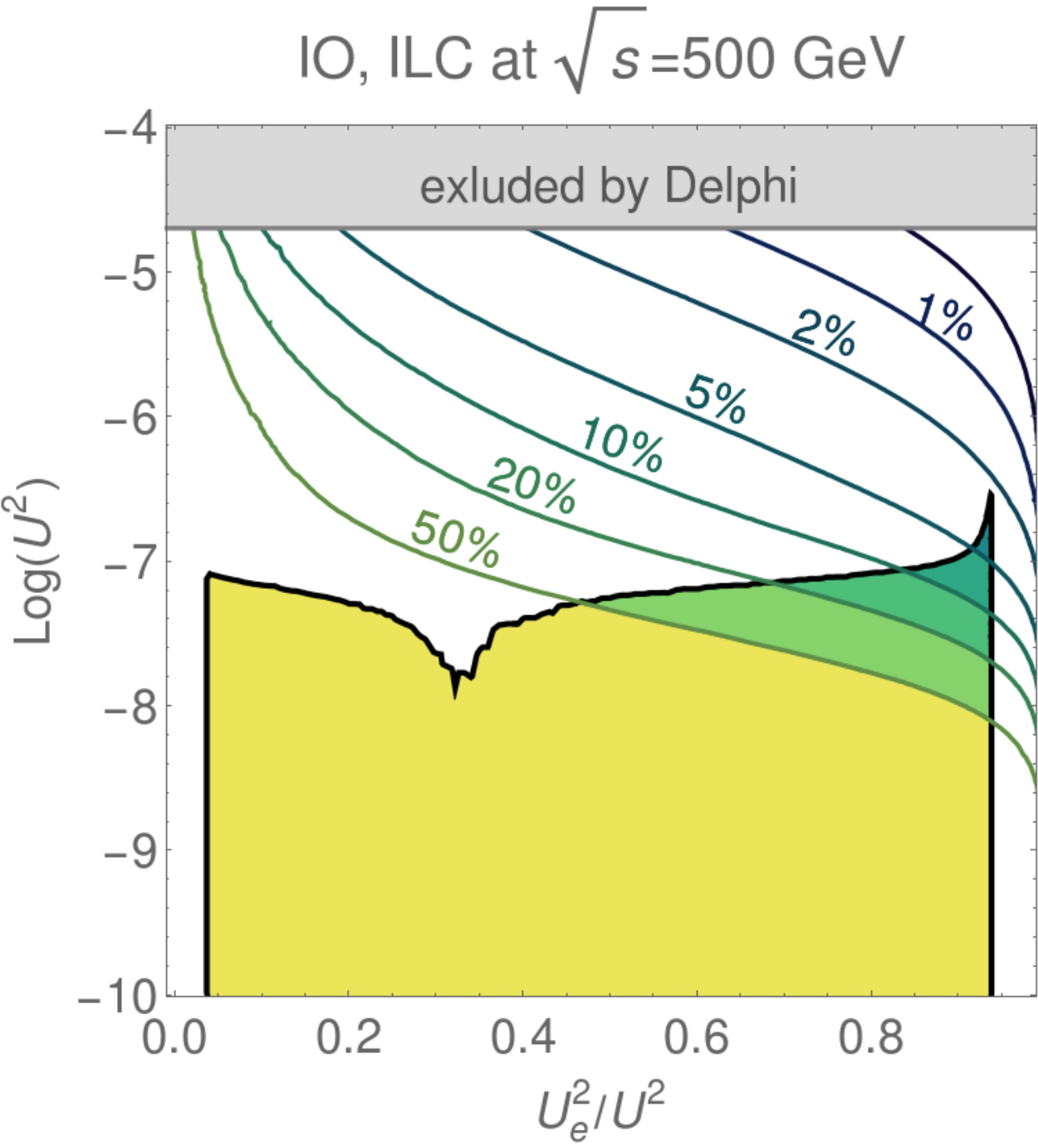}
			&
			\includegraphics[width=0.115\textwidth]{Figures/U2_Ua2_Final/BarLegend.pdf}
			\\			
			\includegraphics[width=0.4\textwidth]{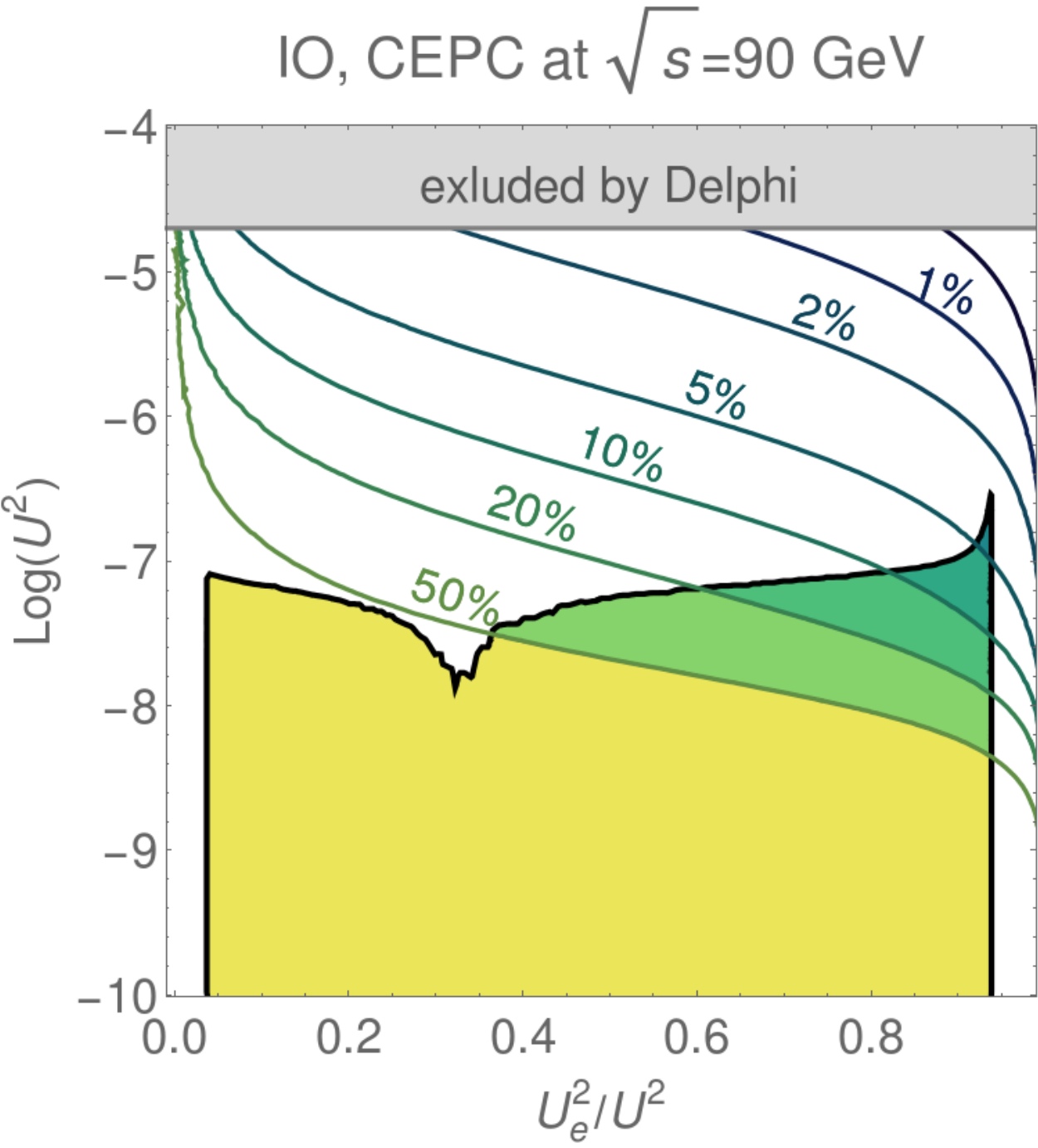} &
			\includegraphics[width=0.4\textwidth]{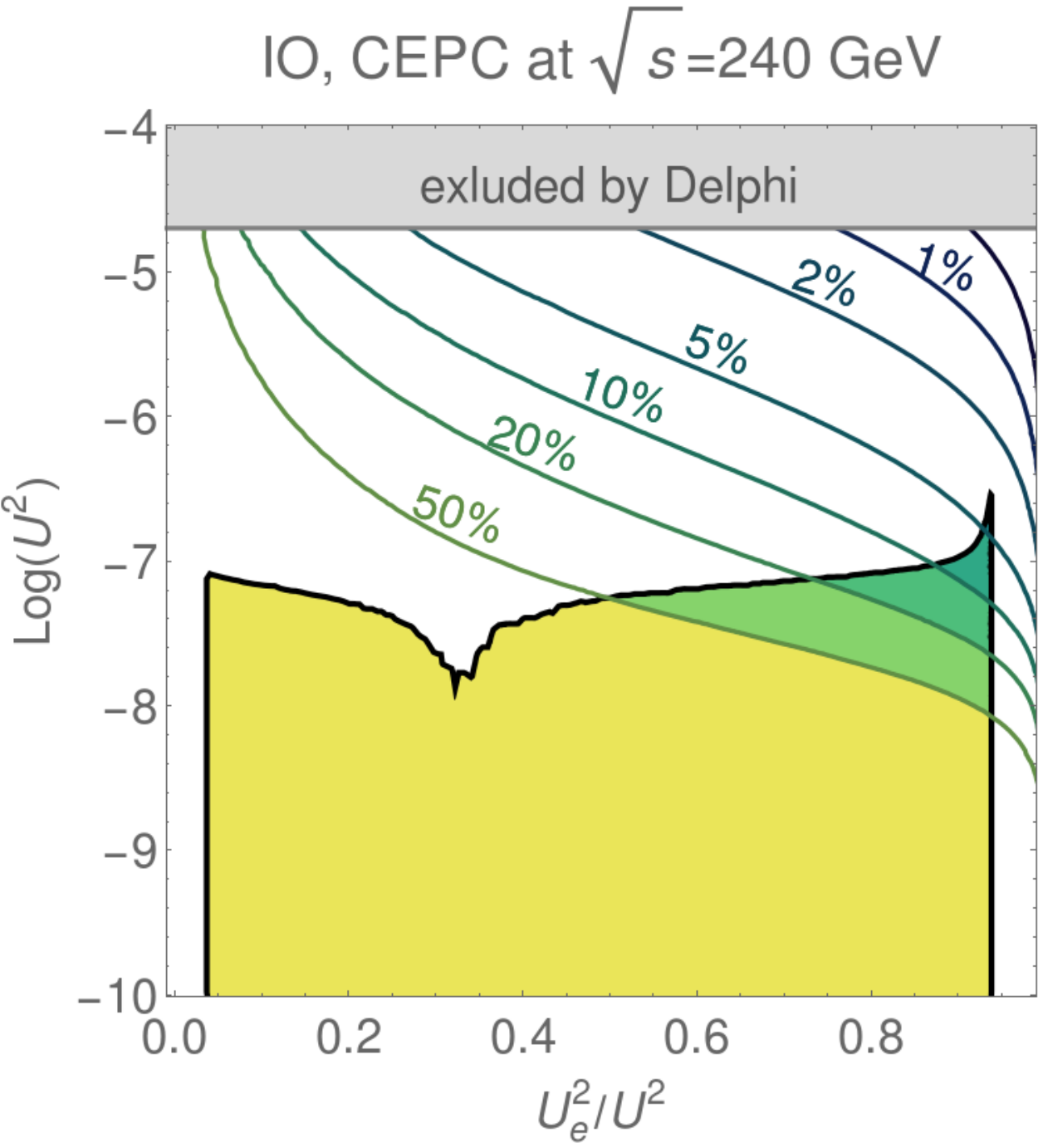}
			&
			\includegraphics[width=0.115\textwidth]{Figures/U2_Ua2_Final/BarLegend.pdf}
		\end{tabular}
\caption{\label{fig:Precision-ILC-CEPC}
The colour indicates the precision that can be achieved for measuring $U_e^2/U^2$ at the ILC with $\sqrt{s}=90\,\text{GeV}$ (top, left) and $\sqrt{s}=500\,\text{GeV}$ (top, right), as well as at the CEPC with $\sqrt{s}=90\,\text{GeV}$ (bottom, left) and $\sqrt{s}=240\,\text{GeV}$ (bottom, right) for the case of inverse ordering (IO) of the light neutrino masses. It turns out that besides the FCC-ee, cf. figure\ \ref{fig:Precision-FCC}, only the four channels displayed here can be tested with  precision better than $50\,\%$. All other channels, e.g. the ones that test $U_\mu^2/U^2$ and $U_\tau^2/U^2$ at the ILC and CEPC are not plotted here because the precision that can be achieved is below $50\,\%$ for all values of $U^2$ consistent with leptogenesis. The two heavy neutrinos are assumed to be almost degenerate in mass at $M=10\,\text{GeV}$. Details are given in the main text, cf. section \ref{Section5.2}.}
\end{figure}

In figures~\ref{fig:Precision-FCC} and \ref{fig:Precision-ILC-CEPC} we show the precision for measuring $U_a^2/U^2$ with $a=e,\mu,\tau$ using the method described in section~\ref{Section4_measurement}.  
Due to the potentially large number of events, the future experiments can not only discover the right-handed neutrinos but also measure their flavour-dependent mixing, i.e.\ $U_a^2/U^2$. Together with a measurement of $M$ and $U^2$, this can be a first step towards checking the hypothesis that the observed right-handed neutrinos are indeed responsible for the generation of the baryon asymmetry of the universe (as well as for the observed light neutrino masses), as we  discuss below.

The coloured regions in figures~\ref{fig:Precision-FCC} and \ref{fig:Precision-ILC-CEPC} correspond to the parameter space where leptogenesis is possible, taking $\bar{M}=30\,\GeV$ as an example. 
The lines in the different colours correspond to the precision that can be achieved for measuring the ratios $U_a^2/U^2$ (with $a=e,\mu,\tau$). The sensitivity depends also on the other flavour ratios not shown explicitly, and we display the most conservative precision estimate here, i.e. for the choice of the other parameters where the precision is lowest.    
For NO, the precision is best for measuring $U_\mu^2/U^2$ and $U_\tau^2/U^2$ since the active-sterile mixing with the $e$ flavour is suppressed in the NO case. For the IO case, on the contrary, the best precision can be achieved for $U_e^2/U^2$. The possible large number of events at the FCC-ee allows for precision to the percent level (cf.\ figure\ \ref{fig:Precision-FCC}). At the ILC and CEPC (cf.\  figure\ \ref{fig:Precision-ILC-CEPC}) a precision up to about 10\% - 5\% could be reached for part of the parameter space for $\bar{M}=30\,\GeV$. We remark that for smaller masses, a larger number of displaced vertex events could be measured, which would improve the relative precision of the flavour mixing ratios.

\begin{figure}
	\centering
	\begin{tabular}{cc}
		\textbf{Normal ordering} & \textbf{Inverted ordering}\\
\includegraphics[width=0.45\textwidth]{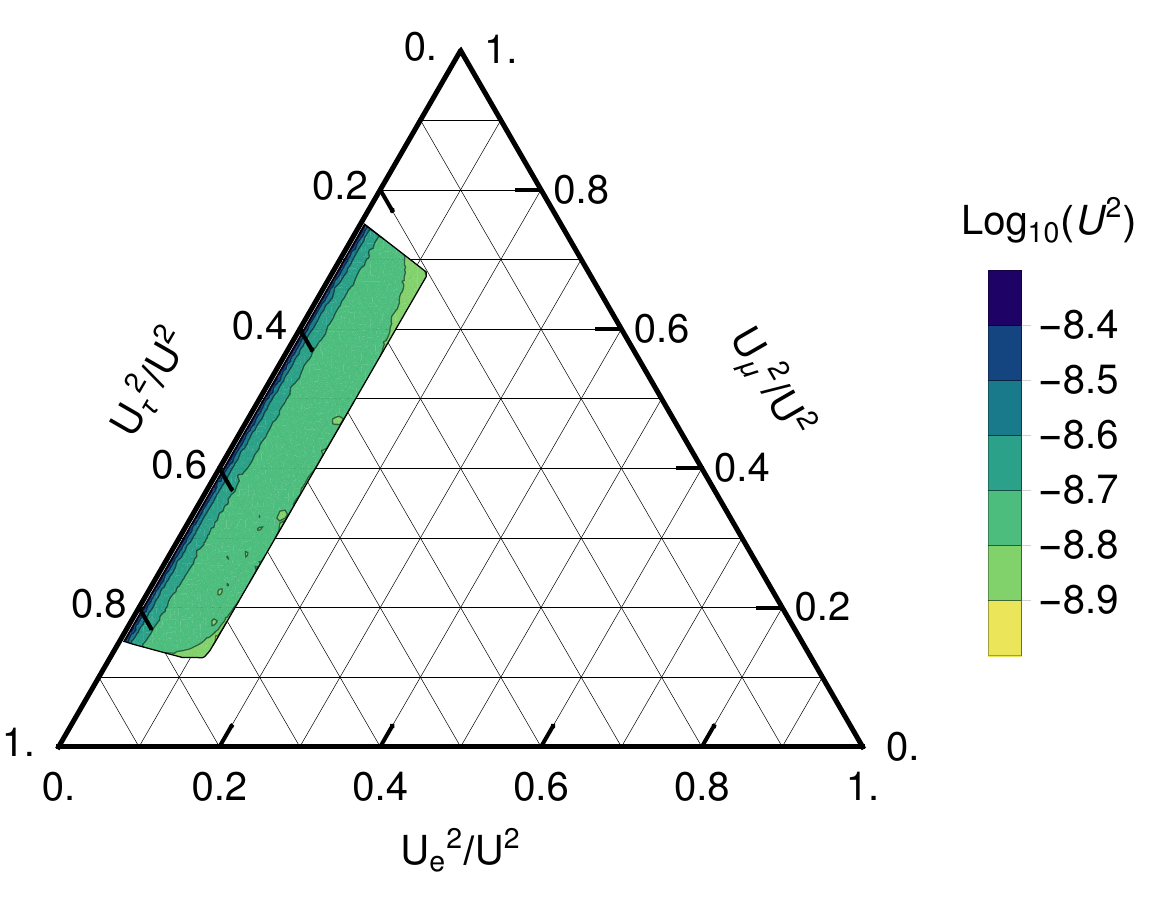}&
\includegraphics[width=0.45\textwidth]{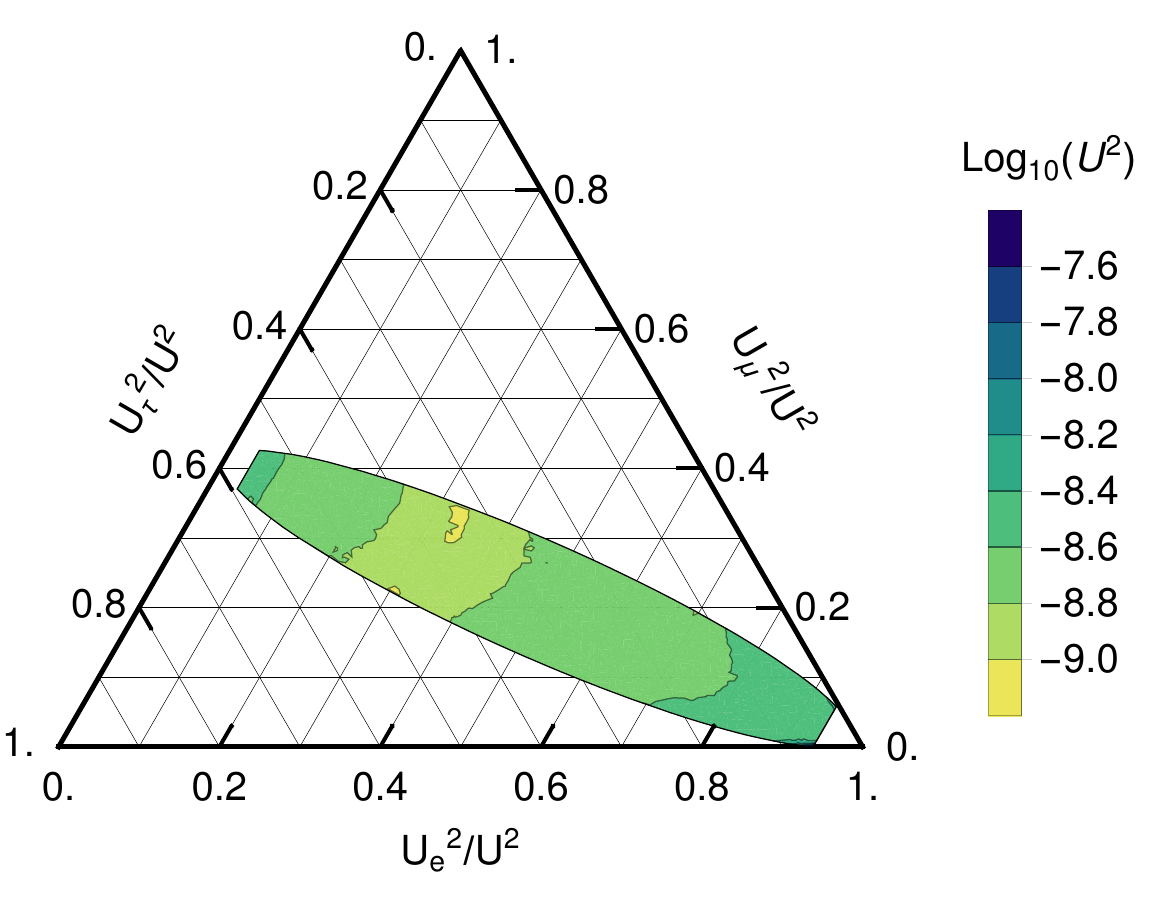}
\end{tabular}
	\caption{The region within the black lines is allowed by light neutrino oscillation data.
	The colour indicates the largest mixing angle $U^2$ consistent with the observed BAU and seesaw constraints for the cases of normal ordering (left) and inverted ordering (right) for right-handed neutrino with an average mass $\bar{M}=30\, \GeV$. Note that the largest viable mixing angles are found in the case of a highly flavour asymmetric flavour pattern, where $U^2_a\ll U^2$ for any of the flavours.}
	\label{fig:triangleplt}
\end{figure}

The plots for inverted ordering feature prominent spikes. 
They are a result of the fact that leptogenesis with the largest $U^2$ requires a flavour asymmetric washout, i.e., a strong hierarchy amongst the $U_a^2$.
These can be understood from the fact that a large $U^2$ implies large Yukawa couplings. 
Larger $Y_{ia}$ increase both, the source and washout terms in the kinetic equations (\ref{sec3:eq:diff_eq_RHN}) and (\ref{sec3:eq:diff_eq_SM}).
For $U^2\gtrsim10^{-8.5}$, a complete washout of the lepton asymmetries prior to the sphaleron freezeout can only be avoided if one of the $U_a^2$ is much smaller than $U^2$, protecting the asymmetry stored in that flavour from the washout.
However, the requirement to explain the observed light neutrino mixing pattern imposes constraints on the relative sizes of the $U_a^2$, cf.~figure\ \ref{fig:triangleplt}.
For IO it turns out that the electron has to couple maximally $U_e^2/U^2 \approx 0.94$. This explains the peak in the bottom left panel of figure \ref{fig:Precision-FCC}. Having a large $U_e^2/U^2$ requires the other ratios to be small: $U_\mu^2/U^2+U_\tau^2/U^2 \lesssim 0.06$. But still there is the freedom to choose which of the ratios is small. It turns out that  the largest possible $U^2$ is given if the muon couples minimally, explaining both the peak in the bottom middle and the bottom right panel of figure~\ref{fig:Precision-FCC}. 
Note that consequently the maximal height of the peaks should be equal for all three flavours. For NO the electron has to couple minimally, $U_e^2/U^2 \approx 0.006$, in order to allow for the largest $U^2$. In this case the range in which the other ratios are allowed is rather large: $U_\mu^2/U^2+U_\tau^2/U^2 \lesssim 0.994$. Consequently, the peaks are not visible in case of NO.

\paragraph{How to read these plots.}
If heavy neutral leptons are discovered at a future collider, the relative size of their mixings $U_a^2$ can be used to test of the hypothesis that these are the common origin of light neutrino masses and baryonic matter in the universe \cite{Hernandez:2016kel,Drewes:2016jae,Caputo:2016ojx,Caputo:2017pit}.
With figures~\ref{fig:Precision-FCC}-\ref{fig:triangleplt} we can estimate how strong this test can be at CEPC, ILC and FCC-ee. 
We use the case $\bar{M}=30\, \GeV$ as an example to illustrate this.
If all three $U_a^2$ could be measured exactly by experiments, then one could simply check whether the point with the observed ratio $U_e^2/U_\mu^2$ lies within the region in figure~\ref{fig:triangleplt} that is allowed for the observed $U^2$. This can either support or rule out the hypothesis that the discovered particle is involved in neutrino mass generation and leptogenesis.
In reality the $U_a^2$ can only be measured with a finite precision.
This smears out the corresponding point in figure. 
Figures~\ref{fig:Precision-FCC} and \ref{fig:Precision-ILC-CEPC}  can be used to estimate the expected uncertainty for any given value of the $U_a^2$. 
Further input which will affect such consistency checks will of course come from measurements of the parameters in $U_\nu$ and of the light neutrino mass ordering by neutrino oscillation experiments. Note that we have completely neglected the experimental uncertainties on these parameters in the plots.

\subsection{Measuring $\Delta M$ at colliders}\label{sec:5.3}
\begin{figure}
	\centering
	\begin{tabular}{cc}
		\textbf{Normal ordering} & \textbf{Inverted ordering}\\
\includegraphics[width=0.45\textwidth]{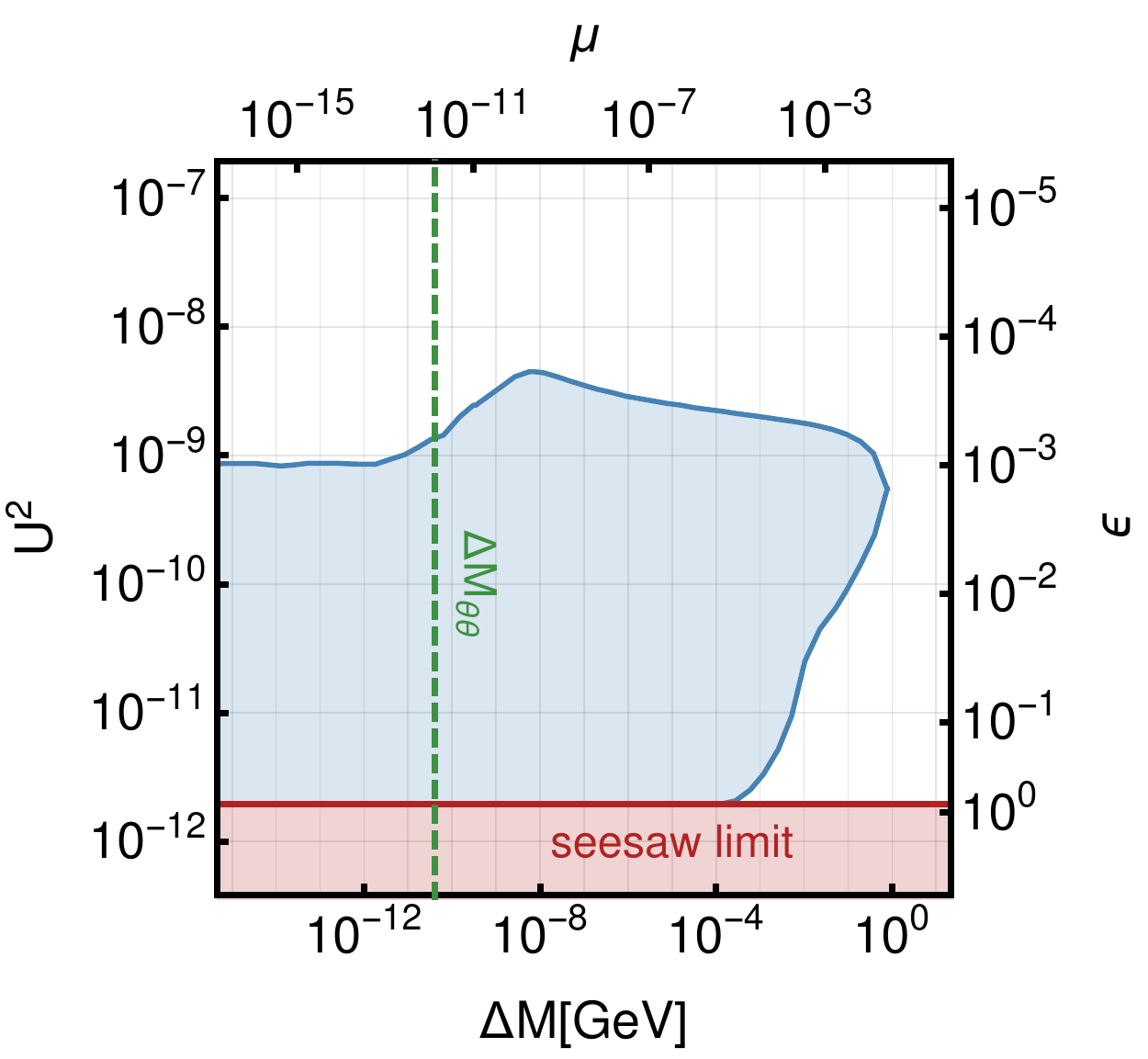}&
\includegraphics[width=0.45\textwidth]{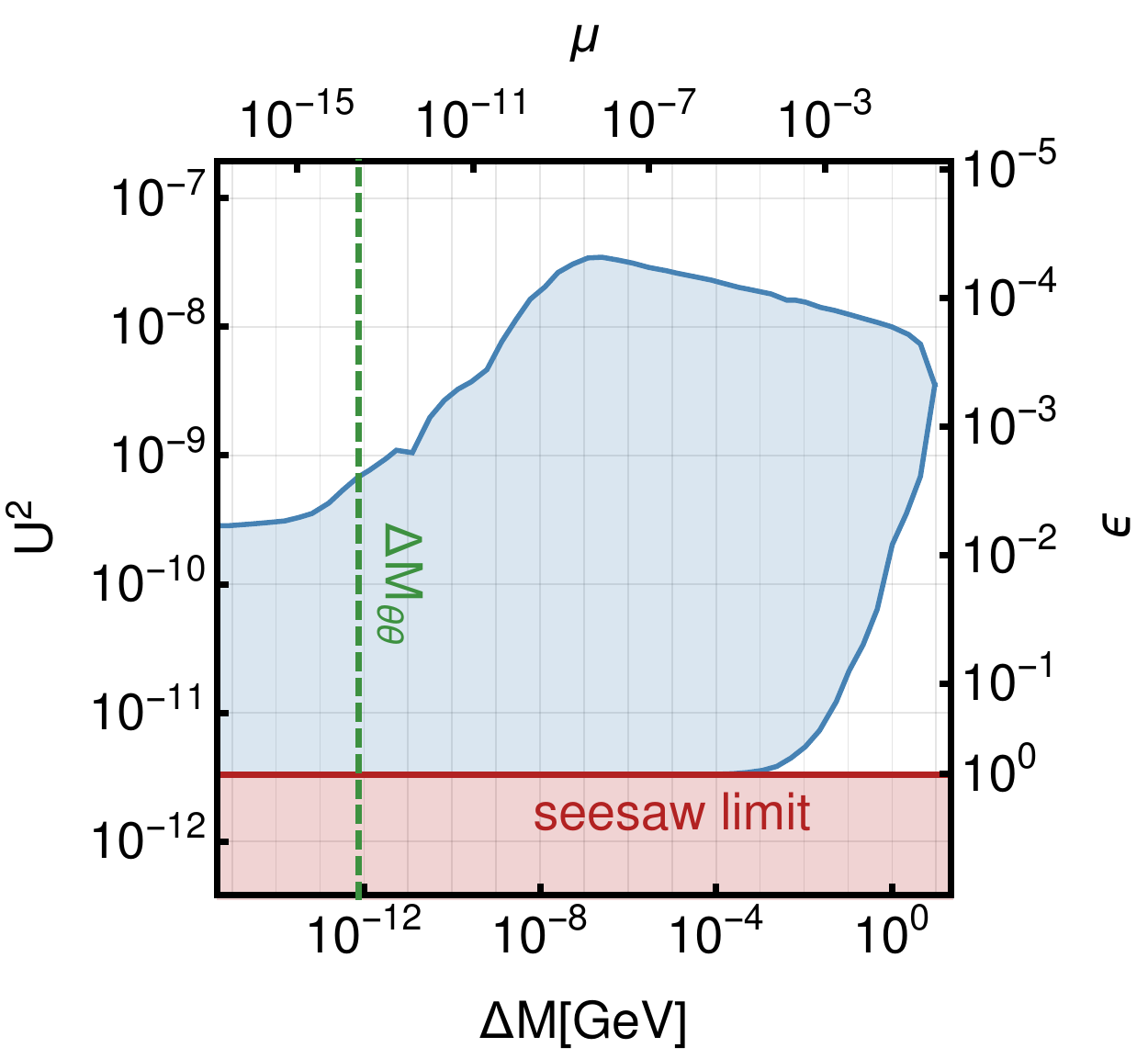}\\
\includegraphics[width=0.45\textwidth]{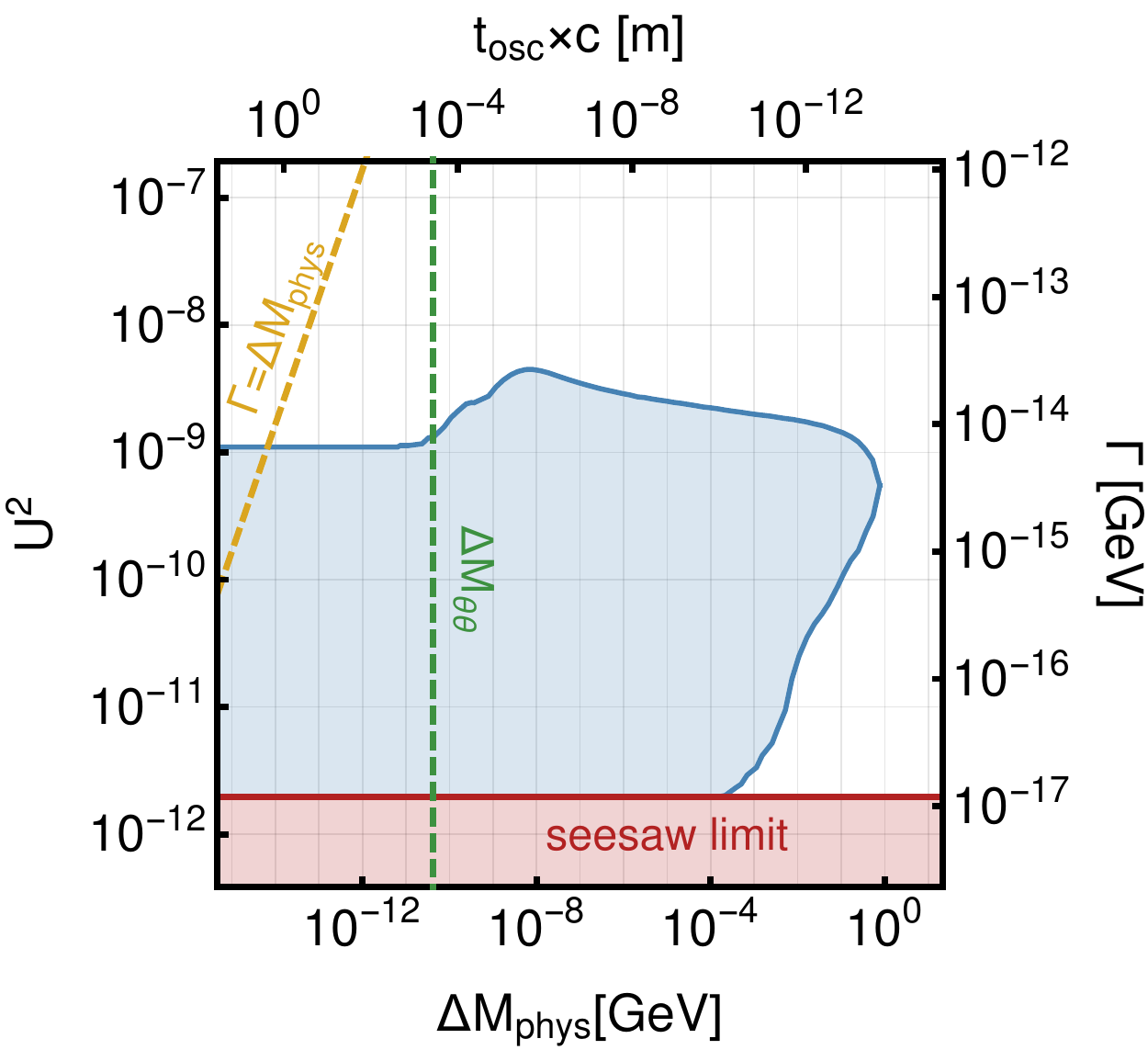}&
\includegraphics[width=0.45\textwidth]{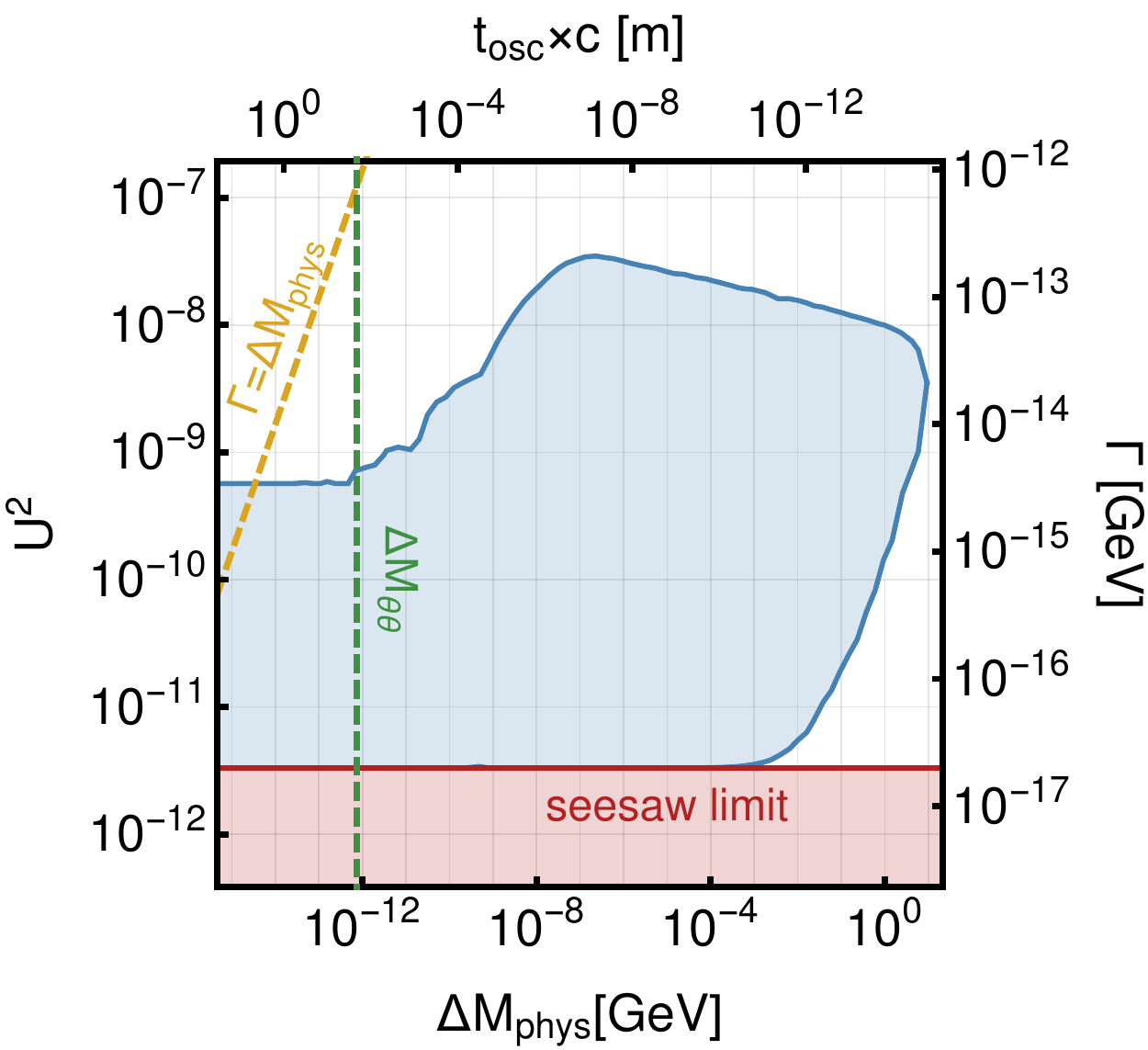}
\end{tabular}
\caption{The allowed mixings $U^2$ in comparison to the Lagrangian mass splittings $\Delta M$ (upper panels) and the physical mass splittings $\Delta M_{\rm phys}$ (lower panels) with an average mass $\bar{M}=30\, \GeV$ are shown in blue for normal ordering (left panels) and inverted ordering (right panels), respectively. The red line represents the seesaw limit, below which the parameter region is excluded by neutrino oscillation data. The vertical, dashed, green lines correspond to the difference $\Delta M_{\theta \theta}$ between the eigenvalues of $M_N$, cf. eq.~(\ref{MNDef}), solely from the coupling to the Higgs field. 
The physical mass splitting $\Delta M_\mathrm{phys}$ acessible in experiments is related to the mass splitting $\Delta M$ in the electroweak unbroken phase and $\Delta M_{\theta \theta}$ by eq.~(\ref{eq:DeltaMphys}). Note that leptogenesis is possible even for $\Delta M = 0$ due to $\Delta M_{\theta\theta} \neq 0$ during the electroweak crossover.
}
\label{fig:mass_splitting}
\end{figure}

It is important to note that, if the parameters of the right-handed neutrinos pass the necessary condition imposed by the previous test (i.e.~the observed flavour-dependent ratios are in agreement with neutrino mass generation and leptogenesis), then this does not yet mean that we can confirm them as the source of the baryon asymmetry.
For this, it is in particular crucial to obtain information on $\Delta M$. 
In figure \ref{fig:mass_splitting} we show the regions of $\Delta M$ which are consistent with leptogenesis and the neutrino oscillation data.

The most straightforward way of obtaining information on the mass splitting is by a direct (kinematic) measurement of $\Delta M_\mathrm{phys}$. 
$\Delta M_\mathrm{phys}$ is related to the mass splitting $\Delta M$ in the electroweak unbroken phase by eq.~(\ref{eq:DeltaMphys}). 
Realistically, however, this may be possible only for $\Delta M_\mathrm{phys}$ in the GeV range (cf.\ section 4). In this regime $\Delta M \simeq \Delta M_\mathrm{phys}$, so this corresponds to the largest $\Delta M$ we found to be consistent with leptogenesis in our scan. 

We note that the indirect measurement in neutrinoless double $\beta$ decay that was proposed in refs.~\cite{Drewes:2016lqo,Hernandez:2016kel,Asaka:2016zib} cannot be used in the range of $M_i$ considered here because the contribution from $N_i$ exchange to this process is strongly suppressed by their virtuality.

Another possible way to probe $\Delta M_\mathrm{phys}$ is via non-trivial total ratios between the rates of LNC and LNV involving the $N_i$. This is possible for instance at proton-proton or electron-proton colliders, where there are unambiguous LNV signatures. Some work in this direction has been done for the LHC \cite{Gluza:2015goa,Dev:2015pga,Anamiati:2016uxp,Dib:2016wge,Antusch:2017ebe,Das:2017hmg}. Non-trivial ratios require a decay rate $\Gamma_N$ which satisfies $\Gamma_N \sim \Delta M_\mathrm{phys}$. With $\Gamma_N \approx 6.0 \times 10^{-6} U^2$ GeV for our benchmark value of $\bar M = 30$ GeV, we obtain that e.g.\ $U^2 = 10^{-9}$ requires $\Delta M_\mathrm{phys} = {\cal O}(10^{-14})$, which is much smaller than $\Delta M_{\theta\theta}$. Such small $\Delta M_\mathrm{phys}$ is in principle possible, but requires a large cancellation between the contributions from $\Delta M$ and $\Delta M_{\theta\theta}$ in eq.~(\ref{eq:DeltaMphys}). 
From eq.~(\ref{eq:DeltaMphys}) we can see that the cancellation happens for a correlation between $\Delta M$ and $\mbox{Re}\,\omega$. 
For larger $\Delta M_\mathrm{phys}$, the ratios between the rates of LNC and LNV processes gets close to one and it is not possible to infer 
$\Delta M_\mathrm{phys}$. 

Furthermore, a direct measurement of the heavy neutrino mass splitting $\Delta M_\mathrm{phys}$ could be possible via resolved heavy neutrino-antineutrino oscillations at colliders, as recently discussed in \cite{Antusch:2017ebe}. This also allows to measure $\Delta M_\mathrm{phys}$ when the total (integrated) ratios between the rates of LNC and LNV processes is close to one, since the heavy neutrino-antineutrino oscillations give rise to an oscillating pattern between the rates of LNC and LNV processes as a function of the vertex displacement. The oscillation time is directly related to the mass splitting $\Delta M_\mathrm{phys}$. Again, from eq.~(\ref{eq:DeltaMphys}) we get that a measurement of $\Delta M_\mathrm{phys}$ implies a non-trivial relation between $\Delta M$ and $\mbox{Re}\,\omega$, which could be used to test leptogenesis. An interesting limit is given by $\Delta M \ll \Delta M_{\theta\theta}$, or even $\Delta M = 0$, which corresponds to the pure (or approximate) linear seesaw scenario. Then, $\Delta M_\mathrm{phys} \approx \Delta M_{\theta\theta}$, with good prospects to resolve the oscillation patterns and measure $\Delta M_\mathrm{phys} $ (cf.\ \cite{Antusch:2017ebe}). But also for $\Delta M$ well above $\Delta M_{\theta\theta}$, such that $\Delta M \simeq \Delta M_\mathrm{phys}$, the heavy neutrino-antineutrino oscillations could be resolved, for instance at FCC-hh where a large boost factor is possible which enhances the oscillation length in the laboratory frame.

\section{Discussion and conclusions}
\label{Section6_Conclusions}

In this paper, we have investigated the question whether leptogenesis, as a mechanism for explaining the baryon asymmetry of the universe, can be tested at future lepton colliders. 

Focusing on the minimal scenario of two right-handed neutrinos, we have estimated the allowed parameter space for successful leptogenesis in the heavy neutrino mass range between $5$ and $50\,\GeV$. We have improved previous calculations in various ways.  The main improvement lies in the consistent inclusion of the lepton flavour violating source from heavy neutrino oscillations and the lepton number violating source from Higgs decays. In addition, we have included the effect of the temperature dependent Higgs expectation value and used updated values for the light neutrino oscillation parameters.

Regarding future colliders we have focused on the FCC-ee, i.e.\ the Future Circular Collider in the electron positron mode with its envisaged high integrated luminosity of $10\,\text{ab}^{-1}$ for the $Z$ pole run, the CEPC (Circular Electron Positron Collider) running at the $Z$ pole and $240\,\GeV$ center-of-mass energy with an integrated luminosity of $0.1\,\text{ab}^{-1}$ and $5\,\text{ab}^{-1}$, respectively, as well as the ILC (International Linear Collider)  running at the $Z$ pole and $500\,\GeV$ center-of-mass energy with an integrated luminosity of $0.1\,\text{ab}^{-1}$ and $5\,\text{ab}^{-1}$, respectively. 

We have confronted the parameter region where the heavy neutrinos can simultaneously explain the observed light neutrino oscillation data and the baryon asymmetry of the universe
with the discovery potential for heavy neutrinos at the above-mentioned future lepton collider options, with results shown in figure~\ref{fig:SensitivityU2M}. Future lepton colliders can be very sensitive in this mass range via displaced vertex searches.\footnote{We remark that also future proton-proton colliders like the FCC-hh and electron-proton colliders like the LHeC and the FCC-eh could be excellent experiments to probe leptogenesis. Estimates for the sensitivities from searches via displaced vertices can be found in \cite{Antusch:2016ejd}.} We found that especially the FCC-ee can cover a substantial part of the heavy neutrino parameter space consistent with leptogenesis. A similar sensitivity could be achieved with the CEPC if more time for the $Z$ pole run is devoted than assumed here, cf. figure~\ref{fig:CEPCcomparison}.

Also the ILC and the CEPC (with the current plan for run times) can discover heavy neutrinos for a significant part of the parameters consistent with leptogenesis (cf.\ figure~\ref{fig:SensitivityU2M}). For an inverse neutrino mass hierarchy we find that the runs with $500\, \GeV$ and $240\, \GeV$ can be competitive with the $Z$ pole run, while for normal neutrino mass hierarchy only the $Z$ pole runs at CEPC and ILC feature a heavy neutrino discovery potential within the leptogenesis parameter region. 

Beyond the discovery of heavy neutrinos, towards testing whether they can indeed generate the baryon asymmetry, we have studied the precision at which the flavour-dependent active-sterile mixing angles can be measured. Due to the possible large number of events at the FCC-ee, measurements with a relative precision at the percent level would be possible (cf.\ figure\ \ref{fig:Precision-FCC}). At the ILC and CEPC (cf.\  figure\ \ref{fig:Precision-ILC-CEPC}) a precision up to about 10\,\% - 5\,\% could be reached for parts of the parameter space.
We also provide a way to check whether the measured flavour-dependent ratios can be the cause of light neutrino masses and the BAU (cf.\ figure\ \ref{fig:triangleplt}). 
Furthermore, we have studied which values of the heavy neutrino mass splitting are consistent with leptogenesis (cf.\ figure\ \ref{fig:mass_splitting}) and discussed how they could be measured at colliders. A detailed study of the prospects for measuring $\Delta M$ at future colliders is highly desirable.

Confronting the ratios of the flavour-dependent active-sterile mixing angles and the heavy neutrino mass splittings measured at future lepton colliders with the parameter space for successful leptogenesis can be a first step towards probing this mechanism of generating the cosmological matter-antimatter asymmetry.


\subsection*{Acknowledgements}
This research was supported by the DFG cluster of excellence 'Origin and Structure of the Universe' (www.universe-cluster.de), by the Collaborative Research Center SFB1258 of  the  Deutsche  Forschungsgemeinschaft, by the Swiss National Science Foundation, by the ``Fund for promoting young academic talent'' from the University of Basel under the internal reference number DPA2354, and it has received funding from the European Unions Horizon 2020 research and innovation programme under the Marie Sklodowska-Curie grant agreement No 674896 (Elusives).


\appendix

\section{Total number of events}
\label{Appendix_1}

\begin{figure}
        \centering
        \begin{tabular}{cc}
			\textbf{NO, FCC-ee at $\sqrt{s}= 90 \,{\rm GeV}$} & \textbf{IO, FCC-ee at $\sqrt{s}= 90 \,{\rm GeV}$} \\
            \includegraphics[width=0.5\textwidth]{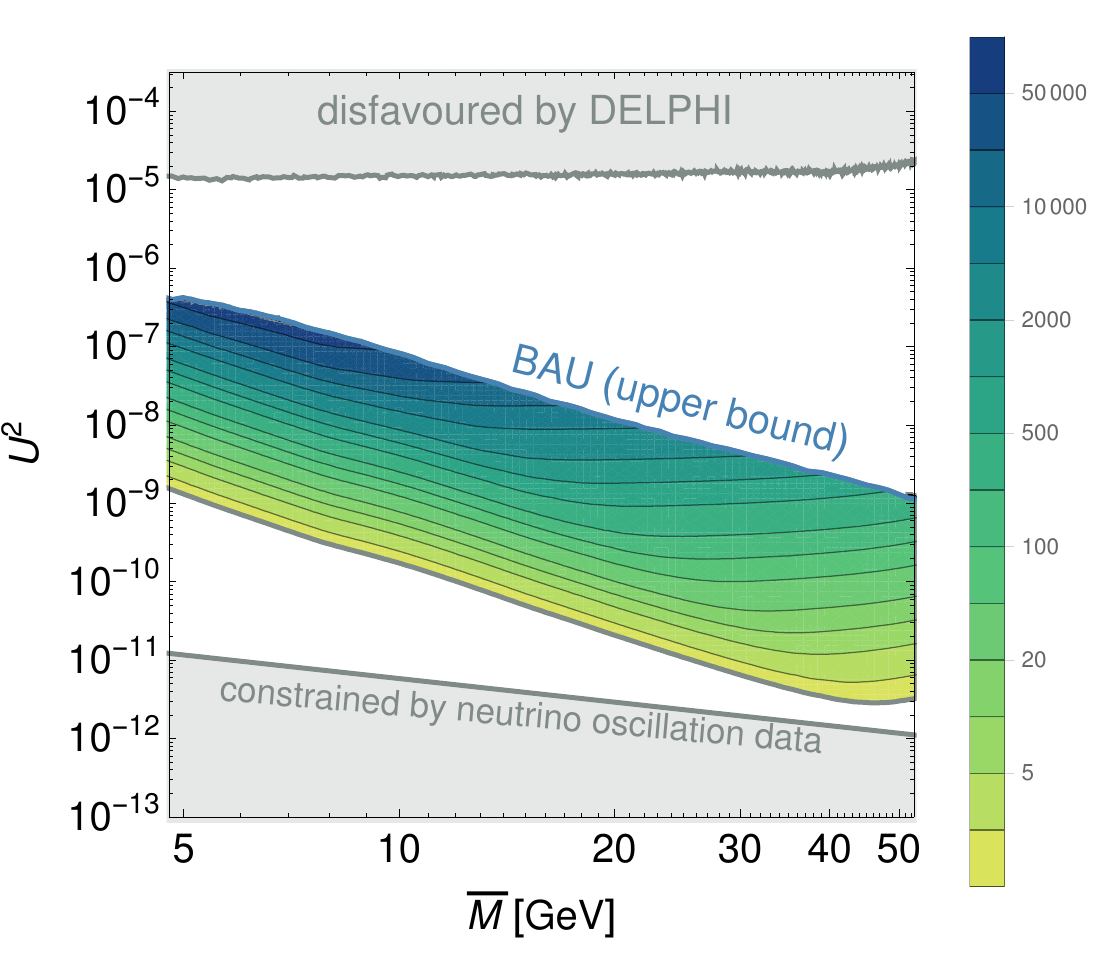} &
			\includegraphics[width=0.5\textwidth]{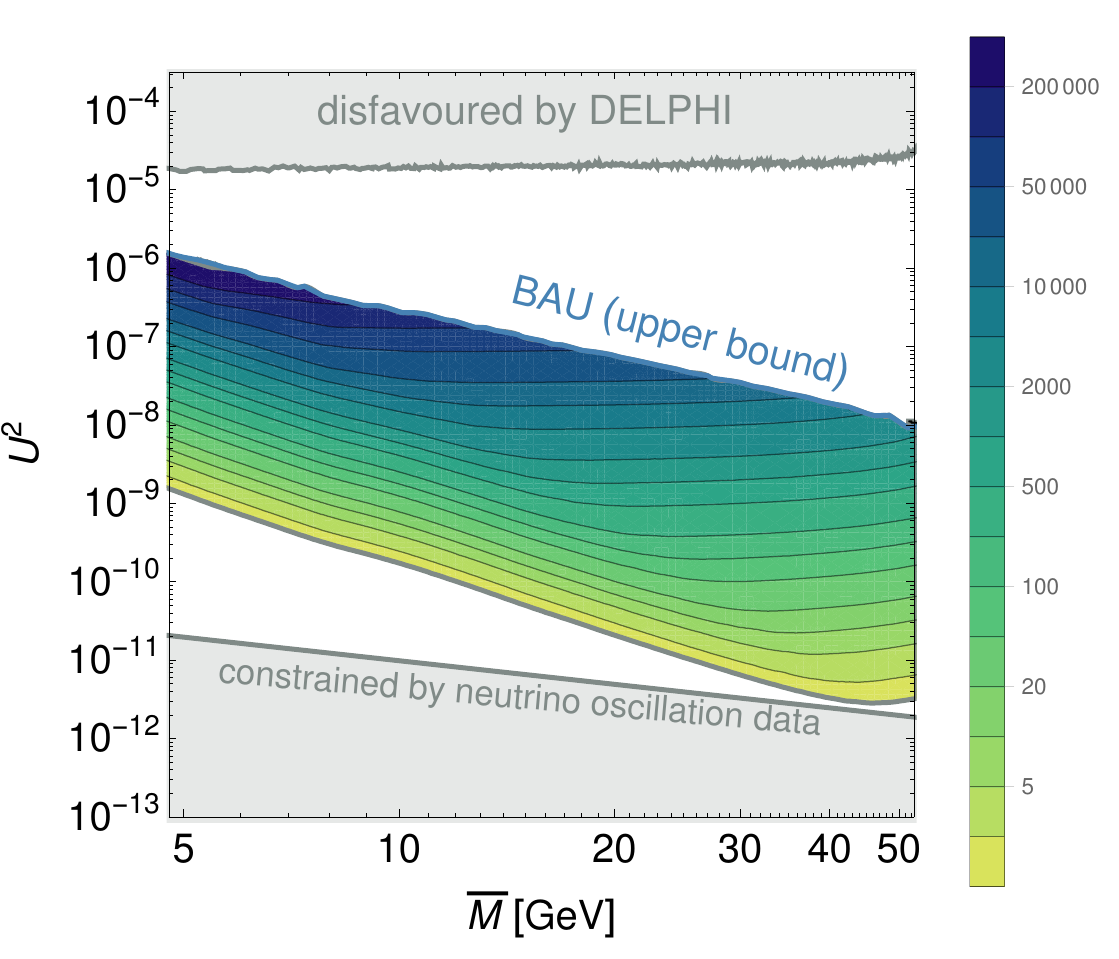}\\
        \end{tabular}
\caption{\label{fig:TotalU2M_FCC}  Number of events expected at the FCC-ee with $\sqrt{s}=90\,\GeV$ for parameter points consistent with leptogenesis. Left and right panel correspond to normal and inverted mass ordering, respectively.}
\end{figure}

\begin{figure}
        \centering
        \begin{tabular}{cc}
			\textbf{NO, ILC at $\sqrt{s}= 90 \,{\rm GeV}$} & \textbf{IO, ILC at $\sqrt{s}= 90 \,{\rm GeV}$} \\
            \includegraphics[width=0.5\textwidth]{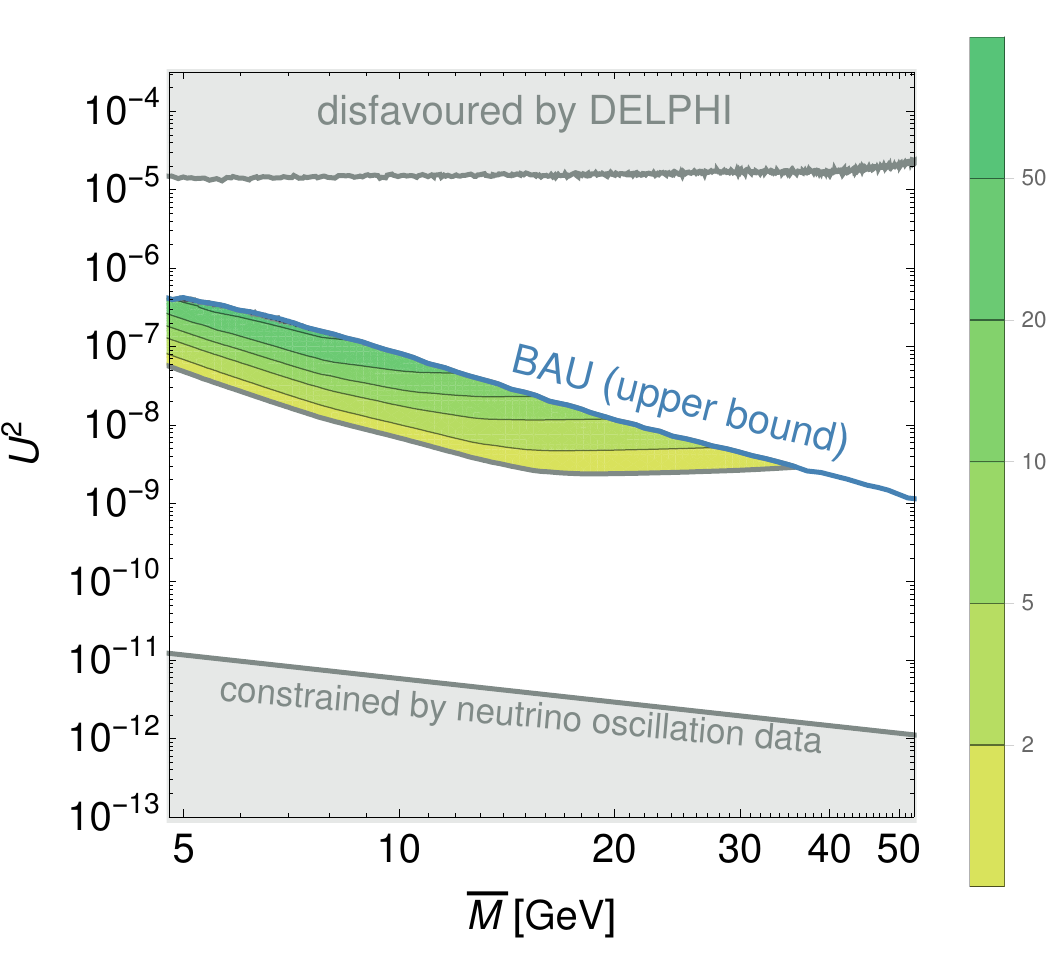} &
			\includegraphics[width=0.5\textwidth]{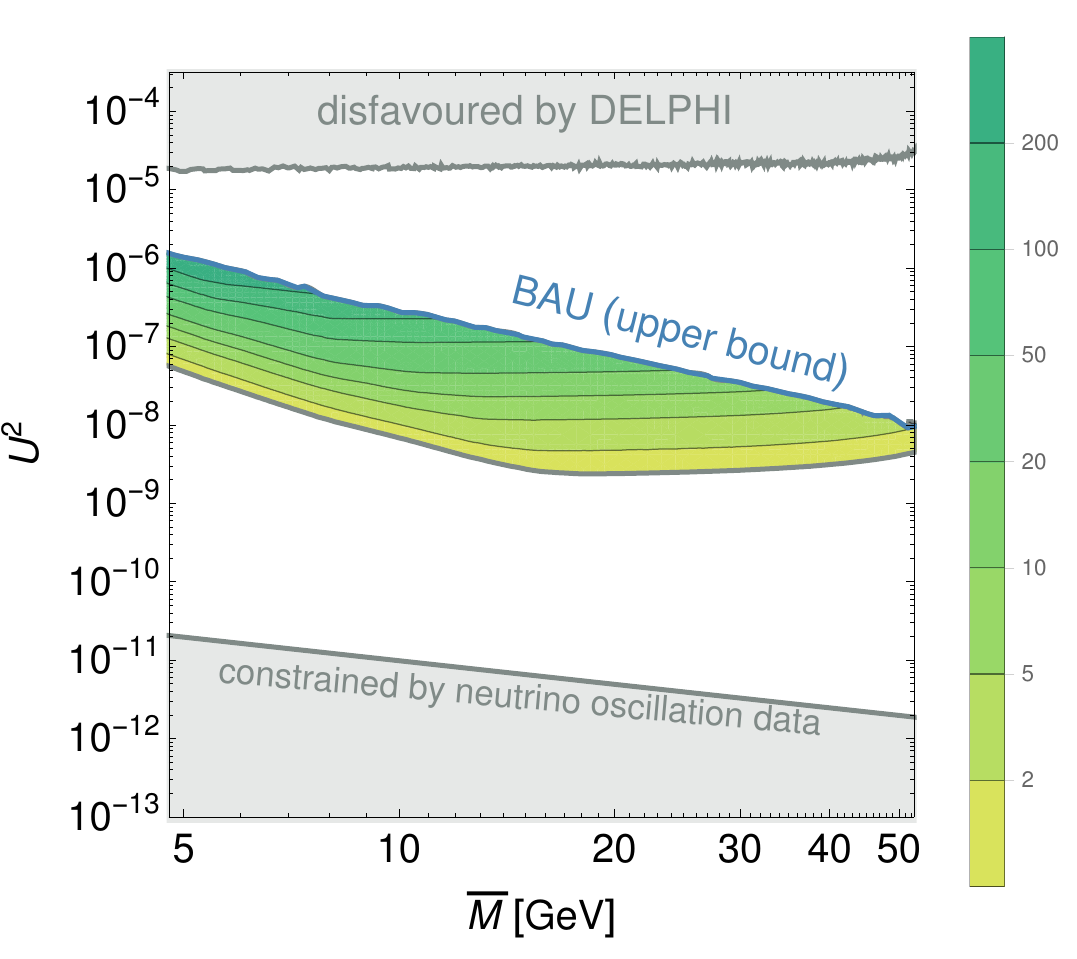}\\
						\textbf{NO, CEPC at $\sqrt{s}= 90 \,{\rm GeV}$} & \textbf{IO, CEPC at $\sqrt{s}= 90 \,{\rm GeV}$} \\
			\includegraphics[width=0.5\textwidth]{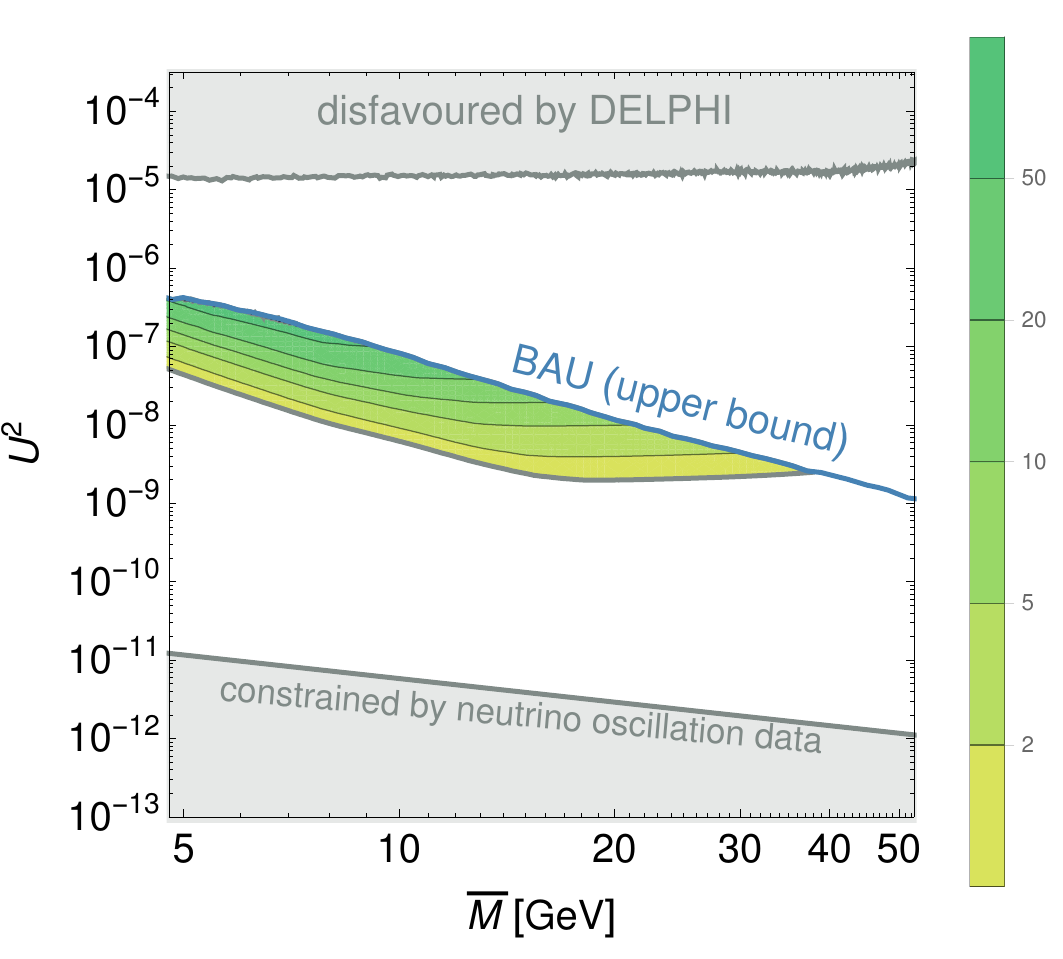} &
			\includegraphics[width=0.5\textwidth]{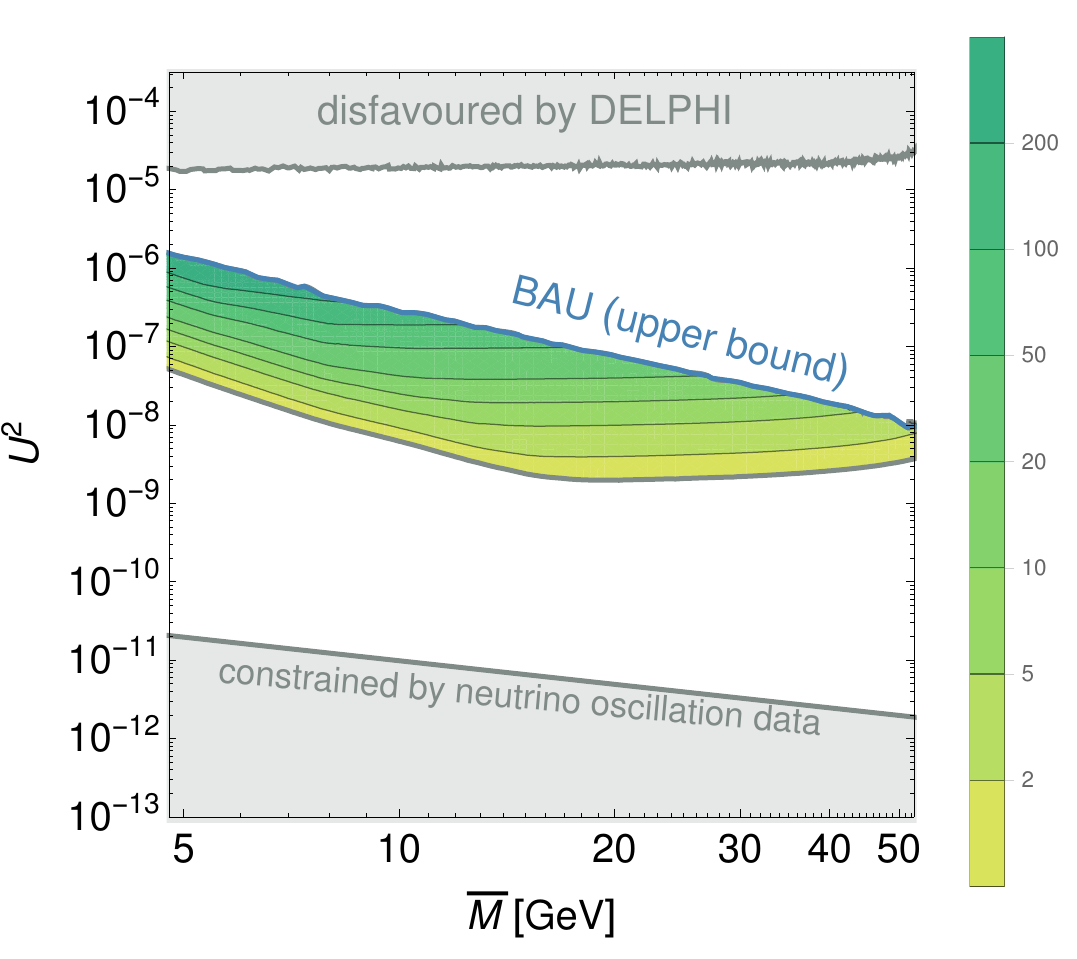}\\
        \end{tabular}
\caption{\label{fig:TotalU2M_ILC_CECP} Number of events expected at the ILC with $\sqrt{s}=90\,\GeV$ (top line) and at the CEPC with $\sqrt{s}=90\,\GeV$ (bottom line) in case of both normal ordering (left column) and inverted ordering (right column) for parameter points consistent with leptogenesis.}
\end{figure}

\begin{figure}
        \centering
        \begin{tabular}{cc}
			\textbf{IO, ILC at $\sqrt{s}= 500 \,{\rm GeV}$} & \textbf{IO, CEPC at $\sqrt{s}= 240 \,{\rm GeV}$} \\
			\rotatebox{90}{\quad\quad Minimal number of expected events}
            \includegraphics[width=0.5\textwidth]{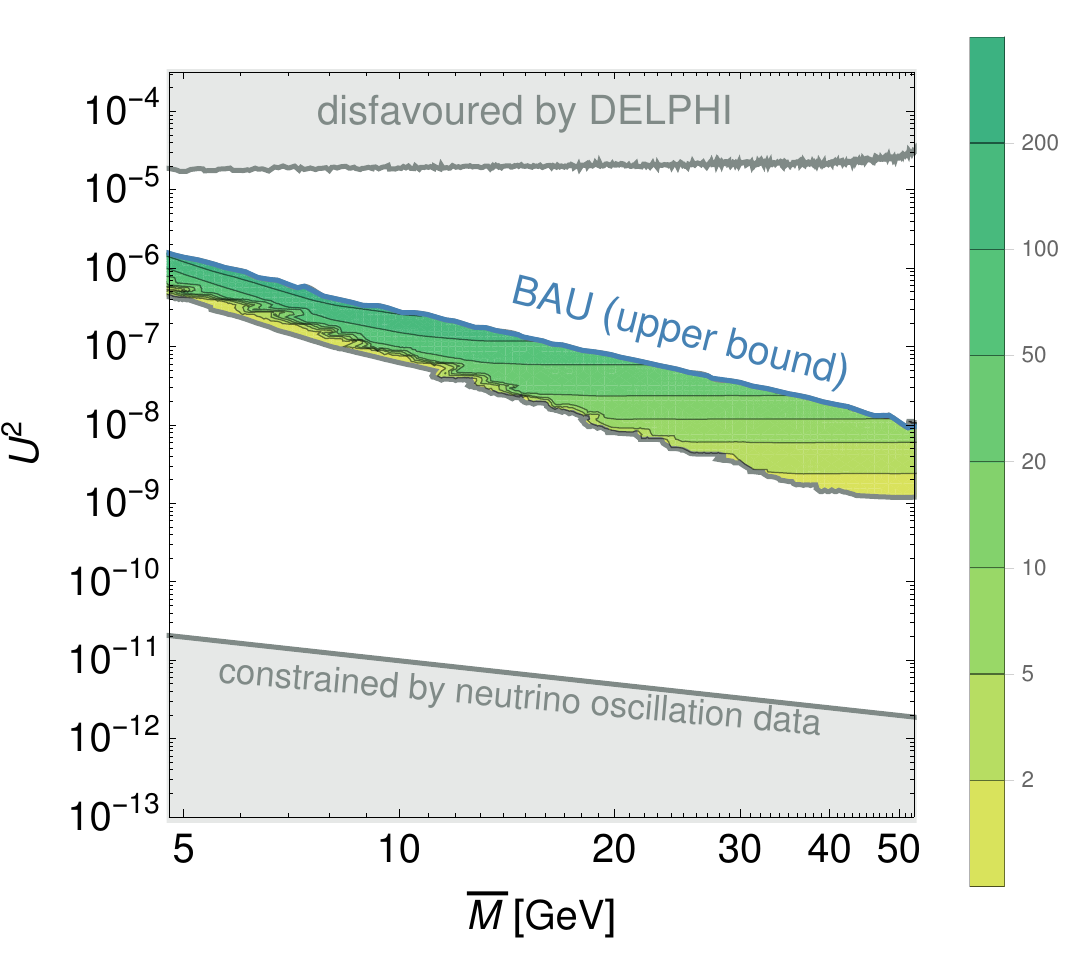} &
			\includegraphics[width=0.5\textwidth]{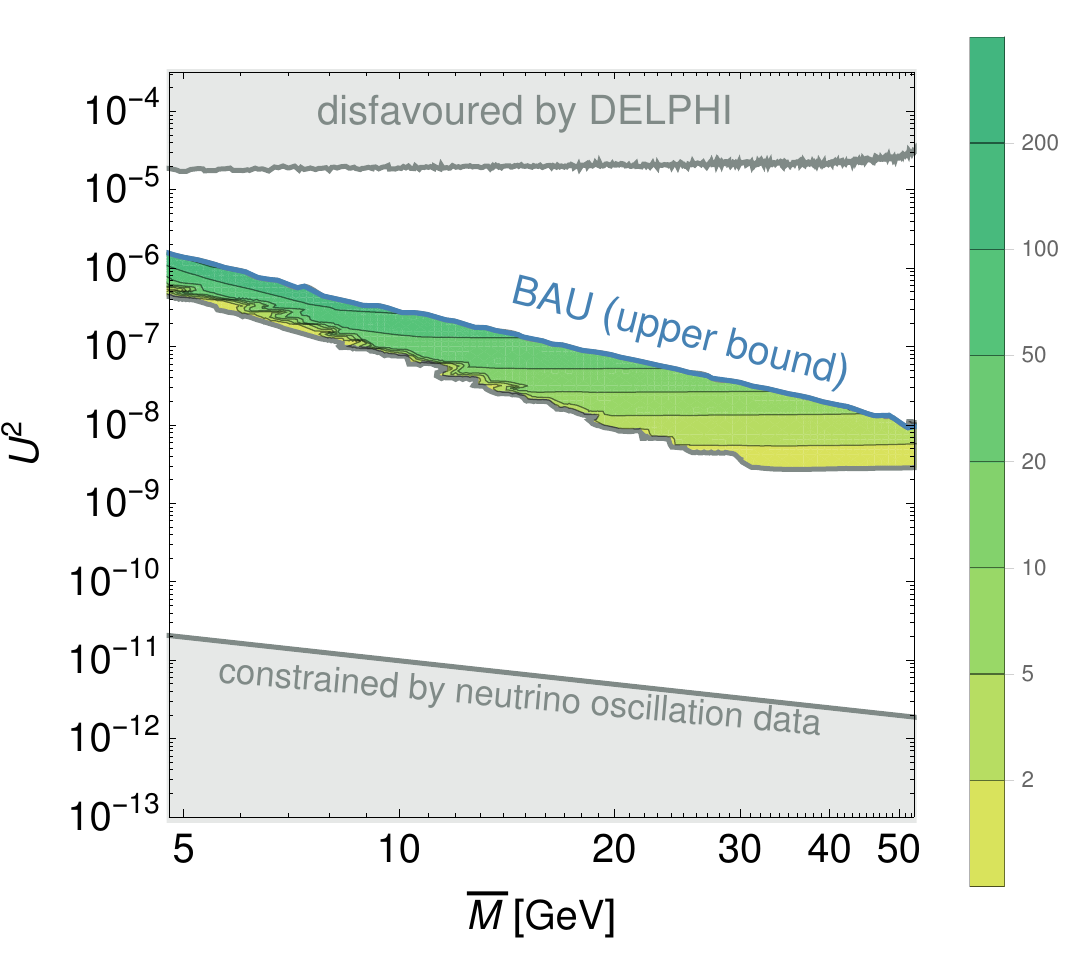}\\
			\textbf{ILC with $\sqrt{s}= 500\,\GeV$} & \textbf{CEPC with $\sqrt{s}= 240\,\GeV$} \\
			\rotatebox{90}{\quad\quad Maximal number of expected events}
                    \includegraphics[width=0.5\textwidth]{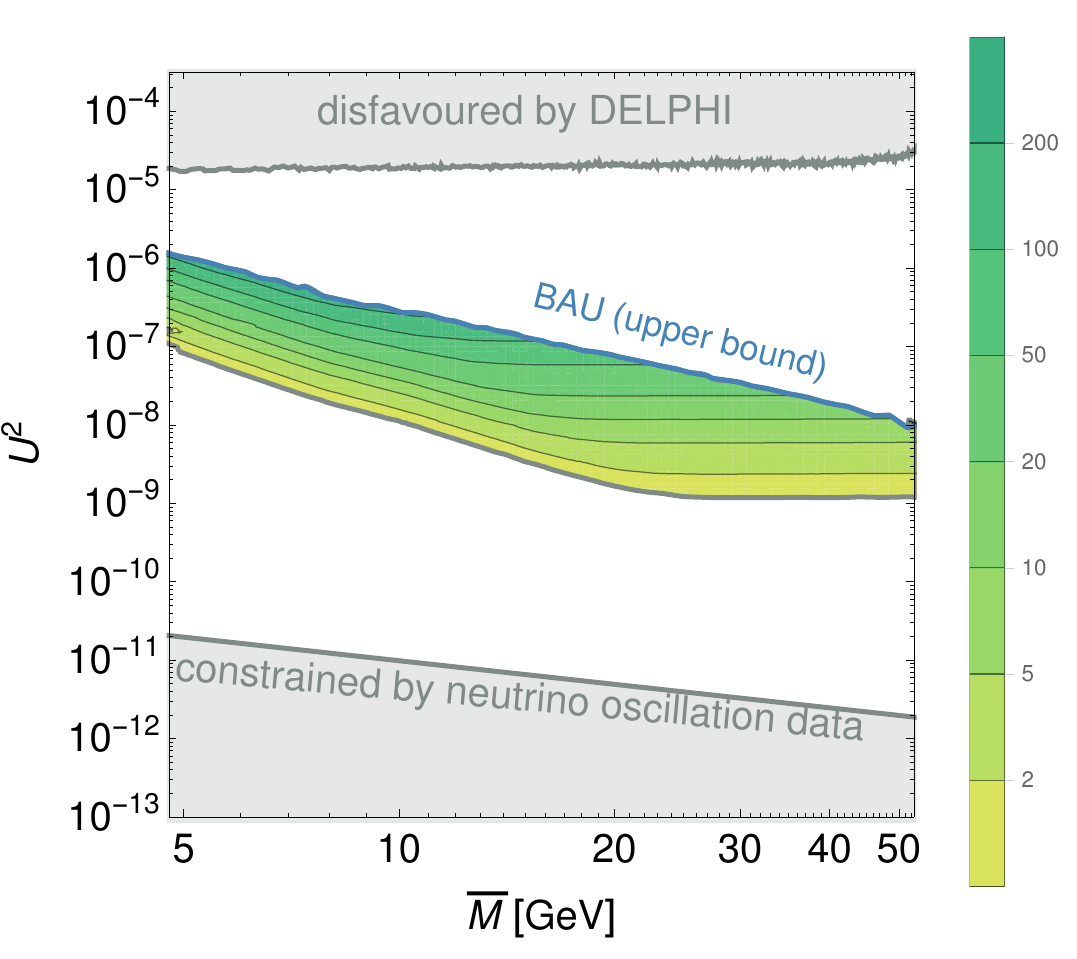} &
			\includegraphics[width=0.5\textwidth]{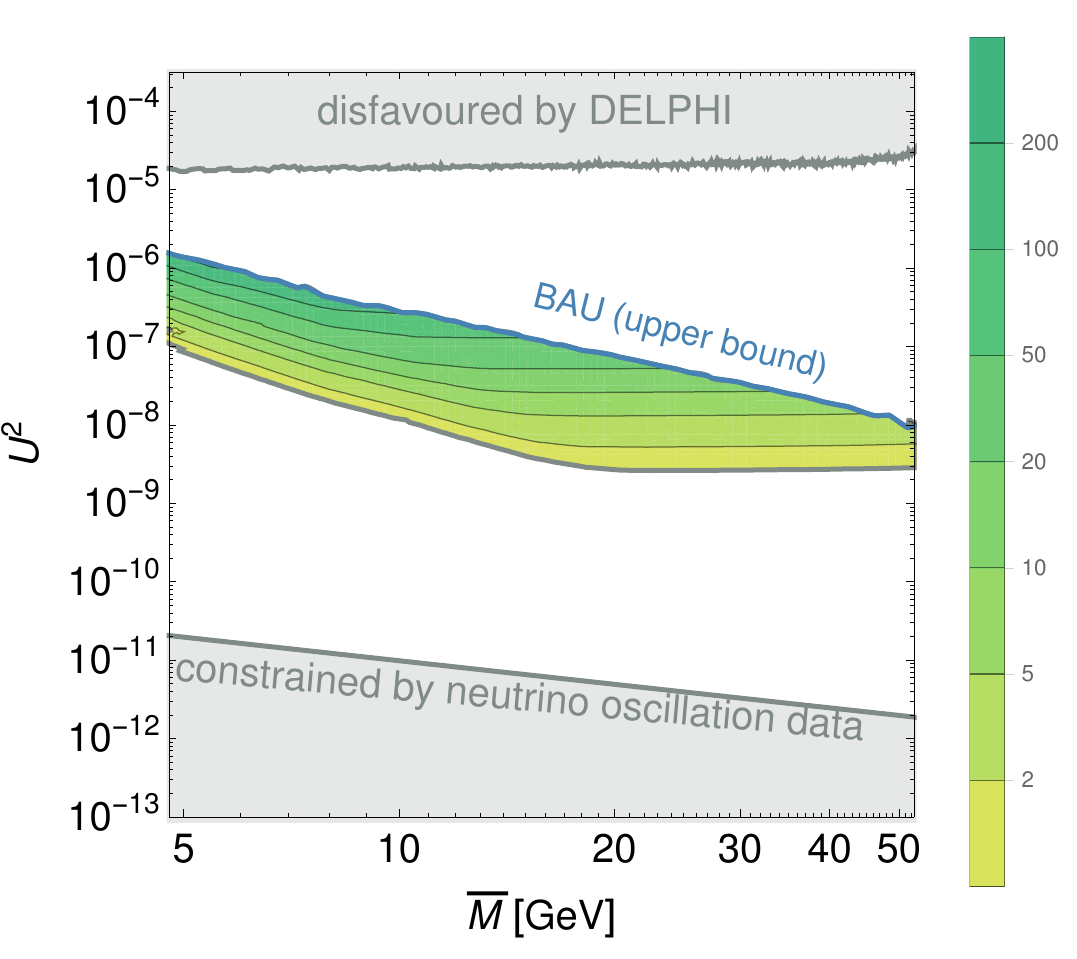}\\
        \end{tabular}
\caption{\label{fig:TotalU2M_nonZ} Minimal (top row) and maximal (bottom row) of expected number of events expected at the ILC with $\sqrt{s}=500\,\GeV$ (left column) and at the CEPC with $\sqrt{s}=240\,\GeV$ (right column) in case of inverted ordering for parameter points consistent with leptogenesis.}
\end{figure}

In this appendix we present detailed information about the number of events expected at the FCC-ee at the $Z$ pole (figure \ref{fig:TotalU2M_FCC}), CEPC and ILC at the $Z$ pole (figure \ref{fig:TotalU2M_ILC_CECP}) as well as  CEPC and ILC at their maximal energy (figure \ref{fig:TotalU2M_nonZ}). The plots have been generated as explained in section~\ref{HowToDo}.

\section{Precision of measuring flavour mixing ratios}
\label{Appendix_2}

\paragraph{Distribution for $N_{\rm sl}$ semileptonic events and $N_a$ in a flavour $a$.}

The probability distribution function (PDF) for $N_{\rm sl}$ semileptonic events with $N_a$ of them being in the flavour $a$ is a product
of a Poisson and a binomial distributions,
\begin{align}
    P(N_{\rm sl},N_a) = \frac{{\rm e}^{-\lambda_{\rm sl}}\lambda^{N_{\rm sl}}_{\rm sl}}{N_{\rm sl}!}
    \binom{N_{\rm sl} }{N_a} {\rm Br}(a)^{N_a} (1-{\rm Br}(a))^{N_{\rm sl}-N_a}\,,
\end{align}
where $\mathrm{Br}(a)=U_a^2/U^2$ is the branching ratio of semileptonic states with $a$ in the final state.
The expected numbers of events are $\langle N_{\rm sl} \rangle=\lambda_{\rm sl}$ and
$\langle N_{a} \rangle=\lambda_{\rm sl} {\rm Br}(a)\equiv \lambda_a$.

However, if one does not keep track of the total number of semileptonic events, and marginalizes over $N_{\rm sl}$,
the PDF reduces to a pure Poisson distribution:
\begin{align}
	\notag
    P(N_a)&=
    \sum_{N_{\rm sl}=N_a}^\infty P(N_{\rm sl},N_a)\\
	\notag
    &=\sum_{k=0}^\infty \frac{{\rm e}^{-\lambda_{\rm sl}}\lambda^{N_a+k}_{\rm sl}}{N_a! k!}
   {\rm Br}(a)^{N_a} (1-{\rm Br}(a))^{k}\\
   &=\frac{(\lambda_{\rm sl} {\rm Br}(a))^{N_a}}{N_a!}  {\rm e}^{-\lambda_{\rm sl}}
   \sum_{k=0}^\infty \frac{[\lambda_{\rm sl} (1-{\rm Br}(a))]^k}{k!} = \frac{{\rm e}^{-\lambda_a} \lambda_a^{N_a}}{N_a!}\,.
\end{align}
The variance of $N_a$ is then equal to its expectation value: ${\rm Var}(N_a)=\langle N_a \rangle =\lambda_a$.

\paragraph{Precision of measuring $U_a^2/U^2$.}
The main quantity to calculate is the expected value of ${\rm Var}(U_a^2/U^2)$.
Here one may use the usual propagation of error, however, with the caveat that $N_{\rm sl}$ is not independent of $N_a$, which results in:
\begin{align}
    \frac{\delta (U_a^2/U^2)}{U_a^2/U^2} = \sqrt{\frac{1}{N_a}-\frac{1}{N_{\rm sl}}}\,.
\end{align}
Since the propagation of error assumes large event numbers, for small $N_{\rm sl}$ one has to calculate the expected value and variance of ${\rm Var}(N_a/N_{\rm sl})$ from the full PDF,
\begin{align}
    \langle N_a/N_{\rm sl} \rangle = \sum_{N_{\rm sl}=0}^\infty \sum_{N_a=0}^{N_{\rm sl}} P(N_{\rm sl},N_a) \frac{N_a}{N_{\rm sl}} = {\rm Br}(a) = \frac{\langle N_a \rangle}{\langle N_{\rm sl} \rangle}
    \,,
\end{align}
and similarly,
\begin{align}
	\notag
    &\langle N_a^2/N_{\rm sl}^2 \rangle = \sum_{N_{\rm sl}=0}^\infty \sum_{N_a=0}^{N_{\rm sl}} P(N_{\rm sl},N_a) \left(\frac{N_a}{N_{\rm sl}}\right)^2\\
    &= {\rm Br}(a)^2
    +{\rm e}^{-\lambda_{\rm sl}} (1-{\rm Br}(a))
    [1-{\rm Br}(a)(-1+\gamma_E+\Gamma(0,-\lambda_{\rm sl})+\log(-\lambda_{\rm sl}))]\,,
\end{align}
with the Euler constant $\gamma_E\approx 0.58$.
The expected sensitivity is then given as:
\begin{align}
	\notag
    \frac{\delta (U_a^2/U^2)}{U_a^2/U^2} &=
    \sqrt{
    \frac{
    {\rm e}^{-\langle N_{\rm sl} \rangle} (\langle N_{\rm sl}\rangle -\langle N_a \rangle)
    [\langle N_{\rm sl}\rangle + \langle N_a \rangle (1-\gamma_E-\Gamma(0,-\langle N_{\rm sl}\rangle)
    -\log(-\langle N_{\rm sl}\rangle) ]
    }{\langle N_a \rangle^2}
    }\\
    &\approx
    \sqrt{\frac{1}{\langle N_a \rangle}-\frac{1}{\langle N_{\rm sl}\rangle}}\,.
\end{align}
In the large $\langle N_{\rm sl} \rangle$ limit this equation agrees with the result obtained through error propagation.

The vanishing uncertainty in the $\langle N_a \rangle \rightarrow \langle N_{\rm sl} \rangle$ limit might seem concerning. However, note that this is the a-priori uncertainty. In other words, provided that the parameters really are such that $\langle N_a \rangle = \langle N_{\rm sl} \rangle$, we do not expect any events in the other channels, i.e. the uncertainty of $\langle N_a/N_{\rm sl} \rangle$ vanishes.


\section{Derivation of the evolution equations}
\label{Appendix_3}

The closed-time-path (CTP) formalism of non-equilibrium quantum field theory~\cite{Schwinger:1960qe,Keldysh:1964ud,Calzetta:1986cq} offers a convenient way to derive quantum kineitc equations for the evolution of charge densities in the early universe. 
A detailed derivation of the rate equations~(\ref{sec3:eq:diff_eq_RHN}, \ref{sec3:eq:diff_eq_SM}) in this formalism is given in the appendix of ref.~\cite{Drewes:2016gmt}.
However, the authors neglected the contribution from LNV processes to the kinetic equations, which we include in the present work. 
In the following we summarise the differences with the derivation given in ref.~\cite{Drewes:2016gmt} if LNV processes are taken into account. An alternative derivation can e.g. be found in ref.~\cite{Ghiglieri:2017gjz}.

\subsection{Quantum kinetic equations for the heavy neutrinos}
\paragraph{Nonequilibrium correlation functions} - In the CTP formalism, all observables can be expressed in terms of correlation functions. For the heavy neutrinos, the relevant correlation functions are the spectral and statistical propagators 
\begin{align}
\ii \mathcal{S}_{N ij}^{\mathcal{A}}(x_{1},x_{2})_{\alpha\beta}&\equiv \frac{\ii}{2}\left(
\langle {\rm N}_{i,\alpha}(x_{1})\bar{\rm N}_{j,\beta}(x_{2})\rangle
+\langle \bar{\rm N}_{j,\beta}(x_{2}){\rm N}_{i,\alpha}(x_{1})\rangle
\right)\,,
\\
\ii \mathcal{S}_{N ij}^{+}(x_{1},x_{2})_{\alpha\beta}&\equiv \frac{1}{2}\left(
\langle {\rm N}_{i,\alpha}(x_{1})\bar{\rm N}_{j,\beta}(x_{2})\rangle
-\langle \bar{\rm N}_{j,\beta}(x_{2}){\rm N}_{i,\alpha}(x_{1})\rangle
\right)\,.
\label{SMinus}
\end{align}
Here ${\rm N}_i$ are the Majorana spinors of the heavy neutrinos in the symmetric phase of the SM, i.e., the eigenvectors of $M_M$. 
In the following, we suppress the spinor indices $\alpha,\beta$, and $i,j$ are heavy neutrino flavour indices.
We can separate $S_N^+$ into an equilibrium part $\bar{S}_N^+$ and a deviation $\delta S_N$, i.e., $S_N^+=\bar{S}_N^+ + \delta S_N$ and decompose 
\begin{eqnarray}
\label{eq:tensor_dec_propagator}
-\ii \gamma^0 \delta S_N = \sum_h \frac{1}{2} P_h \left(g_{0h} + \gamma^0 g_{1h} - \ii \gamma^0\gamma^5 g_{2h} -\gamma^5 g_{3h}\right)\,,
\end{eqnarray}
with the helicity projectors 
\begin{equation}
P_h \equiv \frac{1}{2}\left(1+h\hat{\bf k}\gamma^0\pmb{\gamma}\gamma^5\right)\,.
\end{equation}
The equilibrium propagator $\bar{S}_N$ can be decomposed into functions $\bar{g}_{ih}$ in exactly the same way.
In order to relate the functions $g_{ih}$ to on-shell quasiparticle occupation numbers, a number of approximations are necessary. We follow the same steps as in appendix B of ref.~\cite{Drewes:2016gmt}, with the exception that we keep terms up to first order in $\bar{M}^2/T^2$. 
In the symmetry protected regimes with $\mu\ll1$ that we consider in this work, we can still neglect terms involving $\Delta M$ in the \emph{constraint equation} that determines the quasiparticle dispersion relations $\Omega_i$. In addition, we also neglect ``thermal masses'' and set $\Omega_i^2=\textbf{k}^2+\bar{M}^2$. Physically this corresponds to the assumption that these corrections are kinematically negligible. 
We emphasise that we do not neglect $\Delta M$ and the thermal masses in the \emph{kinetic equations}, where they are absolutely crucial for the flavour oscillations.\footnote{Roughly speaking, this correpsonds to keeping $\Delta M$ and thermal corrections in the numerator of the propagators and neglecting them in the denominator, see also refs.~\cite{Garbrecht:2011aw,Fidler:2011yq,Drewes:2016gmt} for a more detailed discussion.}  
This leads to the approximate relations
\begin{align}
\label{eq:constraint_propagator}
g_{1h}&=\frac{1}{2k^0}\left( \{\Re\, M,g_{0h}\} +
[\ii \Im\, M,g_{3h}] \right)\,, \\
g_{2h}&=\frac{1}{2\ii k^0}\left([\Re\, M,g_{3h}]+\{\ii \Im\, M,g_{0h}\} \right) \,,\\
g_{3h}&=h\frac{|\textbf{k}|}{k_0}g_{0h}\,,
\end{align}
which can be compared to eqs.~(B.30) in ref.~\cite{Drewes:2016gmt}.
They allow to express all Lorentz components in terms of the flavour space matrices $g_{0h}$, 
\begin{align}
	\ii \delta S_{N\,h} &= -\frac{1}{2 k_0} P_h\left(\slashed{k}g_{0h} + \{M,g_{0h}\}/2 - h \frac{|\k|}{2 k_0} [M,g_{0h}]\right)\,.
\end{align}
Using the on-shell approximations 
\begin{align}
\label{eq:quasiparticle_app_bar}
\bar{g}_{0h}(k)_{ij}&\approx
-2\pi\frac{1-2f^{\rm eq}(k)}{2} \delta_{ij} 
2 k^0 \delta(k_0^2-\Omega_i^2)\,,\\
\label{eq:quasiparticle_app}
g_{0h}(k)_{ij}&\approx 2\pi  \delta f_{h}(k)_{ij} 2 k^0 \delta(k_0^2-\Omega_i^2)  \,,
\end{align}
we find in the mass degenerate case
\begin{align}
\notag
	\ii \delta S_{N\,h}
	&=-2\pi \delta(k^2-\bar{M}^2) P_h \left( \slashed{k} \delta f_{h} + \frac{1}{2}\{M,\delta f_{h}\} - h \frac{|\k|}{2 k_0} [M,\delta f_{h}] \right)\\
	&=-2\pi \delta f_{h}\delta(k^2-\bar{M}^2) P_h \left( \slashed{k} + \bar{M} \right) + \mathcal{O}(\Delta M)\,.
\end{align}
In the comoving frame, the equilibrium distribution functions of fermions are given by
\begin{align}
	f^{\rm eq} (k) \equiv \frac{1}{\ee^{|\k|/a_{\rm R}}+1}\,.
\end{align}

\paragraph{On-shell kinetic equations} - The evolution equation for the flavour matrices $\delta f_{h}$ can be obtained by the procedure presented in ref. \cite{Drewes:2016gmt}, using eqs. (\ref{eq:quasiparticle_app_bar}, \ref{eq:quasiparticle_app}). At leading order in the chemical potentials for leptons ($\mu_\ell$) and the Higgs  ($\mu_\phi$) it reads 
\begin{align}\label{KinEqFor_f}
\frac{\dd}{\dd z}\delta f_{0hij}=-\frac{\ii}{2}\left[{\rm H}_N,\delta f_{h}\right]_{ij}
-\frac12\left\{\Upgamma_N,\delta f_{h}\right\}_{ij}
+\frac12\sum_{a=e,\mu,\tau}\frac{\mu_{\ell_a}+\mu_\phi}{T}(\tilde{\Upgamma}_N^a)_{ij}\,,
\end{align}
where the indices $i,j$ are the heavy neutrino flavours.\footnote{
The careful reader may notice that the equation (\ref{KinEqFor_f}) in principle should contain terms that are proportional to $\delta f_h \times (\mu_{\ell_a}+\mu_\phi)/T$, which would lead to terms  $\propto \delta n_h \Delta_a$ in eqns.~(\ref{sec3:eq:diff_eq_RHN},\ref{sec3:eq:diff_eq_SM}).
These terms do not appear in the derivation we follow here because terms $\delta\mathcal{G}^\gtrless \delta\mathcal{S}$ were neglected in eq.~(B.19) in ref.~\cite{Drewes:2016gmt}.
While such terms are "double small" in conventional thermal leptogenesis scenarios, they are not parametrically smaller than the other terms involving the chemical potentials for freeze-in leptogenesis because the helicity even combination $\delta f^{\rm even}$ of the $\delta f_h$  in eq.~(\ref{def:feven}) is not small at early times (while the helicity odd combination in eq.~(\ref{def:fodd}) that defines the "sterile charges" in eq.~(\ref{eq:qNdef}) is indeed small at all times). 
Neglecting the helicity even part practically leads to an overestimate of the backreaction and washout terms at early times.
For the phenomenological analysis presented here, we expect the resulting error to be small.
In the overdamped regime with strong washout the flavour combination of $\delta f^{\rm even}$ that is still or order unity when the chemical potentials become sizeable comes with tiny Yukawa couplings $\epsilon_a$ in (\ref{definitionderepsilona}), while leptogenesis in the experimentally accessible part of the oscillatory regime does not make predictions for the $U_a^2/U_b^2$.
}
The effective Hamiltonian ${\rm H}_N$ can be decomposed into a vacuum mass term ${\rm H}_N^{\rm vac} $ and  a thermal correction ${\rm H}_N^{\rm th}$ from forward scatterings and coupling to the temperature dependent Higgs field expectation value, such that $({\rm H}_N)_{ij} = ({\rm H}_N^{\rm vac})_{ij} + ({\rm H}_N^{\rm th})_{ij}$.
${\rm H}_N$ is responsible for the heavy neutrino flavour oscillations that occur due to the misalignment between their vacuum mass matrix $M$ and $\Upgamma_N$.
The thermal part ${\rm H}_N^{\rm th}$ as well as the thermal damping rates $\Upgamma_N$ and $\tilde{\Upgamma}_N^a$ can be expressed in terms of the Hermitian self-energy $\Sigma_N^{H}$ and the anti-Hermitian (or ``spectral'') self-energy $\Sigma_N^{\mathcal{A}}$, respectively. 
Up to numerical prefactors, these can be identified with the real and imaginary part of the usual retarded self-energy.
We split both self energies into an equilibrium part and a deviation, $\Sigma_N = \bar{\Sigma}_N + \delta \Sigma_N$.
Assuming that all SM degrees of freedom are in kinetic equilibrium, we can express $\delta \Sigma_N$ in terms of chemical potentials $\mu_{\ell_a}$ and $\mu_\phi$ for the SM leptons and Higgs field.  
It is convenient to introduce the quantities
\begin{align}\label{hatSigmaNDef}
\bar{\slashed{\Sigma}}_N&= g_w \hat{\slashed{\Sigma}}_N \left(
Y^* Y^t P_{\rm R}
+
Y Y^\dagger P_{\rm L}
\right) 
\,
\end{align}
for the equilibrium parts.
This allows to express the effective Hamiltonian as
\begin{align}
{\rm H}_N^{\rm vac}& = \frac{z^2 a^2_{\rm R}}{T_{\rm ref}^3|\k|}\left( \Re [M^\dagger M] +
\ii h \Im [M^\dagger M] \right)\,,
\\
{\rm H}_N^{\rm th}& = 2 \frac{g_w}{T_{\rm ref}} \left(
{\rm Re}[Y^*Y^t]\frac{k\cdot\hat{\Sigma}_N^{H}}{k^0}
-{\rm i}h {\rm Im}[Y^*Y^t]\frac{\tilde{k}\cdot \hat \Sigma_N^{H}}{k^0}
\right)
+2{\rm Re}[Y^*Y^t]\frac{a_{\rm R}^2}{k^0}\frac{z^2 v^2(z)}{T_{\rm ref}^3}\,.
\end{align}
Here we have introduced the vector $\tilde{k}=(|\k|,k^0\hat{k})$ that is orthogonal to $k$ in Minkwoski space, $\tilde{k}\cdot k = k\cdot \tilde{k} = 0$, $-\tilde{k}^2 = k^2 =M_i^2$ with $M_i$ the mass of the heavy neutrinos. We approximate the temperature dependent expectation value of the Higgs field as in ref.~\cite{Drewes:2016gmt} by
\begin{align}
\label{app3:eq:Higgs_EV}
\frac{z^2 v^2(z)}{T_{\rm ref}^2} \approx (-3.5+4.4 z) \theta(z-z_v)\,,
\end{align}
where $z_v \approx 0.8$ is the characteristic time where the Higgs expectation value starts to differ from zero. 
Further, $\Upgamma_N$ is the damping rate of the deviations $\delta f_{h}$ towards equilibrium and reads
\begin{align}
\Upgamma_N =2 \frac{g_w}{T_{\rm ref}} \left(
{\rm Re}[Y^*Y^t]\frac{k\cdot \hat \Sigma_N^{\mathcal{A}}}{k^0}
-{\rm i}h {\rm Im}[Y^*Y^t]\frac{\tilde{k}\cdot \hat \Sigma_N^{\mathcal{A}}}{k^0}
\right)\,.
\end{align}
The term
\begin{align}
(\tilde{\Upgamma}_N^a)_{ij} &=  2 h
g_w \left(
{\rm Re}[Y^*_{ia}Y^t_{aj}]\frac{ \tilde{k}\cdot \hat \Sigma_N^{\mathcal{A}}}{k^0}-\ii h {\rm Im}[Y^*_{ia}Y^t_{aj}]\frac{k\cdot \hat \Sigma_N^{\mathcal{A}}}{k^0}
\right)
f^{\rm eq}(k)[1-f^{\rm eq}(k)] \frac{T}{a_{\rm R}}\,,
\label{gamma:backreaction}
\end{align}
describes the backreaction of the matter-antimatter asymmetries in the plasma of SM particles on the evolution of the $N_i$.
We use the notation of helicity-even and helicity-odd parts of the deviations from equilibrium
\begin{align}
\delta f^{\rm even}(k)&=\frac{\delta f_{+}(k)+\delta f_{-}(k)}{2}\,,\label{def:feven}
\\
\delta f^{\rm odd}(k)&=\frac{\delta f_{+}(k)-\delta f_{-}(k)}{2}\,.\label{def:fodd}
\end{align}
The relativistic approximation $|\textbf{k}|/k_0={\rm sign}(k_0)$ that was adopted in ref.~\cite{Drewes:2016gmt} allowed to express all rates in terms of the quantity $\gamma(k)=\frac{2 g_w}{a_{\rm R}} \frac{k\cdot\hat{\Sigma}^\mathcal{A}_N}{k^0}$, which conserves lepton number. When allowing for the next non-vanishing order $\mathcal{O}(M_i^2/|\k|^2)$, one finds that there are both, lepton number conserving and lepton number violating contributions to the damping rates. 
We refer to the lepton number conserving coefficient (which corresponds to $\gamma(k)$ in ref.~\cite{Drewes:2016gmt}) as $\gamma_+(k)$ in the following, and to the lepton number violating coefficient as $\gamma_-(k)$. At zeroth order in $\mathcal{O}(M_i^2/|\k|^2)$, $\gamma_-(k)$  vanishes. These rates are given by
\begin{align}
\gamma_+(k) &= \frac{1}{a_{\rm R}}\frac{g_w}{k^0}\left(k+\tilde{k}\right)\cdot \hat{\Sigma}^\mathcal{A}_N \approx  \frac{2}{a_{\rm R}}\frac{g_w}{|\k|}\left(k\cdot \hat{\Sigma}^\mathcal{A}_N\right)\,,
\\
\gamma_-(k) &= \frac{1}{a_{\rm R}}\frac{g_w}{k^0}\left(k-\tilde{k}\right)\cdot \hat{\Sigma}^\mathcal{A}_N \approx \frac{1}{2a_{\rm R}}\frac{g_w}{|\k|}\frac{M_i^2}{|\k|} \left(\hat{\Sigma}^{\mathcal{A}0}_N+\hat{k}_i\hat{\Sigma}^{\mathcal{A}i}\right)\,.
\end{align}
Analogously we find for the Hermitian part
\begin{align}
\mathfrak{h}_+^{\rm th}(k) &= \frac{1}{a_{\rm R}}\frac{g_w}{k^0}\left(k+\tilde{k}\right)\cdot \hat{\Sigma}^H_N \approx  \frac{2}{a_{\rm R}}\frac{g_w}{|\k|}\left(k\cdot \hat{\Sigma}^H_N\right)\,,
\\
\mathfrak{h}_-^{\rm th}(k) &= \frac{1}{a_{\rm R}}\frac{g_w}{k^0}\left(k-\tilde{k}\right)\cdot \hat{\Sigma}^H_N \approx \frac{1}{2a_{\rm R}}\frac{g_w}{|\k|}\frac{M_i^2}{|\k|} \left(\hat{\Sigma}^{H 0}_N+\hat{k}_i\hat{\Sigma}^{Hi}\right)\,,
\end{align}
and the term accounting for the effect of the Higgs field expectation value
\begin{align}
\mathfrak{h}^{\rm EV}(k)= \frac{2}{k^0}\frac{z^2 v^2(z)}{T_{\rm ref}^2}a_{\rm R}\,.
\end{align}

\paragraph{Evolution equations for number densities.} To allow for a fast numerical exploration of the parameter space, we use momentum averaged rate equations in this work. The equilibrium number density $n^{\rm eq}$ of the $N_i$  and the deviation $\delta n_{hij}$ from it 
which appear in the coupled system of differential equations (\ref{sec3:eq:diff_eq_RHN}, \ref{sec3:eq:diff_eq_SM})
are given by
\begin{align}
n^{\rm eq}&=\int\frac{{\rm d}^3 k}{(2 \pi)^3} f^{\rm eq}(k) = \frac{3}{4 \pi^2}a_{\rm R}^3 \zeta(3)\,,
\\
\delta n_{hij}&=\int\frac{{\rm d}^3 k}{(2 \pi)^3} \delta f_{h ij}(\k)\label{deltanij}\,.
\end{align} 
In order to obtain an equation in terms of $n^{\rm eq}$ and $\delta n_{hij}$, we face the usual problem of approximating an integral over a product by a product of integrals, because of
the fact that the distributions $f^{\rm eq}$ and $\delta f_{h ij}$   appear together with another quantity that depends on the momentum $\k$ on the RHS of the kinetic equation (\ref{KinEqFor_f}).
To describe the two momentum averages, for a rate $X(k)$ we introduce
\begin{align}
	X &\equiv \frac{1}{\delta n} \int\frac{{\rm d}^3 k}{(2 \pi)^3}  \delta f_{hij}(k) X(k)\,,\\
	\tilde{X} &\equiv \frac{1}{n^{\rm eq}}\int\frac{{\rm d}^3 k}{(2 \pi)^3}  f^{\rm eq}(k)[1-f^{\rm eq}(k)] X(k)\,.
\end{align}
We may use two different approximation strategies depending on the dependence on $k$ and whether we are integrating over $f^{\rm eq}$, or $\delta f_{hij}$
to obtain the system of kinetic equations 
(\ref{sec3:eq:diff_eq_RHN}, \ref{sec3:eq:diff_eq_SM}):
\begin{enumerate}[(i)]
	\item {\bf Averaging with equilibrium weights.}
		\label{avgeqweight}
		The expressions for the backreaction term (\ref{gamma:backreaction}), as well as the washout term of the active charges appear multiplied with the equilibrium distribution functions of the right handed neutrino and charged leptons.
		The backreaction and washout rates will therefore be governed by the lepton number conserving rate
		\begin{align}
			\tilde{\gamma}_+ \approx \frac{1}{n^{\rm eq}}\int\frac{{\rm d}^3 k}{(2 \pi)^3}  f^{\rm eq}(k) \gamma_+(k)\equiv \gamma_+^{\rm av}=0.012\,,
		\end{align}
		where we used the observation that one may neglect the contribution of order $[f^{\rm eq}(k)]^2$ if the rates are not infrared enhanced by a power smaller 
		than $k^{-2}$ without introducing an error of more than $\mathcal{O}(40\%)$, which allows us to use the production rate from~\cite{Garbrecht:2014bfa}
		based on refs.~\cite{Besak:2012qm,Garbrecht:2013urw}.
		The lepton number violating rate is infrared enhanced, which means that we have to use the full expression
		\begin{align}
			\tilde{\gamma}_- = \frac{1}{n^{\rm eq}}\int\frac{{\rm d}^3 k}{(2 \pi)^3}  f^{\rm eq}(k)[1-f^{\rm eq}(k)] \gamma_-(k)\equiv \gamma_-^{\rm av} =
			1.9 \times 10^{-2} z^2 \frac{\bar{M}^2}{T^2}\,.
		\end{align}
		Similarly, if one assumes kinetic equilibrium for the right-handed neutrinos, i.e. that the non-equilibrium distribution of the right-handed neutrinos is proportional to the Fermi-Dirac distribution $\delta f \approx \frac{\delta n}{n^{\rm eq}} f^{\rm eq}$, we can calculate the rates
		\begin{align}
		\label{eqavg}
		\nonumber
		\gamma_+ &= \frac{1}{\delta n} \int\frac{{\rm d}^3 k}{(2 \pi)^3}  \delta f_{hij}(k) \gamma_+(k)\\
		&\approx \frac{1}{n^{\rm eq}} \int\frac{{\rm d}^3 k}{(2 \pi)^3} f^{eq}(k) \gamma_+(k) = \gamma_+^{\rm av}\,,\\
		\nonumber
		\mathfrak{h} &\equiv \left\langle\mathfrak{h}(k) \right\rangle = \frac{1}{n^{\rm eq}}\int\frac{{\rm d}^3 k}{(2 \pi)^3}  f^{\rm eq}(k) \mathfrak{h}(k)\,,
		\\
		\left\langle\frac{1}{|\k|}\right\rangle &= \frac{1}{n_{\rm eq}}\int\frac{{\rm d}^3 k}{(2 \pi)^3} f^{\rm eq}(k) \frac{1}{|\k|}=\frac{\pi^2}{18 T\zeta(3)}\,,
		\end{align}
		where $\mathfrak{h}$ is either $\mathfrak{h}^{\rm th}_\pm$ or $\mathfrak{h}^{\rm EV}$.
	\item {\bf Evaluating the rate at the average momentum.}
		The rate $\gamma_-$ requires a separate treatment in the overdamped regime. When one allows for lepton number violating processes, the equilibration of
		the weakly coupled state can proceed not only via the mixing with the strongly coupled state, but also through Higgs decays, which are
		suppressed by a factor $z^2 M^2/k^2$. This causes an enhancement in production for small momenta $k$, due to which the deviation from equilibrium
		$\delta f$ may significantly deviate from the kinetic equilibrium assumption used in the momentum averaging procedure required to calculate~(\ref{eqavg}).
		Therefore, as for a momentum mode which can realistically represent the production of the right-handed neutrinos we instead choose the average momentum
		\begin{align}
		|\k_{\rm av}| = \frac{1}{n_{\rm eq}}\int\frac{{\rm d}^3 k}{(2 \pi)^3} f^{\rm eq}(k) |\k| = \frac{7\pi^4 T}{180 \zeta(3)}\approx 3.15 T\,.
		\end{align}  
		of the heavy neutrinos, such that
		\begin{align}
		\int\frac{{\rm d}^3 k}{(2 \pi)^3} \delta f(k) \,\gamma_-(k) \approx \delta n  \,\gamma_-(|\k_{\rm av}|)\equiv  \delta n \gamma_-^{|\k_{\rm av}|}\,.
		\end{align}
\end{enumerate}
With these approximations we obtain the momentum independent equations (\ref{sec3:eq:diff_eq_RHN}) and (\ref{sec3:eq:diff_eq_SM}) for the heavy neutrinos and the SM charges.

\subsection{Evolution equations for the SM lepton charges}
The gauge interactions among the SM fields are fast and keep them in kinetic equilibrium.
This allows to describe the slowly evolving deviations of the SM fields from thermal equilibrium by chemical potentials, which can be expressed in terms of the charges introduced in section~\ref{Section3_Leptogenesis} through the approximate linear relations 
\begin{eqnarray}
\label{ChargeDefs}
q_X=\left\{\begin{array}{l}\frac{a_{\rm R}^2}{3}\mu_X\;\textnormal{for massless bosons}\\\frac{a_{\rm R}^2}{6}\mu_X\;\textnormal{for (massless) chiral fermions}\end{array}\right.\,,
\end{eqnarray}
cf. e.g. appendix A of ref.~\cite{Laine:2008pg}.
The evolution of the lepton doublets is described by the following differential equation
\begin{align}
\frac{\dd}{\dd z}q_{\ell a}=&-
\frac{\tilde{\gamma}_+ + \tilde{\gamma}_-}{g_w} \frac{a_{\rm R}}{T_{\rm ref}}\sum_{i} |Y_{ia}|^2
\,\left(q_{\ell a}+\frac12 q_\phi\right)+\frac{S_a(\delta n_{hij})}{T_{\rm ref}}\,,
\end{align}
where the first term corresponds to the \emph{washout term} 
and the second term is the \emph{source term} $S_a\equiv S_{aa}$, defined as 
\begin{align}
S_{ab}&=\int \frac{\dd^3 k}{(2\pi)^3}\mathcal{S}_{ab} (k) =
-\sum_{i,j} Y_{ia}^* Y_{jb}
	\int \frac{\dd k^4}{(2\pi)^4} {\rm tr} \left[ P_R \ii \delta S_{N\,ij}(k) 2 P_L \hat{\slashed{\Sigma}}^{\mathcal A}_N \right].
\end{align}
It is fed by the correlations $\delta n_{ij}$. When accounting for lepton number violating effects the source term is slightly modified when compared to the purely flavoured source in ref.~\cite{Drewes:2016gmt}
\begin{align}
\label{app3:eq:source_k}
\mathcal{S}_{ab} (k)
&= \sum_{i,j} Y_{ia}^*Y_{jb}\sum_{s_k=\pm} 2\left[\frac{k\cdot\hat\Sigma^\mathcal{A}_N}{\sqrt{|\k|^2+M^2}}\delta f_{ij}^{\rm even}+\frac{\tilde{k}\cdot\hat\Sigma^\mathcal{A}_N}{\sqrt{|\k|^2+M^2}}\delta f_{ij}^{\rm odd}\right]
\\\notag
&= 2\frac{a_{\rm R}}{g_w}\sum_{i,j} Y_{ia}^*Y_{jb} \sum_{s=\pm} \gamma_s (k) \left[\ii {\rm Im}(\delta f_{ij}^{\rm even})+s  {\rm Re}(\delta f_{ij}^{\rm odd})\right]\,.
\end{align}
As mentioned previously, we evaluate $\gamma_-(k)$ at the average momentum $|\k_{\rm av}| \approx 3.15 T$, while for $\gamma_+(k)$ we use the approximation from ref.~\cite{Garbrecht:2014bfa}. In each case $\delta f$ is replaced by $\delta n$. The momentum independent expression is given by
\begin{align}
S_{ab}&= 2\frac{a_{\rm R}}{g_w}\sum_{i \neq j} Y_{ia}^*Y_{jb} \sum_{s=\pm} \gamma_s \left[\ii {\rm Im}(\delta n_{ij}^{\rm even})+s  {\rm Re}(\delta n_{ij}^{\rm odd})\right]\,.
\end{align}
It is the combination $(k-\tilde{k})\cdot \hat\Sigma^\mathcal{A}_N$ that gives rise to a term that violates lepton number. Note that although the source term, $\gamma_{\rm -}= \gamma_-^{|\k\rm av|}$ seems negligible compared to $\gamma_+= \gamma_+^{\rm av}$, it needs to be included as it violates the generalised lepton number $\tilde{L}$, and the lepton charge generated this way may not be deleted by the lepton number conserving washout as pointed out in refs.~\cite{Hambye:2016sby,Hambye:2017elz}.
\subsection{Effects from spectator fields}
It is known that spectator fields contribute to the chemical equilibration and redistribute charges during leptogenesis, such that some asymmetries are hidden from the washout.
The charge densities of the leptons $q_{\ell}$ and the Higgs field $q_\phi$, as well as the baryon charge density can be expressed in terms of the asymmetries $\Delta_a$: 
\begin{align}
q_\ell = A \Delta
\,,\quad 
q_\phi = C \Delta
\,,\quad
B=D \Delta
\,,
\end{align}
with the matrix $A$ and the vectors $C$, $D$ 
\begin{align}
A=
\frac{1}{711}
\left(
\begin{array}{ccc}
-221 & 16 & 16\\
16 & -221  & 16\\
16 & 16 & -221
\end{array}
\right)
\,,
\quad
C=
-\frac{8}{79}
\left(
\begin{array}{ccc}
1 & 1 & 1
\end{array}
\right)
\,,
\quad
D=
\frac{28}{79}
\left(
\begin{array}{ccc}
1 & 1 & 1
\end{array}
\right)\,.
\end{align}
\subsection{Determination of the transport coefficients}
\begin{figure}
\begin{center}
\includegraphics[width=0.6\textwidth]{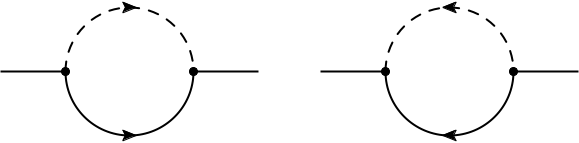}
\end{center}
\caption{Feynman diagrams contributing to the self-energy of the heavy neutrino, both for its thermal part $\hat{\Sigma}^{H}_N$ and its spectral part $\hat{\Sigma}^{\mathcal{A}}_N$. The outer lines indicate the heavy neutrino, while the dashed and the solid lines in the loop correspond to the Higgs field and the SM lepton doublets.}
\label{app3:Feynman}
\end{figure}

In this subsection we provide some details about the computation of the heavy neutrino dispersion relation and damping rates.
\paragraph{Thermal correction to the heavy neutrino mass} - The momentum dependent thermal correction is defined by
\begin{align}
\mathfrak{h}^{\rm th}_\pm(k)=\frac{g_w}{k^0}(k\pm\tilde{k})\cdot \hat{\Sigma}^H_N\,,
\end{align}
with $\hat{\Sigma}^H_N$ the Hermitian reduced self-energy of the heavy neutrinos.
At leading order it can be computed from the Feynman diagram shown in figure~\ref{app3:Feynman}.
We neglect the thermal masses of the leptons  and the Higgs running in the loop.
The analytic expression for the Hermitian self-energy $\hat{\Sigma}^H_N$ within the hard-thermal-loop approximation has been derived in ref.~\cite{Garbrecht:2013urw}
\begin{align}
\hat{\Sigma}^{H0}_N(k)&=\frac{T^2}{16|\k|}\log \bigg|\frac{k^0+|\k|}{k^0-|\k|}\bigg|\, ,\\
\hat{\Sigma}^{Hi}_N(k)&=\frac{T^2k^0k^i}{16|\k|^3}\log \bigg|\frac{k^0+|\k|}{k^0-|\k|}\bigg|-\frac{T^2k^i}{8|\k|^2}\,.
\end{align} 
Consequently, the momentum dependent thermal correction is given by
\begin{align}
\mathfrak{h}^{\rm th}_-(k)=g_w  \frac{T^2}{16} \frac{M_i^2}{|\k|^2}\frac{1}{k^0}\left[-1+\log \bigg|\frac{k^0+|\k|}{k^0-|\k|}\bigg|\right]\,.
\end{align}
For the purpose of the evolution equations of the heavy neutrinos we are interested in momentum averaged expressions. An approximate analytic solution for heavy neutrinos with mass $\bar{M}$ is given by
\begin{align}
\notag
\mathfrak{h}^{\rm th}_-&=\int\frac{\dd^3 k}{(2\pi)^3}\mathfrak{h}^{\rm th}_-(k)f^{\rm eq}(|\k|) 
\\
&\approx
\left[3.50-0.47\log\left(z^2 \frac{\bar{M}^2}{T_{\rm ref}^2}\right)+3.47 \log^2\left(z \frac{\bar{M}}{T_{\rm ref}}\right)\right] \times 10^{-2} \times z^2 \frac{\bar{M}^2}{T_{\rm ref}^2} \,,
\end{align}
where the heavy neutrino mass $M_i$ acts as a regulator to the IR enhancement.
\paragraph{Damping rate} - 
In analogy to the thermal correction to the heavy neutrino masses, their equilibration rates can be defined via
\begin{align}
\gamma_\pm(k)=\frac{g_w}{k^0}(k\pm\tilde{k})\cdot \hat{\Sigma}^\mathcal{A}_N
\end{align}
and can also be computed from the Feynman diagram~\ref{app3:Feynman}.
Here $\gamma_-(k)$ is the momentum dependent lepton-number violating rate, while $\gamma_+(k)$ corresponds to the lepton number conserving rate.
The lepton number conserving rate in the regime $T\gg M_i$ is discussed in detail in refs.~\cite{Anisimov:2010gy,Besak:2012qm,Laine:2013lka,Garbrecht:2013urw,Garbrecht:2014bfa,Ghisoiu:2014ena}

In contrast to processes that contribute to $\gamma_+(k)$, where decays and inverse decays are subdominant with respect to $2\leftrightarrow 2$ scatterings, it has been assumed that for $\gamma_-(k)$ the $1\leftrightarrow 2$ decays and inverse decays dominate over the scatterings \cite{Anisimov:2010gy,Hambye:2016sby,Hambye:2017elz,Ghiglieri:2017gjz}. In preparation of this manuscript we have verified that $2\leftrightarrow 2$ scatterings give a subdominant contribution. In the Schwinger-Keldysh CTP formalism, rates for these $1\leftrightarrow 2$ processes can be derived from the self-energy
\begin{align}
\hat{\slashed\Sigma}^<_N(k)&=\int\frac{d^4p}{(2\pi)^4}\frac{d^4q}{(2\pi)^4} (2\pi)^4\delta^4(q-k+p) \ii \slashed{S}_\ell^<(p)\ii \Delta^<_\phi(q)\,,
\end{align}
when using  propagators in the zero-width limit
\begin{align}
\ii \slashed{S}_\ell^<(p)&\to -2\pi \delta(p^2-m_\ell^2)f_\psi(p^0){\rm sign}(p^0)\slashed{p}\,,\\
\ii \Delta_\phi^<(q)&\to 2\pi\delta(q^2-m_\phi^2)f_\phi(q^0){\rm sign}(q^0)\,,
\end{align}
and where $\hat{\slashed\Sigma}=\gamma_\mu\hat\Sigma^\mu$.
Here it is crucial to use the  modified dispersion relation of the Higgs field and the leptons due the their thermal masses, which give rise to different kinematic regimes in spite of the fact that the intrinsic particle masses vanish in the symmetric phase of the SM.
The thermal masses from the gauge interactions are equal for the leptons and Higgs, but due to the additional contribution that $m_\phi$ receives from the Higgs self-interaction and the couplings to fermions, one finds $m_\phi^2 > m_\ell^2$. For the following discussion it is useful to mention that the non-thermal masses of the Higgs boson and the leptons are zero since electroweak symmetry is still unbroken. Further, the thermal masses for the heavy neutrinos are suppressed by the Yukawa couplings compared to their Majorana masses $M_i$. For that reason, the relevant masses are given by $M_i$ and the thermal masses $m_\phi^2$, $m_\ell^2$. Since low scale leptogenesis occurs at $T\gg M_i$, only the hierarchy $m_\phi > M_i + m_\ell$ can be realised, which allows Higgs quasiparticles to decay into the heavy neutrino and a lepton.\footnote{More precisely: This sets an upper bound on the range of heavy neutrino masses where the treatment described here can be applied. It is worth mentioning  that this upper bound depends on the temperature of the plasma due to the running couplings.} When making use of the delta function, the full loop integral can be expressed  in terms of the one-dimensional integrals
\begin{align}
I_n(y)&=\int_{\omega_-}^{\omega_+} d x\, x^n \left[1-f_\psi(x)+f_\phi(y-x)\right]
\end{align}
with $n=0,1$. At this stage one can solve the remaining integral analytically
\begin{align}
\hat{\Sigma}^{\mathcal{A}0}_N(k)&=\frac{T^2}{16\pi|\k|} \left[I_1(-\omega_+)-I_1(-\omega_-)\right]\, ,\\ 
\hat{\Sigma}^{\mathcal{A}i}_N(k)&=\frac{T^2}{16\pi|\k|}\frac{k^i}{|\k|} \bigg[\frac{k^0}{|\k|}\left[I_1(-\omega_+)-I_1(-\omega_-)\right]-\frac{M_i^2+m_\ell^2-m_\phi^2}{2|\k|T}\left[I_0(-\omega_+)-I_0(-\omega_-)\right]\bigg]\,,
\end{align}
where
\begin{align}
I_0(\omega_\pm)&\equiv\log(1+\ee^{\beta \omega_\pm})-\log(-\ee^{\beta k^0}+\ee^{\beta \omega_\pm})\, ,\\
I_1(\omega_\pm)&\equiv x(\log(1+\ee^{\beta \omega_\pm})-\log(1-\ee^{\beta \omega_\pm-\beta k^0}))+\mathrm{Li}_2(-\ee^{\beta \omega_\pm})-\mathrm{Li}_2(\ee^{\beta \omega_\pm-\beta k^0})\, ,
\end{align}
which has already been derived in ref.~\cite{Garbrecht:2013urw}. 
The  integration limits of $I_{0,1}$ are determined by the kinematics of the on-shell particles in the loop, such that (cf. \cite{Drewes:2015eoa})
\begin{align}
\omega_\pm&=\frac{|k^0|}{2M_i^2}\left|M_i^2+m_\ell^2-m\phi^2\right|\\
&\pm\frac{\sqrt{(k^0)^2-M_i^2}}{2M_i^2}\sqrt{M_i^4+m_\ell^4
+m_\phi^4-2M_i^2m_\ell^2-2M_i^2m_\phi^2-2m_\ell^2m_\phi^2}\,.
\end{align}
Further, we put the heavy neutrino on its mass shell $k^2=M_i^2$. Note that we are interested in a momentum-averaged description of the coupled system of kinetic equations. The momentum averaged rate for heavy neutrinos with average mass $\bar{M}$ can by approximated by
\begin{align}
\gamma_-^{\rm av} \approx 1.9 \times 10^{-2} \times z^2 \frac{\bar{M}^2}{T_{\rm ref}^2}\,,
\end{align}
while the rate at the averaged momentum $\gamma_-^{|\k_{\rm av}|}$ is evaluated to 
\begin{align}
\gamma_-^{|\k_{\rm av}|}= 9.7 \times 10^{-4} \times z^2 \frac{\bar{M}^2}{T_{\rm ref}^2}\,.
\end{align}
Note that these results are in agreement with those derived in~\cite{Hambye:2017elz}.
The major physical difference comes from using the average-momentum result
$\gamma_-^{|\k_{\rm av}|}$ as the equilibration rate of the heavy neutrinos
instead of the momentum averaged value.
The remaining numerical $\mathcal{O}(10\%)$ differences in the rate most likely comes from the running of the couplings in the thermal masses $m_\phi$ and $m_\ell$ which we neglected.
\section{Numerical treatment of the rate equations}
\label{Appendix_4}
The differential equations describing the evolution of charges and number densities of the different particle species~(\ref{sec3:eq:diff_eq_RHN}, \ref{sec3:eq:diff_eq_SM}) is in general known to be stiff and numerically demanding as it can involve many different time scales.
In this section we briefly overview the methods that we used to tackle some of these issues.

\subsection*{Fast heavy neutrino oscillations}
As the universe expands and cools, the number of heavy neutrino oscillations per decay time generally increases. Therefore, the first few oscillations, will dominate the production of the lepton number.
As the temperature of the universe decreases, the vacuum mass starts to dominate, and soon after a critical time for which the thermal and vacuum masses are of the same size
\begin{align}
	z_c^2 |H_{\mathrm{vac}\,11}-H_{\mathrm{vac}\,22}| \simeq H_{\mathrm{th}11}\,,
\end{align}
we may neglect the thermal masses to determine the time of the first oscillation, which is given by:
\begin{align}
	z_\mathrm{osc}^3 |H_{\mathrm{vac}\,11}-H_{\mathrm{vac}\,22}|/3 \simeq 2 \pi\,.
\end{align}
Already for $z>\mathcal{O}(5) z_\mathrm{osc}$, the oscillation frequency becomes much higher than the temperature.
If many such oscillations happen during a decay time, one may replace the rapidly oscillating off-diagonal densities $\delta n_{i,j}|_{i\neq j}$ by their average value, which can be obtained by solving the equation:
\begin{align}
\label{fastosc:avg}
0 &= -\frac{\ii}{2}(H_{N ii}^\mathrm{vac}-H_{N jj}^\mathrm{vac})\delta n_{h\,ij} -\frac12 (\Gamma_{N\,12}\delta n_{h\,22} + \Gamma_{N\,21}\delta n_{h\,11})+\sum_{a,b=e,\mu,\tau}\tilde{\Gamma}_{Nij}^a (A_{ab} + C_b/2)\Delta_b\,,
\end{align}
for $i\neq j$ which is obtained from~(\ref{sec3:eq:diff_eq_RHN}) by neglecting $\delta n_{h\,ij}^\prime$, as well as terms of order $\delta n_{ij} \Upsilon$ for $i\neq j$.

Since $\delta n_{ij}$ now scales as $T\,\Gamma_N/(z^2 \bar{M} \Delta M)$ it is tempting to completely neglect these fast oscillations completely by setting $\delta n_{ij} \rightarrow 0$ for $i\neq j$, as they are subdominant
to the contribution given by the first few oscillations.
Therefore, we can neglect the fast oscillations when both $z > \mathcal{O}(5) z_c$ and $z >  \mathcal{O}(5) z_\mathrm{osc}$ are satisfied.

However, after one includes the lepton number violating rates, which
scales as $z^2 \bar{M}^2/T_{ref}^2$, we can observe that the lepton number violating part of the source which scales as $S_a\sim \delta n_12 \gamma_{-}^{\mathrm{av}}$ becomes constant, and it may not always be neglected.
We have compared the results of a parameter scan obtained by including the average value for $\delta n_{ij}$ from Eq.~\eqref{fastosc:avg} for a benchmark mass $\bar{M}=30 \textit{GeV}$,
and we have not observed a significant deviation from the parameter scan where it is completely neglected.


\bibliographystyle{JHEP}
\bibliography{all}

\end{document}